\documentclass[reprint, NumberedRefs]{JASA} 

 \usepackage{xcolor}

\usepackage[utf8]{inputenc}
\usepackage{mathtools}
\usepackage{tabularx}
\usepackage{chemformula}
\usepackage{comment}
\usepackage{makecell}
\usepackage{graphicx} 
\usepackage{upgreek}
\usepackage{xurl}
\usepackage{psfrag}
\usepackage{array}
\usepackage{booktabs}
\usepackage{amsmath}
\usepackage{siunitx}
\usepackage{chemformula}
\usepackage{adjustbox}
\usepackage{enumitem}
\usepackage{array}

\begin{document}

\title[Review of ultrasonic methods in LIB]{Review of ultrasonic methods for monitoring, damage detection, and processing of lithium-ion batteries throughout their life cycle} 
\author{Simon Montoya-Bedoya}
\email{simonmontoyabedoya@utexas.edu}
\affiliation{Walker Department of Mechanical Engineering,  The University of Texas at Austin, Austin, TX 78712-1591, USA}

\author{Tyler M.~McGee}
\email{tm34485@utexas.edu}
\affiliation{Walker Department of Mechanical Engineering,  The University of Texas at Austin, Austin, TX 78712-1591, USA}

\author{Joong Seok Lee}
\email{jsleecnu@cnu.ac.kr}
\affiliation{School of Mechanical Engineering, Chungnam National University, Daejeon 34134, South Korea}

\author{Sasha Litvinov}
\email{alitvinov@utexas.edu}
\affiliation{Walker Department of Mechanical Engineering,  The University of Texas at Austin, Austin, TX 78712-1591, USA}

\author{Ofodike A.~Ezekoye}
\email{dezekoye@mail.utexas.edu}
\affiliation{Walker Department of Mechanical Engineering,  The University of Texas at Austin, Austin, TX 78712-1591, USA}

\author{Donal P.~Finegan}
\email{donal.finegan@nrel.gov}
\affiliation{National Laboratory of the Rockies, 15013 Denver W Parkway, Golden, CO 80401, USA}
 
\author{Michael R.~Haberman}
\email{haberman@utexas.edu}
\affiliation{Walker Department of Mechanical Engineering,  The University of Texas at Austin, Austin, TX 78712-1591, USA}

\preprint{Montoya-Bedoya \textit{et al.}, Review of ultrasonics for lithium-ion batteries}		

\date{\today} 

\begin{abstract}
Lithium-ion batteries (LIBs) are the leading technology used in consumer electronics, electric vehicles, and grid-level electrochemical energy storage applications. The ever-increasing use of LIBs has highlighted a gap in understanding of their behavior throughout their life cycle. Current monitoring systems rely on electrical and sometimes temperature measurements to assess the internal state which limits information about complex electrochemical processes. In response, ultrasonic testing (UT) has shown promise for non-invasive assessment due to its ease of use and sensitivity to mechanical changes which are correlated with electrochemical changes within the battery. We summarize the research in UT methods applied to LIBs throughout their life cycle. We also discuss physics-based and data-driven modeling approaches used to interpret ultrasonic signals in the context of LIBs, with an emphasis on the existing challenge of establishing rigorous links between electrochemical behavior and elastic and poroelastic wave physics to gain insight regarding physical changes in the LIB that can be directly measured using UT. Finally, we discuss the challenges of implementing UT across the LIB life cycle and identify opportunities for further research. This review aims to provide helpful guidance to researchers and practitioners of UT in the growing field of UT for electrochemical battery systems.
\end{abstract}

\maketitle

\section{Introduction}\label{sec:Intro} 

\begin{figure}[ht]
    \centering
    \includegraphics[width = \reprintcolumnwidth]{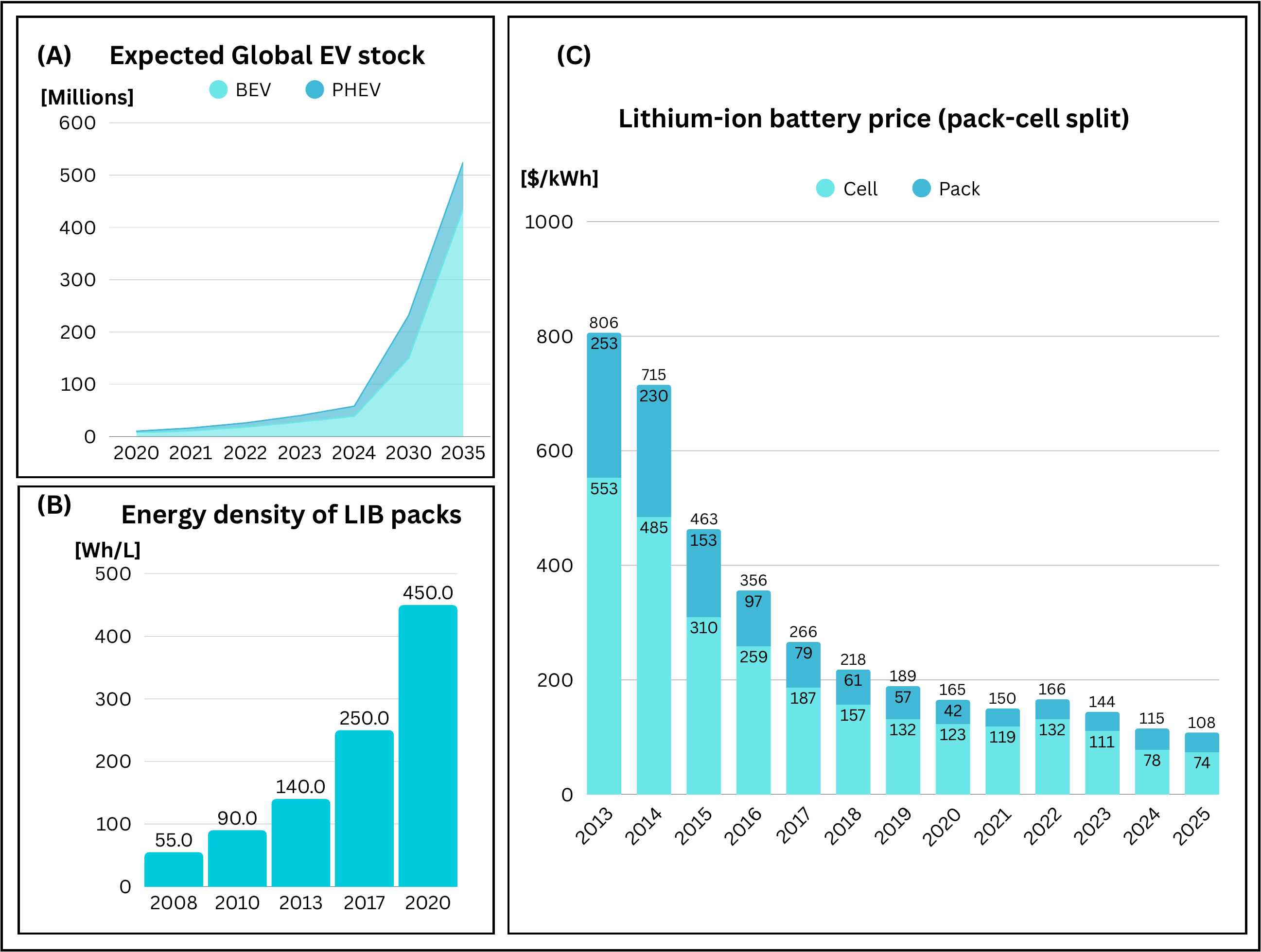}
    \caption{Some of the main drivers for the increased adoption of LIBs in EV applications: \textbf{(A)} forecasted global EV stock (BEV: Battery electric vehicle and PHEV: Plug-in hybrid electric vehicle), \textbf{(B)} energy density of LIB packs in EVs, and \textbf{(C)} price reduction of LIBs at pack and cell level. Data used to generate figures sourced from Refs. \cite{globalEV2025,bnefLithiumIonBattery, energyFOTW1234}}
    \label{fig:batteryMetrics}
\end{figure}

Lithium-ion batteries (LIBs) have been used to power consumer electronics devices such as cellular phones and laptops since their first commercialization by Sony in 1991. With few widely-publicized exceptions \cite{yun2018benefits}, LIBs have achieved a long history of success in these applications. Due to recent advances in energy density \cite{placke2017lithium}, battery management systems (BMS) for safe operation \cite{lipu2021intelligent}, and cost \cite{mauler2021battery}, LIBs have become viable power sources for high-power and high-capacity applications such as electric vehicles (EVs) and renewable energy storage systems (ESS). Their viability in these applications, coupled with their role in transitioning to zero tailpipe emissions, has resulted in rapid growth in EV sales. From 2020 to 2024, the worldwide EV stock grew from \SI{10.2}{million} to \SI{58}{million} and is expected to grow to \SI{525}{million} by 2035 \cite{globalEV2025} (see Fig.~\ref{fig:batteryMetrics}.A). The primary drivers of this growth have been the approximately \SI{720}{\percent} increase in pack-level energy density between 2008 and 2020 and an accompanying battery-pack price decrease of nearly \SI{87}{\percent} between 2013 and 2025 (see Fig.~\ref{fig:batteryMetrics}.B-C).

This growth is projected to continue as governments, companies, and international organizations pledge to transition away from fossil-fuel-powered vehicles. The 2021 United Nations Climate Change Conference of the Parties launched the Zero Emission Vehicles (ZEV) Declaration with over 100 countries, businesses, and organizations committing to the ``sales of new cars and vans being zero-emission globally by 2040" \cite{zevdeclaration}. Indeed, auto manufacturers such as Ford Motor Corporation and General Motors have separately pledged \SI{100}{\percent} EV sales by 2035 \cite{FordEVPledge, GMEVPledge}. The People's Republic of China is aggressively adopting Battery Electric Vehicles (BEV) by setting ambitious sales targets and efficiency mandates and making policies requiring the creation of more widespread EV infrastructure and battery recycling \cite{EYChinaEV}.

The rapid growth of the adoption of BEVs has strained conventional battery management approaches and placed LIBs in more challenging operating environments where thermal, electrical, and mechanical abuse are much more likely to be encountered \cite{sun2020review}. The impacts of battery system failure in EVs are costly in terms of monetary value and occupant safety. To that end, multi-billion dollar product recalls have been enacted to prevent occupant injury and other losses from battery system failures. Safety risk of battery fires is acute because they differ from conventional fires in key ways, namely they release flammable and toxic gases and may even reignite after being transported from the original failure location \cite{sun2020review}. Battery fires therefore spread easily and conventional methods to combat vehicle fires are often ineffective against battery fires. Additionally, battery systems can fail very suddenly. For instance, Maleki et al.~showed that up to \SI{70}{\percent} of a cell's electrical energy can be released in less than \SI{60}{\second} during an internal short circuit, which can cause internal heating that triggers thermal runaway \cite{Maleki2009}. More recent studies have reported that short circuits can range in magnitude; from \SI{100}{\percent} energy release in less than \SI{1}{\second} due to rapid failure \cite{Finegan2017EES}, to the existence of slow-developing internal short circuits that occur over weeks due to slight overcharge throughout long-term cycling \cite{Yang2022}. Early detection and identification of failure, as well as latent defects, are therefore of central importance for the adoption and safe operation of LIB systems. As a result, non-destructive testing (NDT) methods adapted for this purpose have been the focus of recent research\cite{Zuo2024Nondestructive} and growth in battery management system (BMS) development \cite{lipu2021intelligent}. Among the NDT methods, X-ray-, neutron-, and acoustic-based methods, infrared thermography, magnetic resonance, and advanced electrochemical techniques such as electrochemical impedance spectroscopy have been used study LIBs' performance and safety, each with its own advantages and challenges \cite{Chacn2023Review,Gao2024Review,Zuo2024Nondestructive}.

Conventional BMS systems rely on electrical measurements to understand state of charge (SOC) and state of health (SOH). However, since the mechanical properties of LIBs are known to change as a function of charge level and temperature, mechanical methods of inspection have been explored as part of an updated approach to battery management\cite{Popp2020MechReview}. Of those approaches, ultrasonic testing (UT) is particularly promising as different features of ultrasonic waves have been shown to correlate with SOC, SOH, thermal abuse, overcharge, and other cell conditions \cite{Zuo2024Nondestructive,wang2024progress,Williams2024AReviewTR,GervilliMouravieff2024}. Since ultrasonic wave propagation depends on the structure and mechanical properties of the media through which they propagate, and changes in the battery state lead to mechanical and structural changes on the internal materials of the LIBs, there is strong potential that ultrasonic sensors and methods can be developed to detect all of the relevant cell states under all operating conditions, which is likely to reduce the weight and complexity of the BMS. UT methods also show promise for enhancing predictive capabilities of cell cycle-lives and diverging behaviors of cells within battery packs, given the intrinsic link between ultrasonic signals and degradation of the material properties within a cell, and defect detection like gas generation. Furthermore, UT can be performed with very low-footprint sensors that do not need to be placed near the active electrical terminals of the cells to be effective. If UT can be performed with waves propagating through multiple cells, modules, or even full battery packs, the number of sensors in the BMS could be further reduced, which would further reduce BMS weight and cost. 

An additional market for UT methods is in the evaluation of cells for second-life applications. The SOH of EV batteries diverge considerably throughout their cycle-lives\cite{Severson} such that by the time a battery pack reaches its end-of-life, many cells within the pack may still have many more useful cycles in a second-life application. Assessing cell quality for risks using high-throughput approaches presents a tremendous challenge. Ultrasonic techniques have potential to rapidly provide information regarding the internal structural integrity and condition of the constituent materials of cells, which can be used to assess suitability for second-life applications. However, for all of the aforementioned applications of ultrasound-based diagnostics, high confidence does not yet exist in the relationship between ultrasonic signals and relevant physical phenomena within the cell, which limits widespread use of ultrasonic methods for BMS systems and screening for second-life applications\cite{Zhu2021EndOfLife,Fordham2023Correlative}. 

In order to fully understand the propagation of waves through LIBs, one needs a high level of understanding of many different scientific fields, namely: solid mechanics, electrochemistry, materials science, heat transfer, and acoustics. The combination of these scientific domains has made the interpretation of experimental results and the creation of accurate models difficult in this rapidly developing field. In addition, although the use of UT to detect changes in LIBs is a relatively new area of research, there has been a substantial number of studies over the last decade. Despite this research, the complexity and various form factors of LIBs (primarily prismatic, pouch, and cylindrical) and the complex electrochemical-mechanical coupling effects arising during operation, loading, and aging pose significant challenges to up-to-date and accurate ultrasonic state estimation. While these form factors can span a wide range of physical scales, from small pouch cells in consumer electronics to, in some cases, larger format prismatic cells in EVs, this review focuses on UT methods for characterizing individual cells. The fundamental principles of UT remain consistent across scales. Nevertheless, practical implementation details, such as transducer location and frequency ranges, must be customized to cell dimensions and the specific application. This review synthesizes recent advances in ultrasonic techniques and evaluates the experimental UT approaches that have been applied to inspect LIBs for state estimation and damage detection. It also provides a summary and background to modeling approaches that can be exploited to improve the use of UT for LIB monitoring. Finally, it aims to clarify common challenges of ultrasonic measurements and modeling in this rapidly growing area of research, offering valuable insights for researchers new to the field. 

We divide the discussion about ultrasonic methods to characterize and monitor LIBs into three main sections: Section \ref{sec:fundamentals} presents an overview of the fundamentals of ultrasonic wave propagation in lithium-ion batteries and provides a detailed description of the relevant structure and function of LIBs. The latter information is essential for understanding the new and unique challenges of employing UT to inspect and monitor LIBs. Section \ref{sec:USMethods} provides a comprehensive summary of the use of ultrasonic testing for different use cases. Section \ref{sec:ChallengesOpportunities} discusses challenges and opportunities regarding the eventual implementation of UT in the LIB life cycle. We provide a summary and outlook in Section \ref{sec:conclu}. Finally, we provide Supplementary Material with an additional discussion about the mathematics and physics behind wave phenomena in LIBs.

\section{Fundamentals: Wave phenomena in LIBs}\label{sec:fundamentals}

\subsection{Wave phenomena in layered structured media}\label{sec:wavePhenomena}

\begin{figure*}[ht]
    \centering
    \includegraphics[width = \textwidth]{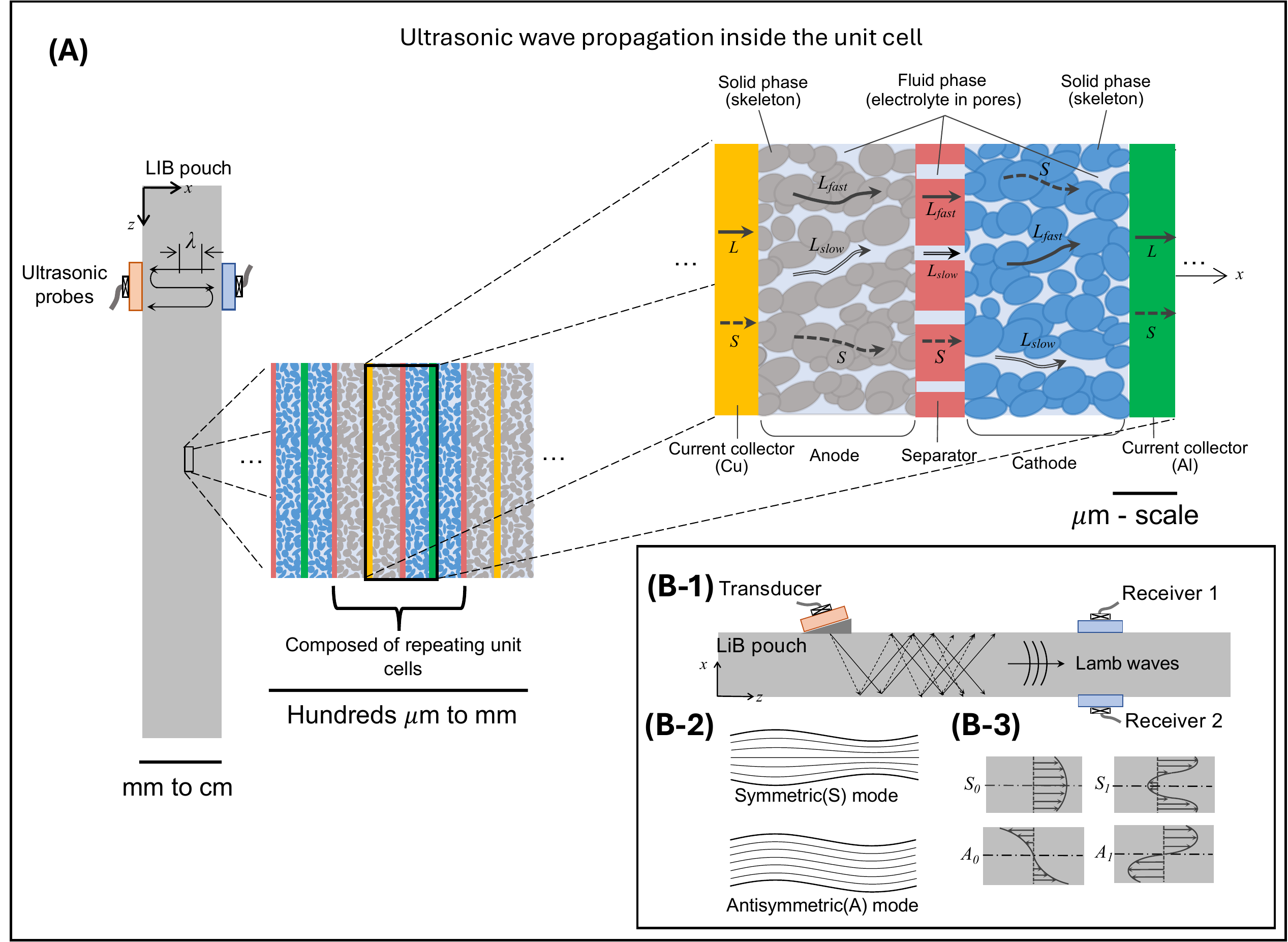}
    \caption{\textbf{(A)} Schematic of ultrasonic wave propagation through a LIB pouch cell, highlighting the layered structure composed of repeating unit cells, which include current collectors (Cu and Al), anode, separator, and cathode. For the porous materials, each layer consists of a solid skeleton and fluid-filled pores (electrolyte). Longitudinal (L) and shear (S) waves propagate through the solids, while only longitudinal waves propagate through the fluid phase. Scale bars are not exact. They are provided as a generic reference scale for microscale (0.5--100 $\mu$m) and macroscale ($>0.1$ mm) features.
\textbf{(B)} Scheme of guided wave generation and detection using ultrasonic probes:
(B-1) Experimental setup showing guided wave propagation along the z-direction.
(B-2) Illustration of symmetric (S) and antisymmetric (A) Lamb wave modes in the LIB pouch.
(B-3) Various modes of Lamb waves ($S_{0}$, $S_{1}$, $A_{0}$, $A_{1}$) and their relative particle displacement profiles across the thickness of the pouch.}
    \label{fig:UnitCellLIBsPoroelastic}
\end{figure*}

Lithium-ion batteries are heterogeneous media constructed with multiple layers of metals, polymers, and slurries filled with a liquid electrolyte. To understand ultrasonic wave propagation in LIBs, we must consider the layered heterogeneity of the cell, including the properties and geometry of each layer (as illustrated in Fig.~\ref{fig:UnitCellLIBsPoroelastic}.A for a pouch cell). One must understand the wave physics of single layers, which can be modeled as equivalent homogeneous elastic media under the long-wavelength assumption, i.e., when the length-scales of the LIB structure are much smaller than the shortest wavelength of US waves used for the measurement\cite{mcgee2024ultrasonic}. In this sense, researchers have primarily employed high-frequency transducers (\SI{1}-\SI{5}{\mega\hertz}) due to their shorter wavelength and associated enhanced sensitivity to internal features \cite{wang2021theoretical}. Note that while the frequency range \SI{1}-\SI{5}{\mega\hertz} is not ``high frequency'' from the perspective of many conventional UT methods, we use this classification in reference to the frequency range commonly employed in the vast majority of UT testing of LIBs. We recognize notable exceptions have employed ultrasonic waves at significantly higher frequencies, notably those employing scanning acoustic microscopy (SAM) \cite{bauermann2020scanning,Bauermann2023,wasylowski2023situ,Wasylowski2024Operando}. Conversely, other works have used low-frequency ultrasonic waves ($<$\SI{250}{\kilo\hertz}) to enable longer propagation distances within the battery due to reduced attenuation\cite{Gao2024TheoreticalGuidedWaves}.

Each layer of the battery is different with distinct electrochemical and mechanical properties. The current collectors (Cu, Al) are solid elastic layers, while the positive electrodes (the cathodes) and the negative electrodes (the anodes), and separators have interstitial pores filled with a liquid electrolyte. Although the microstructure of LIBs is very complex, ultrasonic waves in LIBs have been successfully described using well-known models for elastic/viscoelastic media (see Supplementary Material for a detailed mathematical description). The coupled dynamics of waves propagating in the two-phase solid-fluid structure of anodes, cathodes, and separators can be described using poroelasticity models, namely Biot’s theory \cite{biot1956theory1, biot1956theory2}. Recent research has applied Biot theory to model and understand ultrasonic characterization of LIBs, including the layered structure\cite{gold2017probing, chang2019real, jie2021guided, huang2022quantitative, gold2023ultrasound, huang2023stiffness, zhang2023ultrasonic, binpeng2023ultrasonicreflection, jie2023ultrasonic,zhang2024exploring}. 

Most UT of LIBs have employed bulk waves, which are longitudinal or shear waves that travel in the material without the influence of boundaries. However, researchers have also employed guided wave methods, which are a well-known and powerful tool in UT\cite{chimenti1997guided,chimenti2014review}, to detect changes in LIBs \cite{jie2021guided,Reichmann2023UltrasonicGuided,Gao2024TheoreticalGuidedWaves}. In contrast to bulk waves whose wave speed is not a function of frequency if the material is lossless, guided waves are dispersive, meaning their phase and group velocities are functions of frequency. The frequency dependence is determined by the dispersion relations, which are functions of the waveguide geometry and material properties. For the case of LIB cells, the most relevant geometry is usually the cell thickness. Dispersive wave propagation occurs in bounded elastic media because the wave-field must satisfy boundary conditions at all points along the waveguide. For elastic plate waveguides, this means that one observes standing wave modes in the thickness direction that propagate in the plane of the plate. This can be envisioned as the coherent interference of multiple reflections from top and bottom boundaries of a plate to yield a propagating standing wave pattern as indicated by Fig.~\ref{fig:UnitCellLIBsPoroelastic}.B. Guided waves in elastic plates, commonly known as Lamb waves, are an infinite set of symmetric and antisymmetric modes, each corresponding to specific frequency-wavenumber pairs.  These modes can propagate when excited by a source or scatterer within the waveguide or at its boundaries. Each guided wave mode has its unique deformation pattern and sensitivity to material changes. Thus, Lamb waves may be able to detect variations in the properties of layers and conditions at interfaces within a LIB. It is important to note that the layered structure of LIBs can be represented as a homogeneous anisotropic material on the macroscale when the wavelength of propagating modes is much larger than the thickness of individual layers. In that case, guided wave modes are more complex than Lamb waves in isotropic plates \cite{chimenti1997guided}. To advance research in ultrasonic diagnostics of batteries, it will be critical to understand guided wave modes in battery materials, as they may provide more refined information on the evolution of constituent material properties.

Biot's theory \cite{biot1956theory1,biot1956theory2} describes wave propagation in poroelastic media as consisting of three types of propagating bulk waves: a fast longitudinal wave (mainly transported by the elastic frame), a slow longitudinal wave (caused by the relative movement between fluid and elastic phases), and a shear wave in the elastic frame. These waves are illustrated in Fig.~\ref{fig:UnitCellLIBsPoroelastic}.A. Despite being strongly attenuated and dispersive, the slow wave offers unique sensitivity to changes in porosity, fluid content, and structural integrity. A more detailed development of the governing equations for wave propagation in poroelastic media is discussed in the Supplementary Material. 

Several works have employed poroelastic material modeling of wave propagation in LIBs, the most notable of which is the work by Gold et al.~\cite{gold2017probing}. Although Biot’s poroelastic model was only used in their work to estimate the propagation velocities of the three kinds of waves in the graphite electrodes of a LIB, they were the first to observe and discuss the existence and significance of the slow wave propagating in poroelastic materials in LIBs. Gold et al.~suggested that the slow wave is a key parameter to evaluate SOC, because signal amplitude (SA) and time of flight (TOF) of the slow wave were observed to be linearly correlated with SOC. Namely, they observed that the SA increased linearly and TOF decreased linearly with respect to increases in SOC, while the SA and TOF of the fast wave did not change notably with SOC in their work. In contrast, Chang et al.\cite{chang2019real} reported different results on the sensitivity of fast and slow wave characteristics to SOC. By measuring a directly transmitted signal through a LIB (NiMnCoO$_2$) using two air-coupled transducers, the transmitted amplitudes of the fast and slow waves increased for higher SOC, although the time of arrival of the slow wave decreased as a function of increasing SOC, similar to the observations of Ref.~\cite{gold2017probing}. Zhang et al.~\cite{zhang2023ultrasonic} later measured slow wave propagation through a LIB (LiCoO$_2$) pouch and showed that SA of both the fast and slow waves increased with increasing SOC. While TOF of the slow wave decreased with increasing SOC, they did not observe changes in the TOF of the fast wave. 

The conflicting research highlighted above illustrates that although the existence and properties of Biot's slow wave may be a useful ultrasonic feature for characterizing LIBs, there is currently no consensus on the functional dependence of slow wave properties on LIB SOC. This is likely due to the well-known difficulty in measurement of the slow wave in poroelastic media\cite{johnson1982acoustic,johnson2017impact}. This wave type is strongly dissipative, making it challenging to observe in real-world scenarios. Furthermore, when compared with the fast and shear waves, which are primarily affected by the material properties of the solid phase in the electrodes, the relatively low amplitude of the slow wave originates from the coupling effect of solid/fluid phases, which is highly dispersive and leads to strong distortion of the signal. Nevertheless, the concept of the use of the slow wave in UT of LIBs is still valid, as reported in the previous works \cite{gold2017probing, chang2019real, zhang2023ultrasonic}. 

The key challenge, however, is poroelastic effects as applied to ultrasonic wave propagation within LIB electrodes. Such relative motion introduces complex wave behaviors — including additional modes and increased attenuation — that must be considered in order to interpret ultrasonic data correctly. Thus, ultrasonic studies of LIBs should consider both elastic and poroelastic wave propagation mechanisms in modeling wave phenomena to extract insights from experimentation and increase understanding. The Supplementary Material provides a detailed mathematical description of the bulk waves for elastic and poroelastic media, guided waves, and modeling implementations.

\subsection{Structure and function of LIB}\label{sec:structureAndFunction}

It is essential to understand the structure of LIBs and how changes in electrochemical state during operation, aging, and damage can affect ultrasonic wave motion. Additionally, the wide variety of LIBs, specifically different chemistries and form factors, should be highlighted since they introduce numerous variables that must be considered when interpreting ultrasonic data under the different external loading conditions applied to the battery. 

Commercial lithium-ion cells are manufactured with tens to hundreds of individual component layers in planar or cylindrical arrangements. The simplest representation of an electrochemical system consists of the anode, cathode, separator, and current collectors, as illustrated in Fig.~\ref{fig:UnitCellLIBsPoroelastic}A. Each layer has varying thicknesses: current collectors typically range from \SI{10}-\SI{20}{\micro\meter}\cite{Zhu2021CollecThick}, electrodes from \SI{40}-\SI{150}{\micro\meter}\cite{Boyce2021ElectrodeThick}, and separators from \SI{20}-\SI{25}{\micro\meter} \cite{Lee2014SepThick}. In these systems, design choices such as the chemical composition, thickness of layers, and porosity of electrode coatings and separators, dictate the electrochemical behavior of the cell such as cell voltage, cycle life, and chemical and thermal stability \cite{Larcher2014Toward}. This section provides a detailed description of the mechanical structure and electrochemical function of each cell component, with an emphasis on how the mechanical properties of each component are affected by the state of the cell.

The electrodes are the active components in the cell, and it is the difference in the redox energies between the cathode and anode that dictates the cell's operating voltage. This redox energy is associated with the tendency of the electrodes to be oxidized or reduced to allow the movement of the lithium-ions between both electrodes during the battery operation \cite{manthiram2017outlook}. Notably, this process of lithiation/delithiation leads to changes in mechanical properties, and therefore can be detected with ultrasonic elastic waves. Commercial lithium-ion cells are constructed with metal oxide cathodes, which can be divided into classes based on their chemical structure and operating principle: the layered oxides, the spinel oxides, and the polyanion oxides \cite{manthiram2020reflection}. When performing mechanical inspection of LIBs, it is critical to state the cathode chemistry, as the choice of cathode chemistry affects not only the electrochemical performance, but also the mechanical properties during charge-discharge cycling, aging, and damage mechanisms. When charging under normal operating conditions, lithium ions are removed from the cathode, migrate  through the electrolyte and separator, and are either inserted into or deposited on the anode. Lithium ions follow the same path, but in reverse during discharging. The ionic charge imbalance caused by this electrochemical process initiates the movement of electrons through an external circuit, creating the flow of charge required for electronic devices to work.

\begin{figure*}[ht]
    \centering
    \includegraphics[width = \textwidth]{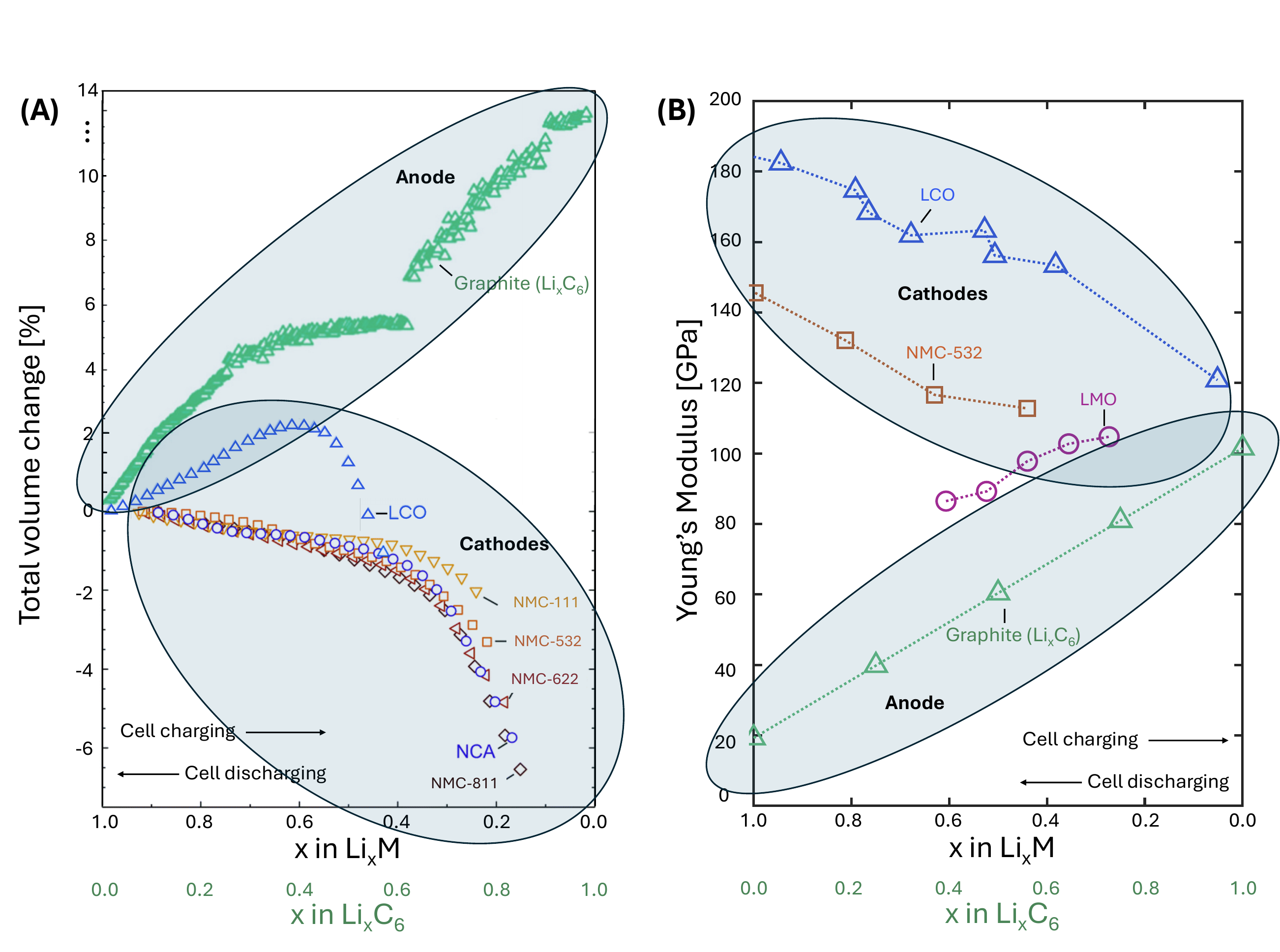}
    \caption{\label{fig:ElectrodesVolumeChange}{Changes in volume and Young's modulus of common commercial LIB cathode and anode materials.  During discharge, cathode materials generally exhibit an increase in volume and Young’s modulus, while the graphite anode typically shows a decrease in volume and modulus. Some exceptions to this behavior are observed for LCO and LMO. This figure is based on raw data for the graphite anode \cite{Schweidler2018VolumeChanges,qi2010threefold} Li$_x$C$_6$, and various cathodes \cite{koerver2018chemo, Stallard2022MechanicalCathode}:  Li$_x$CoO$_2$ (LCO), Li$_x$Ni$\mathrm{_y}$Mn$\mathrm{_z}$Co$\mathrm{_{(1-y-z)}}$O$_2$ (NMC), Li$_x$Ni$\mathrm{_y}$Co$\mathrm{_z}$Al$\mathrm{_{(1-y-z)}}$O$_2$ (NCA), Li$_x$Mn$_2$O$_4$ (LMO). The decrease in volume of cathode materials is generally much smaller than the volume increase of graphite under normal operating conditions.}}
\end{figure*}

The active material of the anode in most commercial LIBs is graphite, although lithium-metal anodes and other anode materials such as silicon are currently being heavily investigated and nearing commercial viability \cite{feyzi2024SiliconAnodes}. Some commercial cells have small quantities of SiO$_x$ and Si mixed into their graphite anodes\cite{Xuemin}, but since graphite is presently the most common commercially available anode material, the present discussion is restricted to the structure and function of anode materials with graphite as the active material to illustrate how charging and discharging affect ultrasonic wave propagation. During charging, lithium ions are inserted into the layered structure of graphite, which causes a strain in the crystal lattice with concomitant changes to the unit cell volume and elastic moduli. The observed changes in the overall mechanical properties of the cell are most often attributed to the changes in the mechanical properties of the graphite anode, as the volume of graphite has been found to increase by approximately \SI{13}{\percent} and the Young's modulus has been found to increase by three times at full lithiation \cite{Schweidler2018VolumeChanges, qi2010threefold}. Lithium ion intercalation in graphite anodes occurs in multiple different stages: stage 4, where lithium ions are sparsely intercalated (ratio of carbon to lithium is $72:1$), stage 3, intermediate lithium insertion (ratio of carbon to lithium is $36:1$), stage 2, where every second graphite layer is replaced by denser lithium layers (ratio of carbon to lithium is $18:1$)  and stage 1, where all layers are fully lithiated with lithium ions (ratio of carbon to lithium is $6:1$). Each stage of lithiation can be identified by specific structural arrangements and voltage plateaus in the negative electrode potential during cycling \cite{sethuraman2010surface,Dahn1991seminalPaper}. Conversely, cathodes experience an opposite behavior, a decrease in both volume and Young's modulus during charging due to delithiation, but to a lesser extent compared to the graphite anode \cite{koerver2018chemo,Stallard2022MechanicalCathode}, see Fig.~\ref{fig:ElectrodesVolumeChange}. Critically, both electrodes contain active materials and conductive additives (typically some form of carbon), and binder materials (most commonly polyvinylidene fluoride (PVDF), styrene butadiene rubber, or carboxymethyl cellulose). In order to create an ionically conductive medium to aid the diffusion of lithium ions between the electrodes, the cell is soaked with a liquid electrolyte. In most commercial LIBs, the electrolyte is a solution of lithium salts dissolved in a carbonate-based solvent, such as ethylene carbonate (EC) or diethyl carbonate (DEC). Thus, the structure of the electrodes has been described as a slurry, which itself retains some level of porosity to allow for electrolyte saturation. This structure leads to poroelastic wave motion in the electrodes as mentioned in Section \ref{sec:wavePhenomena}. A discussion of the impact of this complex porous structure on the coupled acoustic and elastic wave motion equations is provided in the Supplementary Material.

For LIBs to operate, the cathode and anode must be electrically isolated while retaining lithium ion conductivity. This functionality is achieved with the separator. In most commercial LIBs, the separator is a thin and porous polyolefin material, such as polyethylene (PE), polypropylene (PP), or a PP-PE blend. The electrically insulating properties of the separator mitigate the risk of internal short circuits between the electrodes while its porosity allows lithium ion transport. Because the separator is polymeric, it represents a highly compliant, porous, viscoelastic component of LIBs whose properties depend on loading history. Polymeric separators also have the lowest melting temperature of the cell components, and they are known to shrink at elevated temperatures, which can reduce their porosity or lead to internal short circuits. Reductions in separator porosity may be indirectly measured as increases in the internal resistance of a cell or as changes to the effective material properties of the cell, which affect its ultrasonic response\cite{mcgee2024ultrasonic}.

\begin{figure*}[ht]
    \centering
    \includegraphics[width = \textwidth]{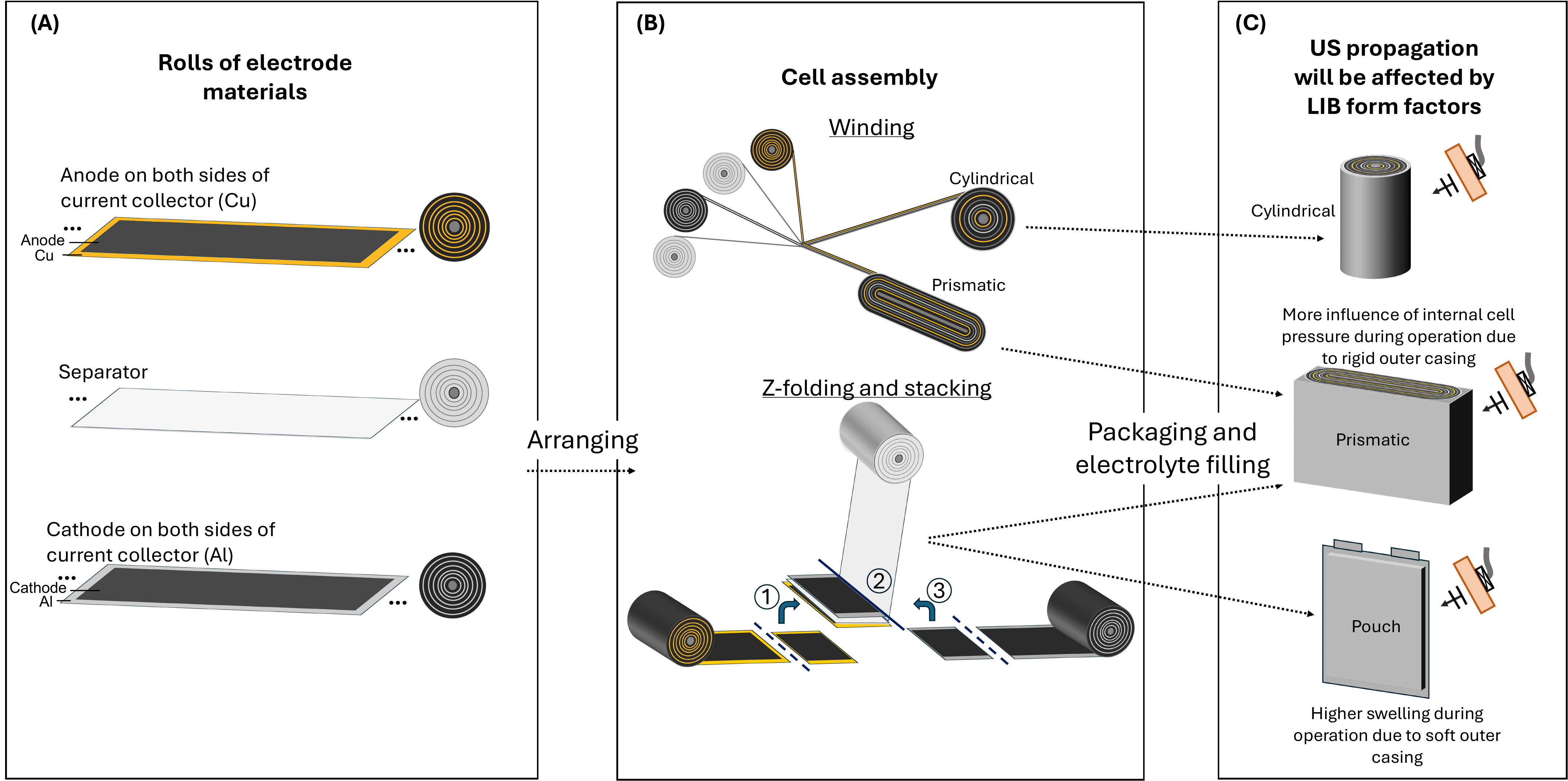}
    \caption{\label{fig:FormFactors}{Principle fabrication steps in the creation of LIB and comparison of LIB form factors, including cylindrical, prismatic, and pouch cells. \textbf{(A)} Assembly process from electrode rolls through various packaging configurations. \textbf{(B)} Winding for cylindrical cells, and stacking or Z-folding for prismatic and pouch cells. \textbf{(C)} Differences in internal structure and mechanical response under operational pressure are highlighted. These structural differences influence swelling behavior and ultrasonic signal propagation within the cells.}}
\end{figure*}

As previously noted, commercial cells consist of hundreds of more basic repeating units consisting of two electrodes in contact with metallic current collectors separated by a polymeric separator, permeated by a liquid electrolyte that fills the pores of the electrodes and the rest of the cell (Fig.~\ref{fig:UnitCellLIBsPoroelastic}). In addition, LIB cells have different form factors as illustrated in Fig.~\ref{fig:FormFactors} depending on the arrangement and packaging of the basic small-scale structure shown in Fig.~\ref{fig:UnitCellLIBsPoroelastic}. Cylindrical cells, such as the common 18650 cell, are constructed by winding these components around a central core. The outside casing of these cells is typically a rigid steel shell. A separate manufacturing approach is to lay the layers on top of each other to produce a planar layered structure. A common approach to this design is the Z-fold, where the separator is folded on top of itself with electrode sheets and conductors are placed in between separator layers. Cells with this arrangement are referred to as pouch cells when the outer casing is a flexible polymer-coated metal.  A third common form factor are prismatic cells, which can have both approaches, wound or Z-folded, in a rectangular rigid steel case. Most commercial prismatic cells are wound\cite{schroder2017comparatively,mcgovern2023review,Wagner2013Current}.

The cell structure can have a large impact on its evolving mechanical properties and therefore ultrasonic measurements. For example when the anode expands with lithiation, the rigid constraint of prismatic and cylindrical cells will resist expansion and cause an increase in the internal cell pressure leading to compression of the porous separator. In the case of thermal abuse, for example, internal cell pressure may increase the boiling temperature of the electrolyte, which may change the point at which gas generation is mechanically detectable \cite{Gulsoy2024,mcgee2023ultrasonic, Owen2024OperandoTemperature}. 

As the electrolyte saturates the porous anode, cathode, and separator, the macroscale effective mechanical properties of the cell change significantly. Given this structure, the electrode materials have been described as slurries \cite{huang2022quantitative}, crushable foams \cite{hsieh2015electrochemical,mcgee2023ultrasonic}, or porous media \cite{binpeng2023ultrasonicreflection,Binpeng2025Ultrasonic}, and each approximation has different approaches to defining effective material properties. Each of these modeling paradigms presents challenges in terms of modeling the physical behavior of LIBs as it pertains to US wave propagation. 

One of the key difficulties in modeling wave phenomena in LIBs is accurately estimating the influence of electrode and separator porosity on the propagation of ultrasonic waves in any given state. This is exacerbated by the fact that the porosity and constituent properties fluctuate over time due to cell wetting or unwetting or ongoing solid-electrolyte interface (SEI) growth. A reliable modeling approach that captures the microscale wave physics and provides insight into experimental data is to use poroelastic material models for these layers. These models include the so-called fast and slow compressional waves which arise when the compression of the fluid (electrolyte) and solid components in each layer are either in- or out-of-phase, respectively \cite{plona1980observation}. These models also capture  attenuation based on fluid viscosity and porosity which can be helpful in monitoring changes in LIB structure. The biggest obstacle in the poroelastic modeling for LIBs is the lack of clear information on the relevant material properties of the constituent solid and fluid phases in LIBs under various conditions. Since the direct measurement of wave-related properties is still a challenge for these LIBs, many mechanical properties used in poroelastic material modeling have been assumed with only minimal validation or have been adopted from a few previous literature in a duplicated manner. 

Additionally, since the cells include hundreds of layers, one must either explicitly model each layer using techniques like the transfer matrix method \cite{ binpeng2023ultrasonicreflection, huang2022quantitative, huang2023stiffness, mcgee2024ultrasonic,Feiler2024ModelingAttenuation, Ma2024USScatteringModel} or simplify the system using effective medium models, i.e.~by assuming periodic boundary conditions\cite{mcgee2024ultrasonic} such as Floquet-Bloch approaches \cite{Ma2024USScatteringModel} (see Supplementary Material for a discussion on modeling implementation through different matrix-based methods). In each of these cases, the microscale structure and properties (constituent material properties, pore size and volume fraction, fraction of active materials, layer thicknesses, etc.) and their evolution when subjected to electrochemical and thermomechanical loading must be considered as a starting point for any model that wishes to capture the complex wave mechanics in these systems. In many cases, the mechanical properties of cell components are not known, especially when considering effects such as degree of lithiation, electrode degradation, formation and growth of the SEI, and the influence of temperature. Without direct knowledge of cell mechanical properties under diverse operating conditions, accurately approximating or modeling cell components is imperative for constructing full cell models which accurately describe the wave phenomena occurring within cells.

Another layer of complexity in LIBs is that they can undergo multiple degradation mechanisms during battery operation, which depends on operating and abuse conditions. Each of these mechanisms can alter the structure of LIBs in different ways. Detailed descriptions of multiple degradation mechanisms with a comprehensive evaluation of their current understanding can be found in Refs.~\cite{Edge2021BatDegradation,deVasconcelos2022ChemoMech,Li2025DegradationModes}. All of these degradation mechanisms have complex mechanical consequences that are still under study, but, more importantly, these degradation mechanisms lead to phenomena like particle expansion, interfacial debonding, loss of stiffness, crack generation and propagation, and porosity changes. All of these changes alter ultrasonic wave propagation through the cell and affecting characteristics of ultrasonic signals by altering propagation velocity, attenuation, and scattering. Table \ref{tab:summaryUSDegradation} reports key features on some of the degradation mechanisms that have been studied with UT, namely as SEI growth, lithium plating, particle cracking/electrode delamination, electrolyte dry-out, separator failure, and gas formation. The latter is not specifically a degradation mechanism but rather a consequence of multiple degradation mechanisms and because of its importance within the study of UT for LIBs, has been included as its own category. In some cases, the same stimulus can cause multiple types of degradation within a battery. However, we have highlighted the most common degradation types investigated using UT and listed them separately for clarity, noting how they are reported in the existing literature. UT observables are similar for many of the battery degradation mechanisms, which can lead to some ambiguity of existing UT methods. Improved UT techniques or integration of UT data with other monitoring methods should be the subject of future work to specify them to known types of damage and degradation in order to make this approach more valuable for all aspects of LIBs life cycle. This review provides an overview of recent research on UT of LIBs with the objective of providing researchers with sufficient understanding of the relevant electrochemistry and elastic wave phenomena to create representative models, design experiments, and interpret results for ultrasonic methods to be useful for research and application of LIBs.

\begin{table*}[]
\large
\caption{Summary of LIB degradation mechanisms detected using UT methods. The following definitions are used within this work. UT sensitivity is qualitatively defined as Low, High, or Very High based on the estimated resolution at common UT frequencies within \SI{0.5}-\SI{5}{\mega\hertz} and relevant LIB length scales. Length scales are defined as: Nanoscale (\SI{1}-\SI{500}{\nano\meter}), Microscale (\SI{0.5}-\SI{100}{\micro\meter}) and Macroscale ($>\SI{0.1}{\milli\meter}$).}
\label{tab:summaryUSDegradation}
\begin{adjustbox}{width = \textwidth}
\begin{tabular}
{p{3cm}>{\raggedright\arraybackslash}p{5cm}>{\raggedright\arraybackslash}p{7cm}>{\raggedright\arraybackslash}p{7cm}>{\raggedright\arraybackslash}p{8cm}>{\centering\arraybackslash}p{3.5cm}>{\raggedright\arraybackslash}p{4cm}>{\raggedright\arraybackslash}p{4cm}}
\hline
\hline
Degradation & Stimulus& Electrochemical characteristic & Hypothesis for UT changes & UT observable & UT sensitivity & Length scale & Relevant refs. \\
\hline
\hline
SEI growth&
High temperature ($>\SI{40}{\celsius}$) \newline
\newline
High current rates ($>1$C) &
Consumes lithium inventory and electrolyte solvent \newline
\newline
Irreversible capacity loss due to lithium confinement in SEI &
Change of impedance at interfaces\newline
\newline
Increased layer thickness\newline
\newline
Stiffness increase with growth of SEI changes wave speed &
Ultrasonic TOF increase due to overall thickness increase \newline
\newline
Slight change in reflection amplitude due to changes in acoustic impedance at anode surface
& Low
& Nanoscale (\SI{10}-\SI{200}{\nano\meter}) \cite{Rah2024,Oyakhire2024} & Bommier et al. (2020)\cite{bommier2020SiGr}, Chai et al. (2021) \cite{Chai2021}\\
\hline
Lithium \newline plating&
Low temperature ($<\SI{10}{\celsius}$) \newline
\newline
High current rate ($>1$C) \newline \newline
High SOC/Voltage &
Lithium plating may generate additional SEI formation through reaction with electrolyte \newline \newline
Lithium plating may cause dendrite growth that short circuit through separator\newline \newline
Accelerated side reactions from reactive lithium plating correlate with gas generation&
Reduced density and stiffness of lithium plating significantly reduces acoustic impedance \newline \newline
Lithium dendrites introduce scattering sites \newline \newline
Dendrites increase surface roughness, which may increases incoherent scattering at boundaries&
Significant decrease in SA (even after correction due to temperature) \newline \newline
TOF at max charge depends on C-rate. Promising indicator of degree of Li plating at low-temperature and fast charge \newline \newline
Attenuation of transmitted signal amplitude at the thickness-mode resonance frequency during low-temperature battery charging
& High
& Microscale (\SI{10}-\SI{100}{\micro\meter} lithium plating) \cite{Rong2017,AghiliMehrizi2025}&
Bommier et al.~(2020)\cite{bommier2020operando}, Chang et al.~(2020)\cite{chang2020understanding}, Wasylowski et al.~(2024)\cite{Wasylowski2024Operando}, Meyer et al.~(2025)\cite{Meyer2025}, Xu et al.~(2025)\cite{Xu2025}\\
\hline
Particle\newline cracking/\newline
Electrode\newline delamination&
High temperature ($>\SI{45}{\celsius}$) \newline $\rightarrow$ Large thermal stress \newline \newline
Low temperature ($<\SI{0}{\celsius}$) $\rightarrow$ graphite embrittlement \newline \newline
Volumetric expansion in cells with Si content &
Exposes pristine electrode surface to electrolyte which may lead to SEI growth and gas generation&
Cracks increase impedance contrast and scattering\newline \newline
Gas-induced delamination\newline \newline
Delamination generates large impedance mismatch, significantly reducing transmission \newline \newline
Scattering increases attenuation &
Significant increases in attenuation yields reduction in waveform center frequency\newline \newline
Loss of transmitted signal due to delamination of the electrode layers
& Very high
& Micro- to Macroscale (\SI{10}-\SI{50}{\micro\meter} cracks and \SI{0.1}-\SI{1}{\milli\meter} delamination)\cite{Pistorio2022} & Pham et al. (2020)\cite{pham2020correlative}\\
\hline
Electrolyte \newline dry-out&
Extended cycling \newline \newline
SEI solvent consumption \newline \newline
High temperature ($>\SI{55}{\celsius}$) &
Gas encapsulation causes loss of ionic contact area \newline \newline
Ionic transport limitation &
Void creation significantly reduces density and stiffness and therefore acoustic impedance \newline \newline
Unwetting reduces layer coupling, significantly reducing transmission amplitude &
Strong reduction of transmitted signals in unwetted regions \newline \newline
Higher reflection in unwetted regions \newline \newline
TOF increase due to reduced wave speed caused by voids
& Very high
& Micro- to Macroscale (Dry regions ranging from \SI{10}{\micro\meter} to \SI{1}{\milli\meter})\cite{deng2020ultrasonic,Eldesoky2022LongTerm}
&  Deng et al. (2020)\cite{deng2020ultrasonic}, Eldesoky et al. (2022)\cite{Eldesoky2022LongTerm} \\
\hline
Separator \newline failure&
High temperature ($>\SI{100}{\celsius}$) \newline \newline
Abusive thermal cycling &
Side electrochemical reactions at high temperature may lead to SEI decomposition and gas generation \newline \newline
Polymer separator melting may lead to internal short circuit &
Elevated temperatures softens, and ultimately melts, polymer separator &
Increased TOF due to increased compliance of polymer separator\newline \newline
Temperature increase reduces SA due to increased viscoelastic losses at moderate temperatures\newline \newline
High temperature gas generation leads to increased impedance contrast and loss of transmitted signal
& High
&  Microscale ($\approx 25$ $\mu$m polymer layer\cite{Lee2014SepThick} with $<1\mu$m internal voids that grow with damage) & McGee et al. (2023)\cite{mcgee2023ultrasonic}, McGee et al. (2024)\cite{mcgee2024ultrasonic}\\
\hline
Gas \newline formation&
Thermal runaway \newline \newline
Overcharge \newline \newline
Overdischarge&
Side electrochemical reactions occur due to high interfacial reactivity\newline \newline
Electrolyte decomposition generates gases \newline \newline
Particle cracking/Electrode delamination &
Gas pockets have low density and stiffness, which significantly reduces acoustic impedance\newline \newline
Strong wave scattering \newline \newline
Increase of internal pressure/stresses may reduce contact between layers in other regions &
Strong reduction of transmitted signals due to acoustic impedance mismatch \newline \newline
Reduced coherence in reflections scattering from gas-rich regions \newline \newline
Ultrasonic imaging can be used to accurately visualize gas pockets
& Very high
& Micro- to Macroscale (\SI{0.1}-\SI{2}{\milli\meter} gas regions and pouch swelling)\cite{Liu2025PhasedArray} & Bommier et al. (2020)\cite{bommier2020SiGr}, Owen et al. (2024)\cite{Owen2024OperandoTemperature}\\
\hline\hline
\end{tabular}
\end{adjustbox}
\end{table*}

\section{UT methods for LIBs}\label{sec:USMethods}
\begin{figure}[ht]
    \centering
    \includegraphics[width=\reprintcolumnwidth]{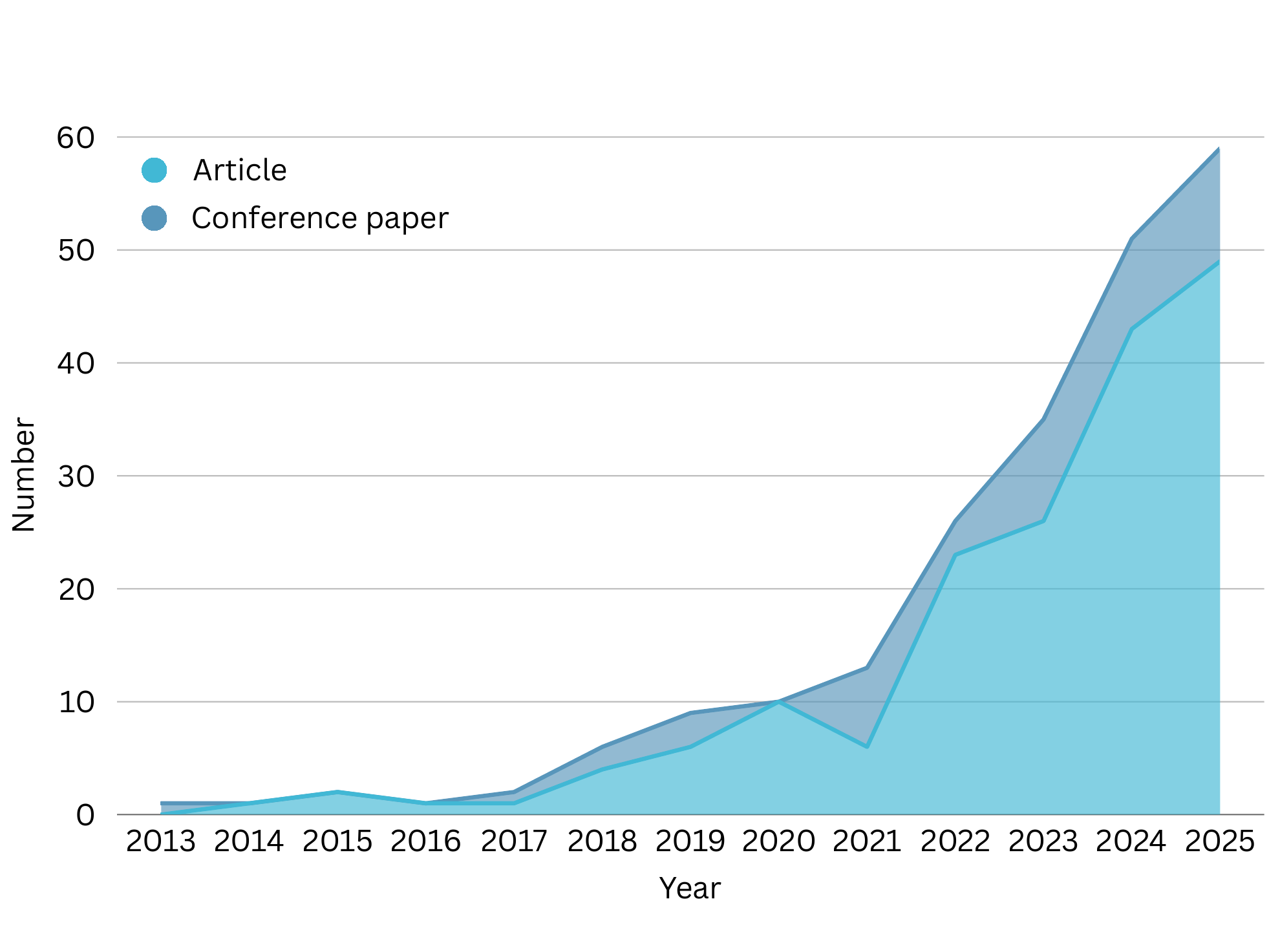}
    \caption{Number of publications by year on the subject of UT of batteries since 2013, using the relevant papers from the following search criteria in Scopus: \small{\texttt{TITLE-ABS-KEY (("Li-ion*" OR "Lithium-ion*") AND (ultraso* OR acoustic*) AND (diagnos* OR estimation OR "state of health" OR "state of charge" OR health OR characterization OR monitoring) AND NOT (welding) AND NOT ("acoustic emission"))}}. Note that these search criteria do not include all contributions to the literature, particularly conference papers, but accurately captures the overall trend of increasing publications on the use of UT for LIB measurements.}
    \label{fig:USPubsPerYear}
\end{figure}

Although previous UT research was conducted on other multilayer membrane systems, such as reverse osmosis membranes \cite{Morra2003OsmMembrane} and solid-state fuel cells \cite{Araki2013FuelCell}. The first publication involving ultrasonic testing of lithium-ion batteries was published in 2013 by Sood et al.~\cite{sood2013health} Since then, interest in UT to detect changes structure and SOH as well as condition and process monitoring of LIBs has grown, as evidenced by the rapidly growing number of research articles published on the subject through 2025, as shown in Fig.~\ref{fig:USPubsPerYear}. To clearly communicate progress in this broad application area of ultrasonics, this review uses the following grouping of UT methods as applied to LIB research: (\textit{i}) UT for state estimation (Sec.~\ref{sec:stateDetection}), (\textit{ii})  UT for damage detection (Sec.~\ref{sec:damageDetection}), and (\textit{iii}) UT to improve manufacturing, extend life cycle, and assess next generation batteries (Sec.~\ref{sec:otherPartsLifecycle}). A life cycle-oriented introduction into Sec.~\ref{sec:USMethods} is given by Table \ref{tab:matchingLifeCycle}.

\begin{figure}[ht]
    \centering
    \includegraphics[width = \linewidth]{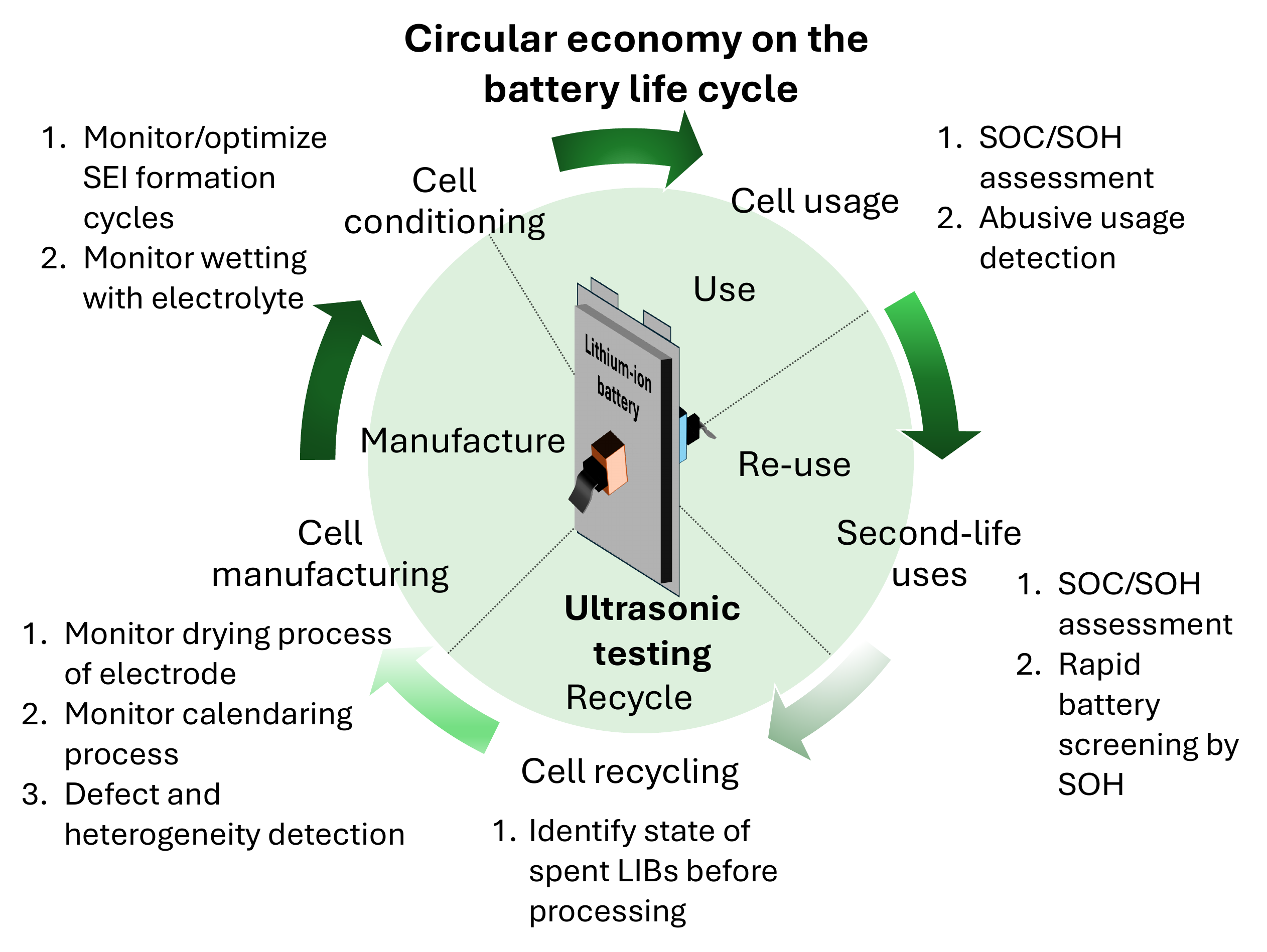}
    \caption{Schematic of the battery life cycle showing some applications of ultrasonic testing at different stages in a circular economy framework.}
    \label{fig:LIBUSlifecycle}
\end{figure}

Generally speaking, ultrasonic testing and monitoring can be used as a non-destructive characterization and monitoring technique for LIBs at the different stages of their life cycle \cite{GervilliMouravieff2024}, which can be identified using a circular economy framework as manufacturing, use, reuse, and recycling processes. Figure~\ref{fig:LIBUSlifecycle} illustrates different applications of UT during the life cycle of a cell and includes cell manufacturing, cell conditioning, cell use, second-life use, and finally cell recycling.

Section \ref{sec:USMethods} is organized around the UT method type rather than the life cycle stages shown in Fig.~\ref{fig:LIBUSlifecycle}. This choice reflects three primary considerations: (\textit{i}) the vast majority of research in UT of LIB has focused on the usage stage of the LIB life cycle, (\textit{ii}) most research to date employs conventional UT techniques at all stages of the LIB life cycle, and (\textit{iii}) the intended readership of this review who have significant knowledge of conventional UT methods but are less aware of the applicability of UT for LIB applications. We believe that a researcher choosing a UT method for a certain application will benefit from understanding how a given UT technique can be employed across life cycle stages, rather than encountering numerous separate discussions of the same technique under each life cycle step. Sections \ref{sec:stateDetection} and \ref{sec:damageDetection} review classical UT methods applied during the usage stage, whether in normal operation or during abusive usage, while Sec.~\ref{sec:otherPartsLifecycle} reviews specific applications of UT methods that have been shown in the remaining life cycle stages (manufacturing, second-life, and recycling). Table \ref{tab:matchingLifeCycle} provides a relation between the life cycle stage and its corresponding subsection in this review, allowing readers primarily oriented around the battery life cycle to guide themselves accordingly.

\begin{table*}[]
\caption{Mapping of battery life cycle stages presented in Fig.~\ref{fig:LIBUSlifecycle}) to the relevant subsection of Sec.~\ref{sec:USMethods}.}
\label{tab:matchingLifeCycle}
\begin{adjustbox}{width = \textwidth}
\begin{tabular}
{p{4cm}>{\raggedright\arraybackslash}p{3cm}>{\raggedright\arraybackslash}p{8cm}>{\centering\arraybackslash}p{2cm}}
\hline\hline
Life cycle stage \newline (see Fig.~\ref{fig:LIBUSlifecycle}) & Circular \newline economy stage & Relevant applications & Section\\
\hline\hline
Cell manufacturing & Manufacture & Monitor electrode drying and calendering; detect fabrication defects and heterogeneity& \ref{sec:UTformanufacturing}\\
\hline
Cell conditioning & Manufacture $\rightarrow$ Use & Monitor and optimize SEI formation cycles; monitor electrolyte wetting & \ref{sec:UTformanufacturing}\\
\hline
Cell usage: State monitoring & Use & SOC/SOH estimation; temperature monitoring; optimize charge/discharge protocols  & \ref{sec:stateDetection}\\
\hline
Cell usage: Abusive \newline conditions & Use & Early thermal abuse and runaway detection; lithium plating detection; overcharge/overdischarge; gas formation; defect detection due to abuse loading & \ref{sec:damageDetection} \\
\hline
Second-life use & Re-use & Rapid SOH screening; defect and heterogeneity detection for repurposing decisions & \ref{sec:secondlife} \\
\hline
Cell recycling & Recycle & Identify state of spent cells before processing; detect residual hazards & \ref{sec:recycling} \\
\hline\hline
\end{tabular}
\end{adjustbox}
\end{table*}

\subsection{State detection during normal operation: SOC and SOH}\label{sec:stateDetection}
The first publication that employed UT techniques to evaluate LIBs by Sood et al.~employed a pitch-catch configuration on an unconstrained lithium-ion pouch cell before and after undergoing a reduction in SOH to approximately \SI{75}{\percent}. They found that the ultrasonic signal received through the aged cell was a ``very weak, delayed pulse'' \cite{sood2013health} and attributed the changes in signal characteristic to various aging mechanisms including gas formation, the formation and growth of SEI layers, and ``electrode ruffling" where the layers within the cell are no longer flat due to the localized stresses associated with repeated cycling. Importantly, their observations of the ultrasonic signal point to two ultrasonic signal characteristics that have been reported in almost every other publication on the subject since: the signal amplitude (SA) and the change in propagation time compared to a pristine cell, which is also known as the time-of-flight shift (TOFS). These signal characteristics are chosen because they are both easy to calculate and indicative of numerous changes within the cell. Linking these metrics to specific internal processes is a challenge that is the subject of a large percentage of publications to date. For example, increases in TOF (positive TOFS) may indicate increased propagation length or decreased wave velocity \cite{sun2022ultrasonic,Ajaereh2024Assessing}. Decreases in SA can be related to increases in attenuation \cite{chang2021operando,meng2022robust}, reduced impedance matching between the transducer and cell \cite{Majasan2021,robinson2020identifying}, or the introduction of frequency-specific band gaps in layered systems \cite{meng2022robust,Feiler2024ModelingAttenuation,Ma2024USScatteringModel}. Following Sood et al.~\cite{sood2013health}, numerous authors have attempted to fully investigate the relationship between changes in ultrasonic signal characteristics and SOC and SOH. That research has been categorized in this review according to whether the work used bulk or guided waves to inspect the cells.

\subsubsection{Bulk Waves for SOC and SOH Detection}\label{sec:bulkwaves}

The use of bulk waves for UT of lithium-ion cells is often employed because the configuration is simple, using either contact or non-contact methods, and the data is easily analyzed and interpreted since dispersion is only a result of material response and layering. As noted in Sec.~\ref{sec:wavePhenomena}, bulk waves can be classified as either longitudinal or shear waves that propagate within a medium without interaction with boundaries. Conversely, the propagation speed of guided waves is determined by the material properties, boundary conditions, and geometry of the sample. Since bulk wave speed is governed by the material properties alone, their behavior is easier to model and interpret. For transducers with large aperture, i.e.~when $k_0a\gg 1$ where $k_0 = 2\pi f_0/c_0$ is the wavenumber at the center frequency, $f_0$, of the generated pulse, $c_0$ is the phase speed in the cell at $f_0$, and $a$ is the descriptive length of the transducer, bulk waves travel as collimated beams from the transmitting transducer through the cell and back. The waveform is detected by a receiving transducer, either positioned at the other side of the cell (known as pitch-catch configuration) or using the same transducer to generate the pulse and receive echoes from inhomogeneities and boundaries (known as pulse-echo configuration). To ensure the beam-like propagation, most studies have employed pulses with $f_0 \ge$ \SI{1}{\mega\hertz}, although a small number of authors have chosen frequencies as low as \SI{300}{\kilo \hertz} \cite{zhang2023ultrasonic} with the intent of using bulk wave methods. The investigation of SOC using ultrasonic propagation in LIBs has employed either ultrasonic measurements concurrent with cell cycling \cite{davies2017state, hsieh2015electrochemical, robinson2019examining, robinson2019spatially, qin2023construction, huang2022quantitative,Galiounas2024Investigation} or at specific SOC levels with the cell at rest \cite{gold2017probing, sood2013health, wu2019ultrasonic, rohrbach2021nondestructive, huang2023precise, farinas2023lithium, Farinas2024Contactless}. Concurrent ultrasonic measurements and cell charging is more practical for real-time monitoring of battery SOC to avoid overcharging or over-discharging. Taking measurements after the cell has reached a certain SOC and rested may be more helpful for understanding SOC-specific structural changes, like the changing stages of graphite intercalation, which is easier to identify in an electrochemically static state. Regardless of the approach, most authors have used SA and TOFS as ultrasonic metrics of SOC. Other authors make use of related measurements like the calculation of wave velocity\cite{sun2022ultrasonic, farinas2023lithium}, time of the maximum pulse envelope, via the Hilbert transform\cite{appleberry2022avoiding}, estimates of the signal energy using the square of the SA\cite{shen2023situ}, signal rise time\cite{Wei2023Risetime}, amplitude of the maximum frequency in the received spectrum\cite{appleberry2022avoiding}, counts of the number of times that the ultrasonic signal exceeds a specified threshold \cite{Sun2022AcousticCount}. Moreover, due to the heterogeneous, multilayered, and dynamic nature of LIBs, some researchers have explored other more advanced signal metrics such as the attenuation of peak resonance frequencies\cite{Feiler2024ModelingAttenuation,Feiler2024Investigation}, tracking shifts in the power spectral density (PSD) of the ultrasonic signals\cite{zhao2021state,Hamann2025psd}, and combined metrics that employ time and frequency features such as the $S$-value proposed in Ref.~\cite{sun2023ultrasonic}. Furthermore, some authors have argued that nonlinear stress-strain relationships may exist in LIB constituent materials due to their highly heterogeneous structure that includes localized damage and components that undergo phase transformations. It has therefore been suggested that UT techniques that exploit the effects of material nonlinearity, such as harmonic generation, may be more sensitive to changes in battery state\cite{Lin2025Quasi,Lin2025SOCNCM,Yuan2024QSCNonlinear,Sun2024SecondHarmonic}. For example, Sun et al.~estimated the nonlinearity parameter, $\beta$, which compares the amplitudes of the fundamental and second harmonic frequencies\cite{Sun2024SecondHarmonic}. In other work considering material nonlinearity, Yuan et al.~measured ``quasi-static'' pulses (QSP) at \SI{0.5}{\mega\hertz} that were generated by a \SI{8.5}{\mega\hertz} bulk longitudinal wave\cite{Yuan2024QSCNonlinear} and Lin and colleagues measured similar QSP at \SI{0.25}{\mega\hertz} generated by a \SI{5}{\mega\hertz} primary guided wave\cite{Lin2025Quasi}. 

In addition, since SA and TOFS measured using contact methods are highly sensitive to the contact pressure between the ultrasonic sensors and the cells, and researchers must be careful to ensure good coupling for reproducibility and accuracy, several researchers have employed non-contact ultrasonic techniques, such as air-coupled ultrasound \cite{chang2019real,farinas2023lithium,Farinas2024Contactless}, laser-generated ultrasound \cite{Sampath2024RealNonContact}, or using contactless electromagnetic acoustic transducers (EMATs) \cite{siegl2022electromagnetic, Li2024BatteryState}.  However, these non-contact techniques present challenges that have been encountered in other applications of ultrasonic testing, namely low transduction efficiency due to the high acoustic impedance mismatch between air and the transducers, and severe roadblocks in the eventual implementation in real-world applications\cite{hosten1996measurement,chimenti2014review}. As a result, most ultrasonic measurements of LIB have employed contact methods, using conventional coupling agents such as liquid couplants (e.g., grease\cite{sun2022ultrasonic}, canola oil\cite{Feiler2024Investigation}, acoustic gel\cite{Chang2024Relating}), solid couplants (e.g., rexolite\cite{Sun2025ChemoMechanics} or epoxy resin for fixed contact approaches\cite{mcgee2023ultrasonic}), or immersion approaches, which employ liquid immersion (e.g., water \cite{Ajaereh2024Assessing}, silicone oil \cite{huo2022evaluating}) to enable mechanical scanning. Still, there is a growing niche for air‑coupled and other contactless methods when handling or contamination is critical.

In contact methods, most authors have found that the TOF tends to decrease with increasing SOC, which allows TOFS to be an indicator of SOC \cite{hsieh2015electrochemical, gold2017probing, davies2017state, kim2020ultrasonic,wu2019ultrasonic, appleberry2022avoiding, bommier2020operando, chang2019real, copley2021measurements, sun2022ultrasonic, owen2022operando, robinson2019examining, knehr2018understanding, chang2021operando, galiounas2022battery, huang2023precise, feiler2023interplay, liu2023decoupling, zhang2023ultrasonic, farinas2023lithium, mcgee2024ultrasonic, Fu2024Overview}. Conversely, the SA tends to increase with increasing SOC, as reported by most studies \cite{chang2019real, davies2017state, kirchev2023part3, gold2017probing, sun2022ultrasonic, ke2022potential, hsieh2015electrochemical, chang2021operando, meng2022robust, xu2022feature, huang2023precise, shen2023situ, zhang2023ultrasonic, Xu2024Dynamic, mcgee2024ultrasonic}. However, some authors have observed a decrease in SA with increasing SOC, which may indicate the presence of non-bulk modes or other factors that were not accounted for in the experimental apparatus \cite{appleberry2022avoiding, sun2022ultrasonic, galiounas2022battery, zhang2023state, kirchev2022part1}. While these latter works were conducted on cells with different cathode chemistries, form factors, and capacities, it is noteworthy that despite these differences, one typically observes the same general trends in ultrasonic measurements, namely a decrease in TOF and an increase in SA with increasing SOC. Under normal operating conditions, this behavior may be ascribed to the graphite anode, which experiences larger changes in volume and Young's modulus than the cathode in LIBs, as shown in Fig.~\ref{fig:ElectrodesVolumeChange}. These results suggest that the anode behavior drives the cell response that is measured when using bulk longitudinal wave propagation \cite{davies2017state,chang2021operando}.

After the initial works described above, authors have tried to delineate the effects of different cell components or experimental variables on the observed ultrasonic signal trends in an effort to improve the specificity of ultrasonic testing methods for LIB monitoring. Liu et al.~ built custom half-cells with extra-thick separators (27 times larger than standard cells) in order to decouple the effects of the anode and cathode from the ultrasonic propagation. They found that the TOF decreased and the SA increased with SOC for the graphite anode, while the lithium iron phosphate (LFP) cathode showed opposite trends for both TOF and SA\cite{liu2023decoupling}. Meng et al.~\cite{meng2022robust} used a frequency sweep excitation from \SI{0.5}{\mega\hertz} to \SI{3.5}{\mega\hertz}, finding that the SA increased for the tested \SIrange{0}{80}{\percent} SOC range for all tested frequencies, although to different extents. They also observed a cell thickness change from \SIrange{4.50}{4.60}{\milli\meter} for the same SOC range. Similarly, Sun et al.~\cite{sun2022ultrasonic} performed experiments at \SI{750}{\kilo\hertz}, \SI{1}{\mega\hertz} and \SI{1.5}{\mega\hertz} excitation frequencies. They identified a direct correlation of SA with SOC for the \SI{1.5}{\mega\hertz} frequency. However, they did not find clear correlations of SA with SOC at the \SI{750}{\kilo\hertz} and \SI{1}{\mega\hertz} frequencies, which they attributed to the low sensitivity of the transducer used for those measurements.  Nonetheless, they showed that the attenuation of the three frequencies were sensitive to battery SOC changes . The choice of input frequency deserves attention for its effect on wave propagation within the cell. Different input frequencies may excite guided types of wave propagation within cells, or may experience frequency-specific attenuation or wave speeds based on the layering structure and constituent properties. McGee et al.~highlighted the importance of understanding the mode of propagation, as different input frequencies were shown to excite different propagating modes within the battery using the same transducers due to the response of the transmitting transducer. Additionally, they noted that through-thickness bulk modes and guided wave modes may have different SA and TOFS trends with cycling and heating\cite{mcgee2023ultrasonic}. Regarding the relationship between ultrasonic metrics and SOC, some authors have explored extending this concept to map SOC as a function of the location within the cell since ultrasonic reflection/transmission scanning allows localized estimation of mechanical property changes, thereby enabling spatially-varying estimates of the SOC\cite{huang2023precise,Gao2024SOCDetection,Ajaereh2024Assessing,Tian2024AISOCdistribution,Tang2025Uneven}. Understanding spatial heterogeneity in the SOC is crucial since it could indicate non-uniform utilization of the lithium inventory and active materials within the cells, which may influence cell degradation\cite{Liu2020ComputationalHeterogeneity}. For example, Xie et al.\cite{Xie2022Inhomogeneous} explored inhomogeneous degradation resulting from low temperatures and fast charging operating conditions and used bulk ultrasonic waves to help identify localized changes in SOC. 

\begin{figure*}[ht]
    \centering
    \includegraphics[width = 0.85\linewidth]{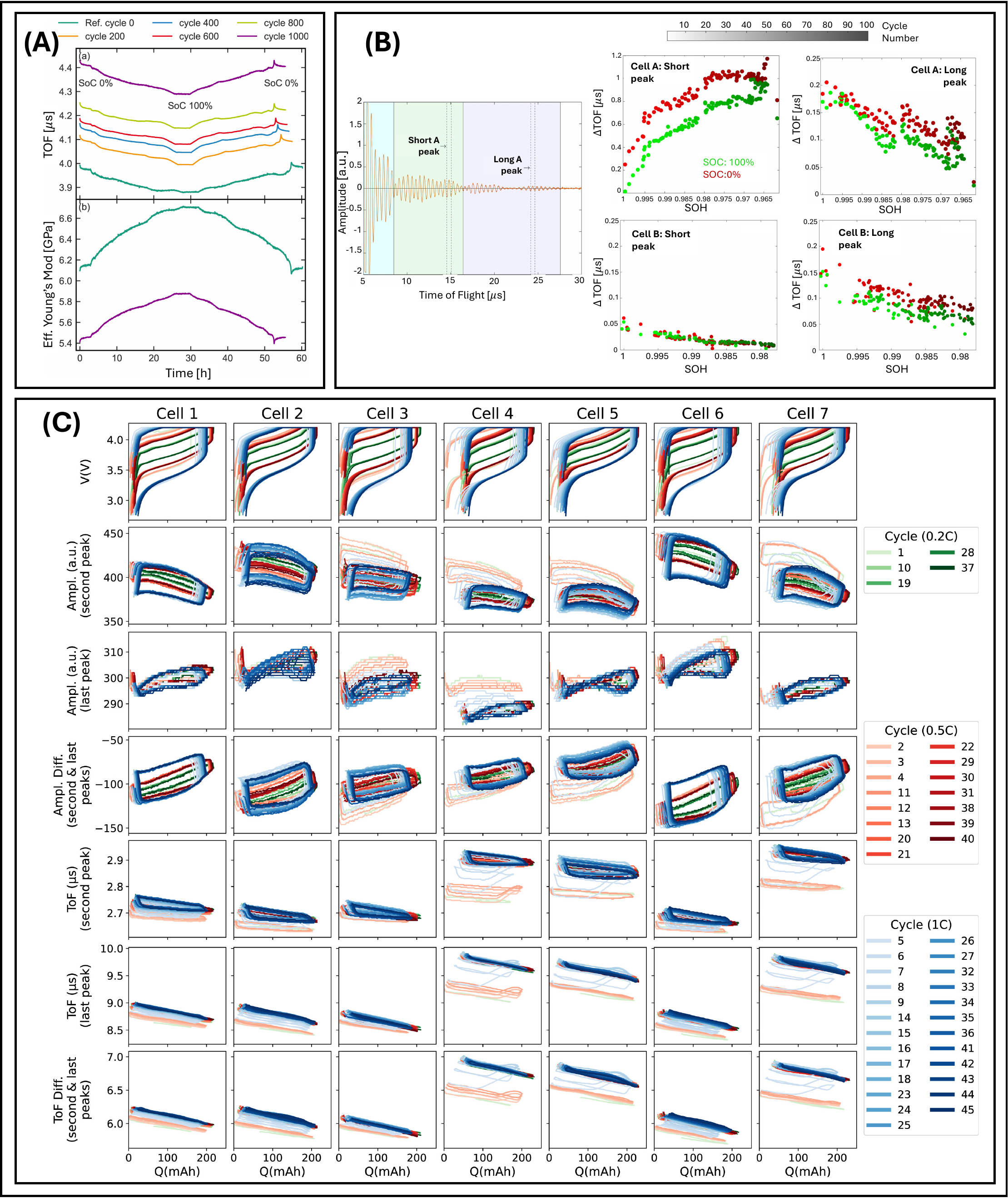}
    \caption{\label{fig:BulkWavesSummary}{Use of ultrasonic bulk waves for SOC/SOH assessment. \textbf{(A)} Feiler et al.~found that the TOFS of bulk longitudinal waves increased for all SOC with decreasing SOH. Reproduced from Ref.~\cite{feiler2023interplay}; licensed under a Creative Commons Attribution 4.0 International (CC BY 4.0) license (\href{https://creativecommons.org/licenses/by/4.0/}{https://creativecommons.org/licenses/by/4.0/}). \textbf{(B)} Trends of earlier and later echo peaks TOF shifts (of about \SI{1}{\micro\second}) with decreasing SOH for two different NMC811 cells. Reproduced from Ref.~\cite{Williams2025BatteryAge}, Copyright (2025), with permission from Elsevier. \textbf{(C)} Voltage and acoustic metrics measured during early aging at different C-rate conditions. Reproduced from Ref.~\cite{Galiounas2025Generalisation}; licensed under a Creative Commons Attribution 4.0 International (CC BY 4.0) license (\href{https://creativecommons.org/licenses/by/4.0/}{https://creativecommons.org/licenses/by/4.0/})}}
\end{figure*}

For bulk wave propagation, the relationship between TOFS and reduced SOH is better understood than that between SA and SOH. The mechanisms that contribute to a change in TOFS are either changes in the propagation path length or changes in the effective stiffness or density of the cell. For example, cells expand as they age which increases the propagation distance and can cause an increase in the TOFS if the sound speed does not change appreciably. It is more difficult to explain how various aging mechanisms that may occur simultaneously can affect the SA. These mechanisms include microcracking of active particles \cite{Edge2021BatDegradation}, continued growth of the SEI \cite{Li2025DegradationModes}, or increasing internal cell pressure \cite{knehr2018understanding, Gulsoy2024}. Perhaps because it is easier to interpret changes in the TOFS of bulk longitudinal waves with aging, at present there is more agreement in the literature on this relationship than between SA and SOH. Many authors have found that the TOFS increases over the entire SOC range with decreasing SOH as shown in Fig.~\ref{fig:BulkWavesSummary}.A\cite{kim2020ultrasonic, davies2017state, wu2019ultrasonic, knehr2018understanding, feiler2023interplay}. This increase has been attributed to a combined effect of electrode thickening and mechanical softening with aging. Likewise, groups who only investigated TOFS at a specific SOC have found that TOF increases with decreasing SOH \cite{cai2024remaining}. A recent study by Williams et al.~\cite{Williams2025BatteryAge} has shown that tracking TOFS from different portions of a single reflected ultrasonic signal using pulse-echo configuration can have different trends with decreasing SOH. As shown in Fig.~\ref{fig:BulkWavesSummary}.B, the TOF trends of the earlier and later echo peaks from A-scans differed between the two tested NMC811 cells. In cell A, earlier peaks consistently shifted to longer TOFS, while long peaks shifted to shorter TOFS as SOH decreased. This behavior was not observed in cell B, where both peaks exhibited matching trends with decreasing SOH. Those authors did not provide a definitive explanation for these differences in the cell and attributed it to possible manufacturing discrepancies between the cells. More research should be conducted on a larger sample of cells and a greater number of cycles to identify the underlying causes of these results.

When investigating the effect of SOH on US propagation, it is important not to extrapolate trends from the first few cycles. Authors have begun to delineate early-life aging mechanisms that occur during the initial charge-discharge cycles from more long-term aging mechanisms that occur progressively as the cell is cycled further since it has been shown that different trends may occur during battery degradation due to SEI formation at a initial stage, followed by stabilization and saturation stages at a higher number of cycles \cite{Edge2021BatDegradation}. Knehr et al.~cycled a cell 100 times and found that the TOFS increased mostly in the first 12 cycles during what they call the ``break-in'' period \cite{knehr2018understanding}, with more modest increases in TOFS occurring in subsequent cycling. They observed that the cell thickness increased most rapidly during the ``break-in'' period, and hypothesized that the increasing internal pressure within the cell may be forcing the electrolyte to wet previously inactive regions of the cathode material, in a process they call ``electrode crosstalk'' \cite{knehr2018understanding}. Furat et al.~\cite{Furat2022} and Wade et al.~\cite{Wade2023} have noted that particle cracking occurs in the first 10 cycles, which may be a contributing factor to the changes associated with the ``break-in'' period in some cases, although the scanning electron microscopy (SEM) images provided by Knehr et al.~\cite{knehr2018understanding} do not suggest particle cracking occurs in their samples. Davies et al.\cite{davies2017state} and Wu et al.\cite{wu2019ultrasonic} also found that the most rapid changes in TOFS occurred in the early cycles. It is also important to note when conducting aging studies that accelerated aging is not entirely representative of aging due solely to charge-discharge cycling. Accelerated aging is often preferred when conducting any experiment on LIBs exploring reductions in SOH due to the extremely long time-scales involved for SOH reduction, for example, Deng et al.~cycled cells 2,500 times over the course of two years to monitor changes in battery properties and ultrasonic metrics \cite{deng2020ultrasonic}. Although faster to perform, caution should be exercised when performing accelerated aging experiments, as the introduction of other variables, such as increased cell temperature or extended voltage ranges, can affect the mechanical properties of the cell over time. Some works have explored the influence of different charge/discharge current rates (C-rate, which is defined as the ratio of the current to the rated capacity, and refers to the speed of charging and discharging) and the temperatures generated for these different operating conditions on the propagation of bulk waves in LIBs \cite{owen2022operando, Xu2024Dynamic}. For instance, Fig.~\ref{fig:BulkWavesSummary}.C shows a summary of the experiments performed by Galiounas et al.~\cite{Galiounas2025Generalisation}, where they cycled seven cells at three C-rates (0.2C, 0.5C and 1C). Other authors showed approaches to account for temperature correction \cite{SoleimaniBorujerdi2024, Owen2024OperandoTemperature}, but this is still an open research question that is being investigated in terms of decoupling the C-rate and temperature conditions on the behavior of bulk ultrasonic waves for different battery states. In this sense, Galiounas et al. \cite{Galiounas2024Investigation} exposed the importance of careful consideration of temperature effects and thicknesses changes over degradation because they can mask degradation markers, generating misinterpretations of the ultrasonic insights.

As previously noted, the relationship between SA and SOH is less clear. Some authors have found that the SA increased globally with decreasing SOH \cite{appleberry2022avoiding, wu2019ultrasonic}, while others found the opposite trend \cite{davies2017state, sun2023ultrasonic, wasylowski2023situ, qin2023construction, cai2024remaining, deng2020ultrasonic, chang2020understanding, kirchev2022part1}. Wu et al.~found that the SA increased consistently for the first 50 cycles, noting a potential leveling off after that point, perhaps indicating an early-life aging mechanism \cite{wu2019ultrasonic}. Sun et al.~ found that SA decreases with decreasing SOH \cite{sun2023ultrasonic}, which they attribute to loosening structure within the cell and the loss of contact between active particles. This hypothesis would also be consistent with the results from others that found an increasing TOFS with cell aging. There are many different aging mechanisms that occur in cells, both due to charge-discharge cycling and from calendar-aging, which may affect the mechanical properties of the cell. In order to improve the understanding of the effect of SOH on SA, we suggest that authors include the full history of cells when discussing UT results, including time from purchase to experiment. When possible, authors should additionally identify aging mechanisms either via advanced imaging techniques or through other analyses such as electrochemical impedance spectroscopy or incremental capacity analysis. 

Regarding SA trends, Chang et al.~charge-discharge cycled a cell for over \SI{800}{\hour} at \SI{60}{\celsius} noting that the TOFS and SA decreased with cycle number \cite{chang2020understanding}. Cai et al.~conducted an accelerated aging study using bulk wave ultrasonic inspection on a \SI{10}{\ampere\hour} pouch cell cycled 200 times at \SI{40}{\celsius}. The authors calculated two measures of SA and one measure of TOFS. However, their TOFS measure appears to be susceptible to errors in the automated calculation, as indicated by un-physical jumps in the TOFS values. Such inconsistencies are common in automated TOFS calculations due to limitations in time resolution \cite{cai2024remaining}. Similarly, Feiler et al.~cycled a cell 1000 times, taking ultrasonic measurements every 200 cycles and reported a decrease in SOH to \SI{95}{\percent} after 1000 cycles and noting that the SA consistently decreased with aging\cite{feiler2023interplay} (see Fig.~\ref{fig:BulkWavesSummary}.B). During LIBs degradation, sometimes LIBs can exhibit a rapid nonlinear degradation drop commonly referred to as ``knee'' points, where the SOH will suddenly decrease due to various ``knee'' degradation mechanisms such as lithium plating, electrode saturation, resistance growth, electrolyte and additive depletion, percolation-limited connectivity, and mechanical deformation, as reviewed by Attia et al.~\cite{Attia2022Knee}. In this regard, Sun et al.~\cite{sun2023ultrasonic} used \SI{5}{\mega\hertz} US bulk waves to detect this nonlinear degradation decay on NMC811 pouch cells. For this purpose, they calculated a joint time-frequency metric based on the SA that exhibits distinct characteristic peaks that change throughout the cycling process. The disappearance or shift of these peaks correlates strongly with the onset of nonlinear capacity decay that occurred at about the 70$^\mathrm{th}$ cycle, providing a precursor signal to detect the ``knee'' point. They induced this behavior using high, uneven C-rates of 0.5C for charge and 1C for discharge, contrary to the 0.2C value recommended by the manufacturer of these cells. 

\subsubsection{Modeling Bulk Wave Propagation}
We can define two broad categories of modeling approaches for UT of LIBs: wave physics models and data-driven models. Wave physics models draw from multiscale wave mechanics to derive models that predict ultrasonic propagation, which enables inversion of ultrasonic signals to determine changes in constituent properties and geometry. In contrast, data-driven models rely on statistical methods like machine learning to solve for empirical relationships between wave propagation and battery state.  

\begin{figure*}[]
    \centering
    \includegraphics[width = 0.6\linewidth]{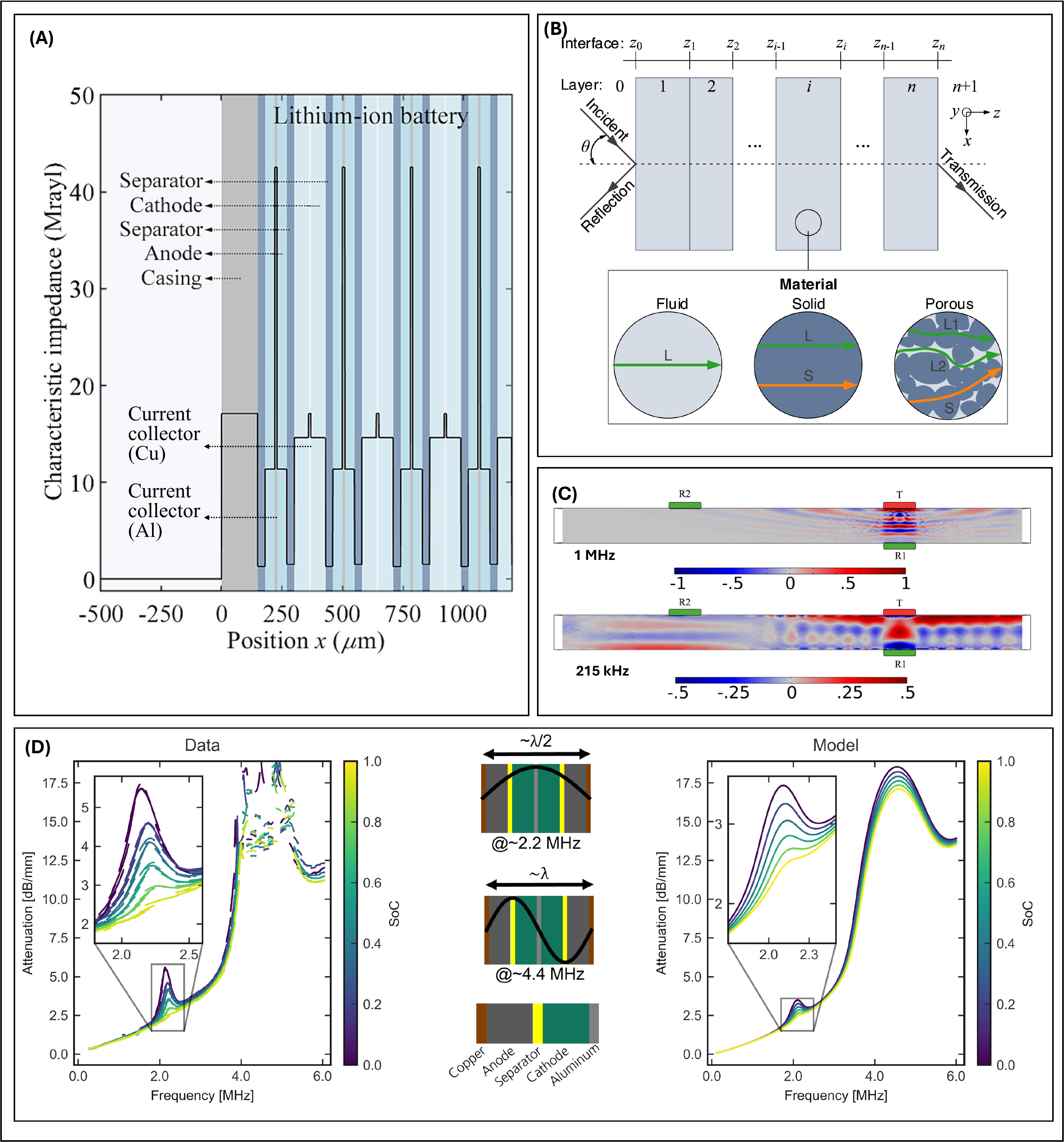}
    \caption{\label{fig:BulkWavesModeling}{Modeling as a key tool to understand ultrasonic bulk waves in LIBs. \textbf{(A)} Representation of the acoustic impedance mismatches between different layers of LIBs. Adapted from Ref.~\cite{Ma2024USScatteringModel}, Copyright (2024), with permission from Elsevier. \textbf{(B)} Each layer of the multilayered structure of LIBs can be modeled as a fluid, solid, or porous material, which will determine the number and type of propagating waves (L: longitudinal and S: shear). Electrodes and separators are modeled as porous materials consisting of elastic active particles in a fluid electrolyte, while the current collectors are assumed to be solid elastic materials. Reproduced from Ref.~\cite{huang2023stiffness}; licensed under a Creative Commons Attribution 4.0 International (CC BY 4.0) license (\href{https://creativecommons.org/licenses/by/4.0/}{https://creativecommons.org/licenses/by/4.0/}). \textbf{(C)} Simulation of US bulk waves (top) that are localized in space due to a beam-like field generated by transducers with a large aperture compared to the transmitted wavelength and guided waves (bottom) that can be used to interrogate larger cell volumes. The selected excitation frequency is a critical parameter that determines the type of ultrasonic propagating waves.  Reproduced from Ref.~\cite{mcgee2023ultrasonic}, Copyright (2023), with permission from Elsevier. \textbf{(D)} Measured and modeled attenuation per length on multilayered LIBs. High attenuation near \SI{2}{\mega\hertz} and \SI{4}{\mega\hertz} is due to stop bands associated with the periodic structure of pouch cells, as indicated by the unit cell shown between data and model results. Panel D has been Reproduced from Feiler et al.~\cite{Feiler2024ModelingAttenuation}; Feiler et al.~\cite{Feiler2024ModelingAttenuation} modified left panel from and reproduced middle panel from Feiler et al.~\cite{Feiler2024Investigation}; all licensed under a Creative Commons Attribution 4.0 International (CC BY 4.0) license (\href{https://creativecommons.org/licenses/by/4.0/}{https://creativecommons.org/licenses/by/4.0/}).}} 
\end{figure*}

\textbf{Wave Physics Models:} When constructing a physical model, researchers have many important questions to answer. For example, what is the best approach to accounting for the high number of layers within cells? Approaches taken to address this have included creating simplified models with reduced numbers of component layers \cite{li2019numerical, sun2023ultrasonic, guorong2019application, song2022ultrasonic, huang2022quantitative}, neglecting individual component layers \cite{sun2023ultrasonic, huang2022quantitative, mcgee2023ultrasonic}, or creating a repeating unit-cell of specific component layers and invoking Floquet-Bloch periodicity\cite{sun2023ultrasonic, huang2022quantitative, mcgee2024ultrasonic}. In general, the propagation of ultrasonic waves in this constantly changing, heterogeneous medium is heavily influenced by the impedance mismatches associated with the characteristic impedance of each layer within the cells, as indicated in Fig.~\ref{fig:BulkWavesModeling}.A. Other questions include how to model the dynamic mechanical response of electrode layers, which are composed of active materials, binders, conductive additives, and are soaked with liquid electrolyte. Researchers have approximated their dynamics by simply modeling them as being composed of only active materials \cite{sun2023ultrasonic}, or by using effective medium theory \cite{hsieh2015electrochemical, mcgee2023ultrasonic, mcgee2024ultrasonic}. For the latter case, the electrode materials have been modeled as either poroelastic \cite{gold2017probing, chang2019real, huang2022quantitative, gold2023ultrasound, binpeng2023ultrasonicreflection, huang2023stiffness, zhang2024exploring} or viscoelastic materials \cite{mcgee2023ultrasonic, mcgee2024ultrasonic}. The choice of assumed constitutive behavior dictates the equations of wave motion and required material properties. Finally, in terms of implementation, previous work has employed finite-elements \cite{li2019numerical, copley2021measurements, zhang2023ultrasonic, mcgee2023ultrasonic, binpeng2023ultrasonicreflection} or transfer matrix methods \cite{meng2022robust, guorong2019application, song2022ultrasonic, binpeng2023ultrasonicreflection, huang2022quantitative, huang2023stiffness, mcgee2024ultrasonic,Feiler2024ModelingAttenuation, Ma2024USScatteringModel} to simulate the wave propagation. The paragraphs that follow provide a summary of relevant physical modeling approaches for understanding and approximating ultrasonic wave phenomena in LIB.

Wave physics models of bulk wave propagation in LIBs can be further divided into two approaches: (\textit{i}) layer-by-layer approaches which employ 1D wave equations and transfer matrix-based methods and (textit{ii}) effective medium theories (EMT). Layer-by-layer models consider the acoustic impedance contrast of each layers and account for phase accumulation across layers. Transfer matrix methods can predict frequency dependent transmission spectra as a function of loading conditions (i.e.~SOC), including band gaps in the transmission when layer thicknesses are on the order of propagating wavelengths. However, these models require knowledge of SOC-dependent densities and elastic moduli for each layer, which are difficult to estimate and measure as discussed in \ref{sec:ChallengesOpportunities}. Conversely, EMTs represent the the stack of layers and a single, homogeneous layer composed of an effective medium using volume averaging schemes\cite{christensen2012mechanics}. While EMTs are less computationally expensive, they cannot capture inter-layer dynamics, such as band gaps. Table \ref{tab:summaryTableModelBulk} presents a summary of key modeling studies on bulk wave propagation in LIBs. It presents the model description and dimensionality, cell representation, a description of properties modeled as SOC-dependent (as necessary), and primary findings.

\begin{table*}[]

\caption{Summary of key modeling approaches for bulk wave propagation in LIBs}
\label{tab:summaryTableModelBulk}
\begin{adjustbox}{width = \textwidth}
\begin{tabular}{p{3cm}>{\centering\arraybackslash}p{3cm}>{\raggedright\arraybackslash}p{7cm}>{\raggedright\arraybackslash}p{7cm}>{\raggedright\arraybackslash}p{8cm}>{\centering\arraybackslash}p{3.5cm}}
    \hline\hline
  Model type & Dimensions & Cell model & SOC-dependent property & Key finding/limitation & Relevant refs.\\
  \hline\hline
  1D propagation via coupled linearized conservation laws & 1D & 5-layer (casing, Cu-conductor, anode, separator, cathode) & SOC-dependent densities for LCO and graphite. & Acknowledged the difficulty in obtaining accurate material properties of cell components and that the model simplified the internal structure of the cell & Hsieh et al.~(2015)\cite{hsieh2015electrochemical}\\
  \hline
  Finite Element (FE) & 2D & 21-layer representation with electrolyte modeled as a discrete layer on both sides of the separator & No SOC-dependent properties & Used to detect void defects at different layer depths and simulate wave propagation generated via air-coupled excitation & Li and Zhou (2019)\cite{li2019numerical}\\
  \hline
  Finite-difference 1D wave equation & 1D & N cathode-separator-anode-separator unit cells & SOC-dependent densities of LCO and graphite. All properties from Hsieh et al.~\cite{hsieh2015electrochemical}. & Accurately simulated through-thickness transmission ultrasonic signals between 0.5 and 3 MHz. No claims regarding simulated TOFS or SA with SOC & Copley et al.~(2021)\cite{copley2021measurements}\\
  \hline
  Global matrix method & 1D & 127 layers & SOC-dependent densities and elastic moduli for NCM and graphite. & Identified bandgaps centered at 0.456, 1.37, 2.28, and \SI{3.19}{\mega\hertz} associated with the layer periodicity. Found that bandgaps shifted by $\approx$\SI{10}{\kilo\hertz} when the cell SOC increased from 0 to 0.8 SOC. & Meng et al.~(2022)\cite{meng2022robust}\\
  \hline
  Pseudo-two-dimensional FE & 1D & 5$\times$NMC811-separator-graphite unit cells & Model includes stress induced via lithium insertion or removal to simulate changes in modulus and density and includes SEI. & Higher TOF and reduced amplitude of longitudinal wave for the aged cell compare to pristine state suggests a reduction in the effective modulus and increase in attenuation with cell aging due to SEI damage & Sun et al.~(2023)\cite{sun2023ultrasonic}\\
  \hline
  Legendre orthogonal polynomial method (LOPM) & 1D & 11- and 51-layer & Energy-strain approach from \textit{ab initio} models in conjunction with Ref.\cite{qi2010threefold} to calculate the SOC-dependent Young’s modulus of graphite. & Good agreement between LOPM model and FE validation to calculate reflection coefficient cell immersed in a fluid at different angles of incidence, frequencies, and SOC  & Song et al.~(2022) \cite{song2022ultrasonic}\\
  \hline
  2D FE and global matrix with poroelasticity & 1D and 2D & 59-layer with Voronoi polygons to account for porosity & Simulated material property values based on lithium ion concentrations via COMSOL electrochemical module. & TOF increased and the SA decreased with decreasing SOC. Identified pass-bands at frequencies around \SI{650}{\kilo\hertz} and \SI{900}{\kilo\hertz} that shifted to higher frequencies with increasing SOC. & Zhang et al.~(2023)  \cite{zhang2023ultrasonic,binpeng2023ultrasonicreflection}\\
  \hline
  Transfer and stiffness matrices with poroelasticity & 1D & Porous-solid-porous anode & Not specified. They used values of the literature at a fixed SOC. &  Slow wave speed is multiple orders of magnitude slower than that of the fast wave and slow wave attenuation coefficient six orders of magnitude larger than that of the fast wave. Demonstrated experimental method to measure angle-dependent transmission coefficient for model verification. & Huang et al.~(2023)  \cite{huang2023stiffness}\\
  \hline
  FE with homogenized properties \newline and periodic transfer matrix (TM) & 2D (FE) \newline and 1D (TM) &5-layer unit cell & SOC-dependent elastic moduli and densities. & Guided or bulk wave excitation based on excitation frequency. TOF decreases with increasing SOC. TOF increases and SA decreases with temperature due to separator softening & McGee et al.~(2023\cite{mcgee2023ultrasonic} and 2024\cite{mcgee2024ultrasonic})\\
  \hline
  Transfer matrix & 1D &  17$times$9-layer unit cell  & SOC-dependent sound speed and density & Calculated group velocity and attenuation of transmitted ultrasonic waves showing potential for fitting ultrasonic spectroscopy data to determine individual material properties. Obtained good agreement between model and experiment for different values of SOC. & Feiler et al.~(2024)  \cite{Feiler2024Investigation,Feiler2024ModelingAttenuation}\\
  \hline\hline
\end{tabular}
\end{adjustbox}
\end{table*}

Some findings reported in Table \ref{tab:summaryTableModelBulk} lead to additional insights regarding current modeling approaches for bulk wave propagation in LIB. The key modeling challenges are associated to understanding the existence of multiple length scales present in LIB (see Fig.~\ref{fig:UnitCellLIBsPoroelastic}) which yields frequency- and direction-dependent phase speed and attenuation. The frequency-dependence is a function of constituent material properties (i.e.~viscosity and viscoelasticity), the periodic layering of electrodes, conductors, and separator, and microscale heterogeneity due to the existence of particles in viscous fluid electrolyte within the electrode layers. Microscale heterogeneity has been the subject of recent work trying to understand dispersive effects introduced by the structure of the electrodes. Huang et al.\cite{huang2023stiffness} developed a multilayer model that included poroelasticity whose results provided a rationale for why Biot's slow wave has not been reliably detected in LIBs experiments, despite the fact that it has been observed in other areas of wave physics\cite{plona1980observation, johnson1982acoustic,johnson2017impact}. Measurement of the slow wave is of interest because modeling suggests that it would demonstrate high sensitivity to electrode porosity and electrode dry out as indicated in Fig.~\ref{fig:BulkWavesModeling}.B. Works by McGee et al.~\cite{mcgee2023ultrasonic} and Feiler et al.~\cite{Feiler2024ModelingAttenuation} recognized the importance of source frequency selection in two different ways. McGee et al.~used a finite element model for wave propagation in a homogenized anisotropic pouch cell to shows that excitation by a small source at different frequencies enabled the selection of either through-thickness bulk wave propagation (1 MHz) or in-plane guided wave propagation (215 kHz), as shown in Fig.~\ref{fig:BulkWavesModeling}.C. Feiler et al.~\cite{Feiler2024ModelingAttenuation} investigated frequency-dependent behavior using a transfer matrix approach to model the attenuation of transmitted waves as a function of frequency, showing frequency ranges with elevated attenuation associated with stop bands due to the periodic layered structure. They obtained good agreement between modeled and experimentally-measured attenuation in LIBs for different values of SOC in the frequency range of \SIrange{0.5}{6}{\mega\hertz} as shown in Fig.~\ref{fig:BulkWavesModeling}.D. 

As indicated by the work of Feiler et al.\cite{Feiler2024ModelingAttenuation}, band gaps in the transmission coefficient of LIB cells as a consequence of their periodically layered structure. As described by Ma et al.~\cite{Ma2024USScatteringModel}, constructive interference of reflected waves at each interface occurs at these frequencies. This increases the reflection, lowers transmission, and can be observed as an increase in signal attenuation as has been reported in Refs.~\cite{meng2022robust,Feiler2024Investigation,Ma2024USScatteringModel}. Meng et al.~\cite{meng2022robust} reported band gaps at approximately \SI{0.5}{\mega\hertz} and \SI{1.3}{\mega\hertz} and developed a transfer matrix model with Floquet-Bloch periodic boundary conditions to describe their observations. As noted above, Feiler et al.~\cite{Feiler2024Investigation} also experimentally observed a band gap, but the structure of the LIB they studied resulted in a gap between \SIrange{2}{2.5}{\mega\hertz}, where they observed enhanced signal attenuation. Likewise, the work by Ma et al.~\cite{Ma2024USScatteringModel} used a transfer-matrix model with Floquet-Bloch periodicity to show that LIBs exhibit a band gap for longitudinal waves exists from \SIrange{2}{3}{\mega\hertz} in the LIB they studied, which was validated with experiments. These results and others like them illustrate the critical importance of frequency selection when using ultrasonic methods to measure or model LIBs, since these band structures are dependent on both the layer properties and unit cell dimensions and can therefore be used as metrics of changes in layer properties, geometry, or both as discussed by Ren et al.~\cite{Ren2025Decoding}.

\textbf{Data-Driven Models:}
Many research groups have opted to use statistical or data-driven models including limited or no physical behaviors to find correlations between US signals and internal battery states including SOC, SOH, and moderate changes in operating temperature \cite{knehr2018understanding, davies2017state, gold2017probing, qin2023construction, cai2024remaining, wu2019ultrasonic, galiounas2022battery, huang2023precise, zhang2023state, copley2023prediction, Xu2024Dynamic, Yang2024SOCNeural,Wang2024ApplicationML,Liu2024SOHWave-based,Liu2025ExplainableAI}. These methods have generally been successful, but remain limited as the creation of a model is specific to its training data. A model created based on a specific training set of experimental data using given excitation frequencies, cell geometry, anode and cathode chemistry, etc, may not be adaptable enough to be applied to scenarios where the cells or experimental setups differ. Additionally, since these models do not consider the underlying physical behaviors, it is possible for them to draw unphysical conclusions\cite{willcox2021imperative}. For example, some models claim that specific data points in a received signal carry more weight in correlating with cell state. However, any individual point in a received waveform does not correlate with any particular component or position in the cell. The point may be statistically correlated, but this signifies nothing of physical interest. This limitation is of particular importance, as it reduces the ability of a researcher to say with certainty which cell component or cell location is the cause of the perceived cell state. Despite this complication, data-driven methods are of significant interest to this area of research and a significant number of publications have shown promising results, which are summarized below.

Regarding the challenge of generalization for data-driven models, Galiounas et al.~ \cite{Galiounas2025Generalisation} examined six features extracted from reflected bulk ultrasonic waves from seven different cells tested over $~40$ cycles for three different C-rates (0.2C, 0.5C, and 1C), with the main goal of finding general patterns between acoustic features and SOC that can be used between different cells (Fig.~\ref{fig:BulkWavesSummary}.C). They found a lack of generalization for the ultrasonic features selected, suggesting the presence of cell-specific patterns rather than generalizable trends for the acoustic features. In this sense, the results call for caution in evaluating the efficacy of acoustic signals for battery diagnosis, as previous studies may have overlooked the difficulty of generalization in their claims. However, this is still an open research question that future experiments and modeling will address.

Notable initial work in this area includes that of Davies et al.~who applied Support Vector Regression (SVR) to develop a model capable of estimating SOC and SOH based on various combinations of input data, including TOFS, SA, or the total waveform, in conjunction with cell voltage. Their model achieved an impressive accuracy, predicting SOC and SOH within approximately \SI{1}{\percent} \cite{davies2017state}. Building on the integration of multiple data sources, Wu et al.~utilized the Mahalanobis distance data fusion method to combine cell temperature measurements with ultrasonic signals. When applying this approach to a cell overcharged to \SI{5}{\volt}, they found that their model provided an earlier indication of damage compared to relying on temperature data alone \cite{wu2019ultrasonic}. Further exploring deep learning techniques, Galiounas et al.~developed a feed-forward neural network that incorporated entire received waveforms, specific waveform data points exhibiting high Pearson correlation with SOC, and Fourier coefficient magnitudes obtained via fast Fourier transform (FFT). Their model achieved highly accurate SOC estimation, with an error as low as \SI{0.75}{\percent} when analyzing C-rates between 0.2 and 1C \cite{galiounas2022battery}. Similarly, Huang et al. applied deep learning techniques, including a feed-forward neural network, a convolutional neural network, and a fully connected neural network, using the full received waveform as input for SOC estimation, achieving a root mean squared error (RMSE) of \SI{3.02}{\percent} \cite{huang2023precise}.

Additional notable data-driven methods include the work of Zhang et al.~who explored a different neural network architecture, developing a backward propagating neural network that leveraged six time-domain features of the received waveforms: time-domain peak, envelope peak, energy integral, waveform index, Kurtosis coefficient, and shape coefficient—as inputs\cite{zhang2023state}. Their model predicted SOC with an RMSE of \SI{7.42}{\percent} and cell temperature with an RMSE of \SI{0.4}{\celsius}. Taking a regression-based approach, Qin et al.~employed nonlinear multivariable regression to model the relationship between two SA-based measures and the reduction in capacity of a single cell cycled 200 times. Their model achieved an RMSE of \SI{0.89}{\percent} \cite{qin2023construction}. Expanding on this work, their research group incorporated machine learning techniques, integrating the Random Forest and extreme gradient boosting (XGBoost) algorithms to enhance SOH estimation based on ultrasonic characteristics similar to SA and TOFS. Their improved model achieved a mean squared error of \SI{0.4}{\percent} when estimating capacity from the same cell’s data \cite{cai2024remaining}.

Beyond conventional machine learning approaches, Copley and Dwyer-Jones developed a genetic algorithm based on a finite difference solution to the 1D wave equation. Unlike other studies that relied on ultrasonic signal characteristics such as SA and TOFS, their model directly processed raw received waveforms. Given material thicknesses, their algorithm predicted wave speed of individual layers with an error ranging from \SI{3} to \SI{29}{\percent}, while, given material properties, it estimated layer thickness with an error between \SI{3}{\percent} and \SI{13}{\percent} \cite{copley2023prediction}. Xu et al.~introduced an adaptive Extended Kalman Filter and an adaptive H-infinity filter, using the ultrasonic signal feature they termed the ``energy of the oscillatory wave,'' which corresponds to the squared SA of the received signal before the first echo. These models accurately predicted SOC during discharge under varying temperature conditions (\SI{-5}{\celsius},\SI{25}{\celsius}, and \SI{40}{\celsius}) with an RMSE of approximately \SI{1}{\percent} \cite{Xu2024Dynamic}.

Another work by Zhang et al.~reported a strong correlation between the 1$^{\mathrm{st}}$ and 2$^{\mathrm{nd}}$ transmitted pulses parameters and battery state, where the amplitude of both of them increased with SOC, while the TOF of the 2$^{\mathrm{nd}}$ transmitted pulse decreased for the same SOC behavior \cite{Binpeng2025Ultrasonic}, indicating an overall increase in stiffness and reduction in losses with increased SOC. Ultrasonic signals showed increasing attenuation and TOF shifts as aging progressed over 183 charge/discharge cycles. To accurately model these relationships, a hybrid sparrow search algorithm (SSA) and relevance vector machine (RVM) approach was developed, optimizing kernel width for precise state estimation. The model achieved a maximum relative error of \SI{1.51}{\percent} for SOC and an absolute error of \SI{0.79}{\percent} for SOH.

Despite these advances, questions remain regarding the direct correlation between AI-based battery state predictions and physical changes within the battery. Addressing this challenge, a recent study applied explainable AI techniques to analyze selected ultrasound signal features such as TOF, max-min values, peak-peak values, and the maximum amplitude of frequency spectra\cite{Liu2025ExplainableAI}. This research provided quantitative insights into the influence of these features on different regions of the predicted SOH, shedding light on the interpretability and physical relevance of AI-driven battery state estimations. However, it is still an open research question how different physical mechanisms (such as gas generation, electrode delamination, voiding, and electrode expansion) are related to these ultrasonic features during battery operation and aging.

\subsubsection{UT for SOC/SOH Detection: Guided Waves}

\begin{figure*}[ht]
    \centering
    \includegraphics[width=\linewidth]{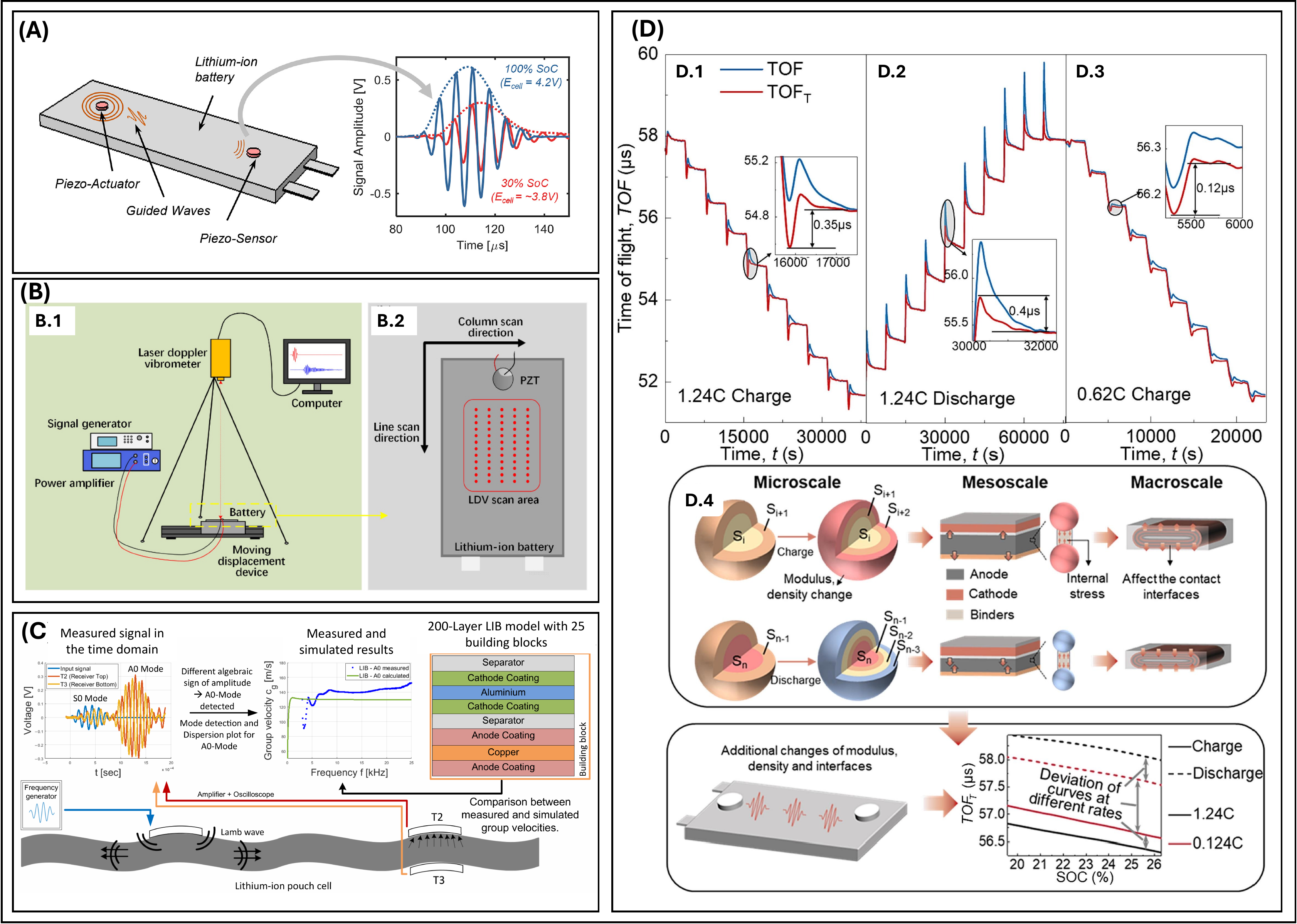}
    \caption{Use of ultrasonic guided waves for SOC and/or SOH assessment. \textbf{(A)} Transducer positioning for guided wave measurements using disk-shaped contact transducers. The receiving transducer is placed on the same surface of the cell as the transmitter. Reprinted from Ref.~\cite{ladpli2018estimating}, Copyright (2018), with permission from Elsevier. \textbf{(B.1-B.2)} Experimental setup using a scanning laser Doppler vibrometer to monitor out-of-plane displacement due to guided wave propagation. Reproduced from Ref.~\cite{Li2022StateUGW}; licensed under a Creative Commons Attribution 4.0 International (CC BY 4.0) license (\href{https://creativecommons.org/licenses/by/4.0/}{https://creativecommons.org/licenses/by/4.0/}). \textbf{(C)} Schematic to determine Lamb wave modes generated in a multilayered LIB. Reproduced from Ref.~\cite{koller2022determination}; licensed under a Creative Commons Attribution 4.0 International (CC BY 4.0) license (\href{https://creativecommons.org/licenses/by/4.0/}{https://creativecommons.org/licenses/by/4.0/}). \textbf{(D.1-D.3)} TOF and $\mathrm{TOF_{T}}$ measurements during galvanostatic intermittent titration experiments for charge and discharge at various C-rates. \textbf{(D.4)} Illustrations of possible factors affecting TOF changes with different C-rates. Panel D reproduced from Ref.~\cite{Liu2024UGWCRate}, Copyright (2024), with permission from Elsevier}
    \label{fig:GuidedWaveSummary}
\end{figure*}

As discussed in Sec.~\ref{sec:wavePhenomena}, ultrasonic guided wave methods offer a unique approach to assess characteristics of LIBs. Guided ultrasonic waves can propagate longer distances and can be selected to probe specific regions within a battery \cite{chimenti2014review} and therefore may be more suited for detecting specific types of internal mechanical changes in LIB. Furthermore, the combination of bulk wave and guided wave ultrasonic testing may allow for more complete mechanical characterization of the cell and its components. Previous use of guided wave modes for UT has been achieved by placing the transmitting and receiving transducers on the same surface of the cell (see Fig.~\ref{fig:GuidedWaveSummary}.A) and using frequencies below \SI{1}{\mega\hertz}. The same ultrasonic signal characteristics, i.e.,~ SA and TOFS, can be used to quantify changes in guided waves associated with changing cell conditions.

Many groups have used longitudinally polarized piezoelectric transducers as their actuators to excite guided wave propagation within LIBs \cite{ladpli2018estimating, popp2019state, zhao2021state, zheng2021guided, li2022state, mcgee2023ultrasonic, Reichmann2023UltrasonicGuided}. While these transducers are not specifically designed to maximize the energy of guided wave modes, they can still excite and detect guided wave propagation within LIBs, providing insight into physical changes within cells under different loading scenarios. When using these non-specialized transducers to excite guided waves, it becomes necessary to prove that the observed waveforms are indeed due to guided wave motion either through experiments together with signal processing techniques \cite{koller2022determination} or via simulation \cite{mcgee2023ultrasonic}. Other groups have opted to design and/or operate transducers that have been optimized to transmit guided wave motion \cite{siegl2022electromagnetic, siegl2023sensor, An2024Construction}. While employing a receiving sensor on the cell face may be more appropriate for \textit{in situ} applications, monitoring guided wave propagation via non-contact methods including the use of a scanning laser Doppler vibrometer (SLDV), \cite{zhao2021state, zheng2021guided, Li2022StateUGW} as indicated by the schematic in Fig.~\ref{fig:GuidedWaveSummary}.B where an SLDV was employed to measure spatial information of the ultrasound guided wave propagation, or a contactless electromagnetic acoustic transducers \cite{Li2024BatteryState}.

While much less work has been done employing guided waves for UT of SOC compared to bulk waves, most authors find that the TOF of guided waves tends to decrease with increasing SOC. This observation has led to the use of TOFS as an indicator of SOC \cite{ladpli2018estimating, popp2019state, zhao2021state, jie2023ultrasonic,Tian2024StateOfCharge,Reichmann2023UltrasonicGuided}. However, Li et al.~\cite{li2022state} found the opposite trend, suggesting that more work is required to understand this discrepancy. These differences are likely due to both microstructural differences in each case and guided wave selection, both of which can have a significant impact on the microscale deformation fields and therefore changes in the observed TOF. All authors who measure guided wave SA find that this metric tends to increase with increasing SOC \cite{zhao2021state, ladpli2018estimating, li2022state, jie2023ultrasonic,Reichmann2023UltrasonicGuided}. This increase may be due to improved impedance matching between the ultrasonic source and guided wave mode with increasing SOC, however no physical rationale or experimental investigation have been been provided to explain these observations. Table \ref{tab:guidedWaveSum} summarizes relevant studies that have used ultrasonic guided waves to study LIBs. The table provides information about the UT observables identified of each study and summarizes key findings.

\begin{table*}[]
\caption{Examples of UT methods for LIB measurements using guided waves.}\label{tab:guidedWaveSum}
\begin{adjustbox}{width = \textwidth}
\begin{tabular}
{p{3.5cm}>{\raggedright\arraybackslash}p{4cm}>{\raggedright\arraybackslash}p{5cm}>{\raggedright\arraybackslash}p{8cm}>{\centering\arraybackslash}p{3cm}}
\hline\hline
Excitation \newline frequency/method & Transducer configuration & Observable(s) & Key finding & Relevant refs.\\
\hline\hline
\SI{100}-\SI{200}{\kilo\hertz}, \newline pitch-catch & Two contact PZT elements on same face of cell & SA and TOFS & First guided-wave study in LIBs. SA and TOFS sensitive to SOC/SOH. & Ladpli et al.~(2018)\cite{ladpli2018estimating}\\
\hline
\SI{9}{\kilo\hertz}, square wave \newline excitation &  Two contact PZT elements on same face of cell & TOF  & TOF and SOC correlation validated with various operating conditions including a real drive cycle. & Popp et al.~(2019) \cite{popp2019state}\\
\hline
5-cycle tone burst \newline at \SI{110}{\kilo\hertz} & Single contact PZT element excitation, single-point laser Doppler vibrometer (LDV) for reception & TOF, SA, and PSD & SA sensitivity reduces progressively with degradation. PSD trends mirror those of SA. Attribute PSD and SA reductions to changes in the lattice structure of the cell electrodes. & Zhao et al.~(2021) \cite{zhao2021state}\\
\hline
5-cycle tone burst \newline at \SI{125}{\kilo\hertz} & Single contact PZT element excitation, 2D scanning LDV (SLDV) receive & Waveform measured at each point on cell surface. ``Point parameters,'' such SA and TOF, extracted for all points. Measured ``line'' characteristic parameters, such as the slope of the TOF during charging/discharging. & Clear relationship observed between SOC/SOH and the ``line'' characteristic parameters derived from the guided wave line scanning data. Discussed limitations of the UT method due to environmental noise. & Li et al.~(2022) \cite{Li2022StateUGW}\\
\hline
Hanning-windowed \newline 5-cycle tone burst \newline at 100 kHz & Contact PZT transmit elements with 2D SLDV for receive & 2D surface of guided wave field & Identified large defect (22$\times$18 mm$^2$) though defect was not confirmed with other methods. & Zheng et al.~\cite{zheng2021guided}\\

\hline
25-105 kHz, tested \newline at \SI{5} to \SI{45}{\celsius} & Contact PZT element excitation on both sides of cell  & $A_{0}$ group velocity & $A_{0}$ velocity increases with SOC and decreases with temperature increase. BMS look-up table SOC estimation with an error of \SI{3.48}-\SI{5.32}{\percent}. & Koller et al.~(2022\cite{koller2022determination} and 2023\cite{koller2023ultrasonic})\\
\hline
\SI{1}-\SI{15}{\kilo\hertz} $A_{0}$ mode & Contact PZT element transmit-receive pairs and Electromagnetic Acoustic Transducer (EMAT) source, accelerometer receive & $A_{0}$ group velocity & Measured $A_{0}$ velocities of \SI{119}{\meter\per\second} and \SI{69}{\meter\per\second} at different SOC. Model required \SI{30}{\percent} reduction in electrode Youngs' moduli to match experiment, highlighting sensitivity to material properties. & Siegl et al.~(2022\cite{siegl2022electromagnetic} and 2023\cite{siegl2023sensor})\\
\hline
\SI{25}-\SI{200}{\kilo\hertz} \newline Hanning-windowed \newline 5-cycle stepped-sine & Contact PZT element pitch-catch & 7 signal features including SA and TOF centroid & TOFS increases with SOC, in contrast to most bulk wave studies. State estimation using an adaptive, fully-connected neural network. \SI{2.24}{\percent} SOC & Li et al.~\newline(2022) \cite{li2022state}\\
\hline
Hanning-windowed \newline 5-cycle tone burst \newline at 150 kHz & Contact PZT, same face pitch-catch & SA and TOFS & SA increases, TOF decreases with increasing SOC with difference C rate (0.75-1.25C) and \SI{20}-\SI{30}{\celsius}. & Jie et al.~(2023) \cite{jie2023ultrasonic}\\
\hline
Hanning-windowed \newline 5-cycle tone burst \newline at 105 kHz & Contact PZT, same face pitch-catch & Temperature-compensated TOF ($\mathrm{TOF_{T}}$) & $\mathrm{TOF_{T}}$ remains C-rate dependent after temperature compensation. GITT reveals asymmetric electrode phase-transitions kinetics between charge and discharge. & Liu et al.~(2024) \cite{Liu2024UGWCRate}\\
\hline\hline
\end{tabular}
\end{adjustbox}
\end{table*}

Some key guided wave studies merit further discussion for their unique contributions. Koller and colleagues used three piezoelectric transducers for guided wave UT in a configuration similar to Ref.~\cite{zappen2020operando}, with one receiving transducer on the same face as the transmitter and the second receiving transducer opposing the top-mounted receiving transducer some distance away from the transmitting transducer \cite{koller2022determination}. The use of this transducer configuration, enables comparison of the phase of the received waveform at the top and bottom cell surface, which they use to determine symmetric or anti-symmetric Lamb modes, as illustrated in Fig.~\ref{fig:GuidedWaveSummary}.C. They measured the A0 group velocity in the range of \SI{60}{\meter\per\second} to \SI{170}{\meter\per\second} for different tested batteries up to \SI{25}{\kilo\hertz}. The same authors extended this work by conducting a separate experiment on \SI{32}{Ah} NMC/graphite pouch cells and piezoelectric transducers as transmitters and receivers \cite{koller2023ultrasonic}. Using this setup, they performed guided wave inspection at nine cell voltages between \SI{3.59}{\volt} and \SI{4.20}{\volt} and at temperatures between \SI{5}{\celsius}, \SI{25}{\celsius}, and \SI{45}{\celsius} and found that group velocity of the A0 increased with increasing SOC and decreased with increasing temperature. They further demonstrated their technique in a realistic scenario, by implementing the ultrasonic measurements into a BMS showing SOC estimation with a mean absolute error of \SI{3.48}{\percent} to \SI{5.32}{\percent} for SOC estimation \cite{koller2023ultrasonic}. 

More recently, Liu et al.~\cite{Liu2024UGWCRate} investigated how the charge/discharge rate affects ultrasonic guided wave propagation launched and received by angle-beam ultrasonic transducer, though they did not specify the guided wave mode employed during the study. They reported a nearly linear decrease in TOF with increasing SOC at a constant charge/discharge rate. To decouple the influence of temperature and C-rate on TOF variations, they introduced a temperature-compensated TOF metric, denoted as $\mathrm{TOF_{T}}$. However, even after temperature compensation, $\mathrm{TOF_{T}}$ remained dependent on the C-rate. To analyze this dependency, they performed a galvanostatic intermittent titration technique (GITT) experiment, evaluating the stationary and dynamic responses of $\mathrm{TOF_{T}}$ with C-rate changes, as shown in Fig.~\ref{fig:GuidedWaveSummary}.D.1-3, as a means to assess the underlying physical phenomena associated with C-rate that alters ultrasonic wave speed. Their results showed that rapid changes in C-rate cause simultaneous shifts in $\mathrm{TOF_{T}}$, suggesting changes to the material properties on times scales that are much faster than can be measured using UT. Liu et al.~made three key observations based on these results: (\textit{i}) the $\mathrm{TOF_{T}}$ decreased more after discharging stopped than it increased after charging stopped, indicating that discharging induces more pronounced electrode phase transitions under equivalent C-rates; (\textit{ii}) $\mathrm{TOF_{T}}$ required longer stabilization times after discharging stopped compared to when charging stopped; and (\textit{iii}) higher C-rates induced larger $\mathrm{TOF_{T}}$ positive and negative shifts after charge and discharge respectively stops. Based on these observations, the authors suggested that electrochemical phenomena at the micro-, meso-, and macroscales contribute to these results. A current understanding is that at the micro-scale, phase transitions occur on the active particles within electrodes during charging/discharging processes. These phase transitions lead to changes in the elastic moduli and density of the particles. The microscale-induced changes of active particles generates internal stresses between the active particles and binder particles, yielding mesoscale changes stiffness changes. Finally, the stresses induced through the processes at the micro- and mesoscales create a cumulative macroscale deformation that affects the contact interfaces between the different electrode layers (see Fig.~\ref{fig:GuidedWaveSummary}.D.4). These multiscale changes can be probed using UT of different frequencies.

A key observation regarding published research that makes use of guided ultrasonic waves in LIB is the challenge of determining the nature of the fields that have been analyzed using the reported experimental parameters. Specifically, existing literature has rarely provided a detailed analysis of whether an observed response was in fact due to a traveling guided wave rather than standing-wave patterns along in-plane directions, which may include the interaction of multiple guided wave modes. The latter occurs when the wavelengths of the field generated by the source are on the order of the in-plane dimensions of the cells, leading to constructive and destructive interference between waves traveling directly from the source to a field point and those reflected from in-plane cell boundaries. For instance, it is possible that the observation that TOFS increases with increasing SOC that was reported by Li et al.\cite{li2022state}, which is counter to those reported by others who used bulk wave approaches (see Sec.~\ref{sec:bulkwaves}), is due in part to the low frequencies used in that study ($<70$ kHz). Guided waves in LIB pouch cells at this frequency have a wavelength on the order of the cell dimensions, yielding interference patterns reminiscent of standing wave fields rather than traveling guided wave packets as indicated by the results showing Fig.~3 of Ref.~\cite{li2022state} and explained the associated discussion provided by the authors. The difficulty in resolving ambiguity in guided wave results can be addressed in part through the design of ultrasonic sources and sensors. For example, Siegl et al.~\cite{siegl2023sensor} showed selective excitation of the leading-order anti-symmetric (A0) Lamb mode through the design of an electromagnetic acoustic transducer (EMAT). The results suggested that the in-plane Young's modulus of both the anode and cathode \textit{in situ} would have to be \SI{30}{\percent} lower than the values measured in the dry state to match the experimental observations.
The complexity of guided wave modes in LIB is complicated by the fact that their mode shapes are a function of both longitudinal and shear deformation\cite{rose2004ultrasonic}, meaning that changes in all moduli due to external loading, i.e.~SOC, are  relevant to observed changes in the wave field. Future research should therefore clearly analyze the dispersion and deformation of guided wave modes, including potential effects of the finite size of the cell being tested, to draw physically meaningful conclusions of the origin of signal changes as a function of SOC or SOH.


The state of the literature on using guided wave US for SOH estimation is even less developed than that for guided wave UT for SOC estimation. To the author's knowledge, only experiments by Ladpli et al.\cite{ladpli2018estimating} and Zhao et al.\cite{zhao2021state} have been conducted to correlate decreasing SOH with signal characteristics of guided ultrasonic waves. Ladpli et al.~considered accelerated aging by cycling cells 200 times at a 0.8C with the cells at an elevated temperature of \SI{45}{\celsius} in a laboratory oven \cite{ladpli2018estimating}. Through this cycling protocol, they achieved a reduction in SOH to \SI{96.5}{\percent}. Zhao et al.~aged a cell by cycling it 165 times, taking UT measurements every 30 cycles \cite{zhao2021state}, leading to a \SI{5}{\percent} decrease in SOH. Presently, in electric vehicle or portable device applications (i.e., high power demand), LIBs are removed from service when their SOH reaches \SI{80}{\percent} \cite{S2024SOHSLBs}. Since both studies only achieved minor reductions in SOH, questions remain about ultrasonic guided wave signal trends with continued aging. Regardless, their results provide the only studies of this type and are therefore summarized below.

Both groups found that the TOFS decreased globally with decreasing SOH \cite{ladpli2018estimating, zhao2021state}, but provide conflicting results regarding the SA trend with SOH. Ladpli et al.~found that the SA increased globally with decreasing SOH \cite{ladpli2018estimating}, while Zhao et al.~found a decreasing trend \cite{zhao2021state}. Zhao et al.~found that the effect of SOH on SA was not the same for all SOC. Instead they found that the SA at \SI{100}{\percent} SOC experienced a greater decrease with aging than the SA at \SI{0}{\percent} SOC \cite{zhao2021state}. Notably, a separate study by Li et al.~, which only cycled their cell 10 times, observed a decreasing SA at \SI{100}{\percent} SOC with each cycle \cite{li2022state}. There is much room for continued exploration of the effect of SOH on guided wave and bulk wave ultrasonic signal characteristics. Future work should be conducted on cells as they undergo isolated, individual aging mechanisms such as the aforementioned calendar-aging and cycle-aging. At the present, one can only extrapolate in order to assume the effect of more severe reductions in SOH on ultrasonic signal characteristics. We also stress that while Ladpli et al.~\cite{ladpli2018estimating} and Zhao et al.~\cite{zhao2021state} found different trends of SA with SOH, this does not automatically suggest that their results are in disagreement. Without understanding the exact guided wave mode measured by their experimental apparatus and how the mode might change with reduced SOH, one cannot draw a conclusion regarding which trend is correct. Here, a topic that may arise is the apparent contradiction in some of the reported trends of ultrasonic signals from bulk and guided waves regarding their changes with SOH. In this sense, degradation has been reported to both increase \cite{feiler2023interplay} and decrease \cite{ladpli2018estimating} the TOF (for experiments on the same NMC chemistry). Bulk wave experiments suggest a decrease in longitudinal stiffness in the thickness direction, while guided waves suggest an increase in the stiffness that is more related to shear wave propagation. This inconsistency may be due to the evolving behavior of the cell over aging and must be a topic of future research.

\subsubsection{Guided wave modeling approaches}
Guided wave propagation can be modeled using similar approaches as bulk wave propagation, namely matrix methods employing standard equations for Lamb waves  \cite{ladpli2018estimating, popp2019state}, matrix methods employing modified equations considering poroelastic effects \cite{jie2021guided, jie2023ultrasonic, Gao2024TheoreticalGuidedWaves}, finite element approaches \cite{jie2023ultrasonic, siegl2023sensor,Gou2024StudyGuidedWavesFE}, or data driven approaches \cite{li2022state,Liu2022EvalSOCGuidedNN, Ji2025AIModelGuidedWaves}.

\textbf{Wave Physics models:} Ladpli et al.~created an analytical model that considered the layered structure of the cell and two-dimensional elastodynamic equations of motion of each layer. Those equations were solved using the global matrix method to find the dispersion curve of the cell \cite{ladpli2018estimating}. By using SOC-dependent material properties of electrodes, they used the dispersion curves to calculate the group velocity and then estimate the TOF as a function of SOC. They observed a good agreement between the simulated and measured TOF of the US guided waves. The difference in TOF between \SI{0}{\percent} and \SI{100}{\percent} measured was about \SI{7}{\micro\second} while the simulation predicted \SI{11}{\micro\second}, without aging considerations. Nonetheless, the \SI{0}{\percent} SOC conditions showed higher mismatches between the simulation and experiments compared to the \SI{100}{\percent} SOC condition. Additionally, the simulation failed to capture some nonlinearities exhibited by the measured TOF.

\begin{figure*}[]
    \centering
    \includegraphics[width=\linewidth]{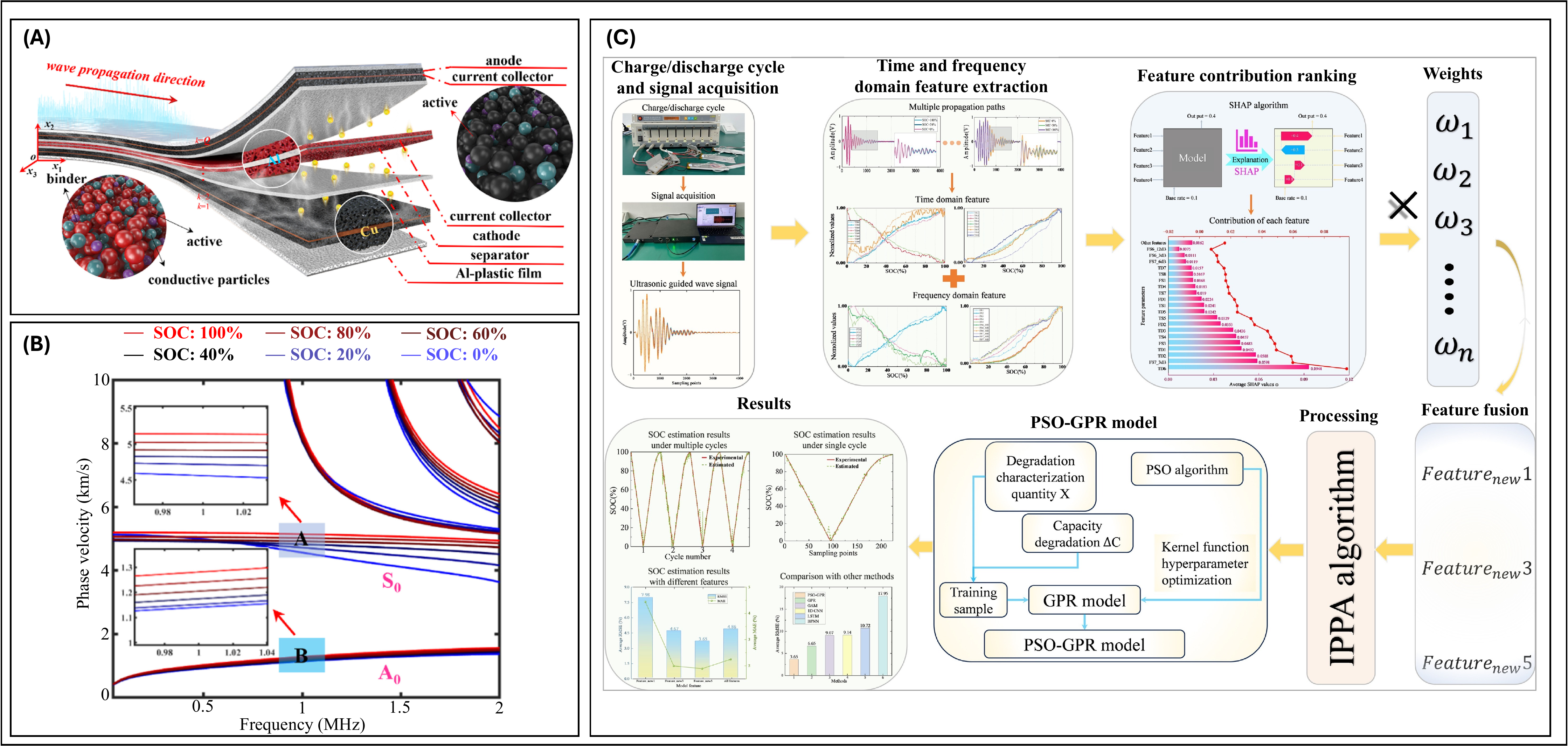}
    \caption{Guided wave modeling approaches in LIBs. \textbf{(A)} Geometry used to model guided wave propagation on multilayered LIB, including porous electrode materials. \textbf{(B)} Dispersion curves calculated using simulations at different SOC values of a unit cell of a LIB. Clear changes in phase velocities of the $A0$ and $S0$ modes as a function of SOC are highlighted in the insets. Panels A and B reproduced from Ref.~\cite{Gao2024TheoreticalGuidedWaves}, Copyright (2024), with permission from Elsevier. \textbf{(C)} SOC estimation framework based on PSO-GPR model. Time and frequency domain features from guided waves were used as inputs to the model. Multidimensional features were ranked and weighted with SHAP. The IPPA algorithm merged and developed a new feature sequence to feed the PSO-GPR model, which then estimated the SOC. Reproduced from Ref.~\cite{Ji2025AIModelGuidedWaves}, Copyright (2025), with permission from Elsevier.}
    \label{fig:GuidedWaveModeling}
\end{figure*}

Jie et al.~also employed a global matrix approach, but used Biot's theory of propagation in poroelastic media to model the wave propagation through a multi-layered structure of graphite/copper/graphite. Using this approach, they calculated dispersion curves for this structure and identified the lowest order symmetric and anti-symmetric Lamb modes. They assumed that the current collector had a porosity of 0.01 in order to use Biot's theory for all layers, thereby exactly satisfying boundary conditions between layers of otherwise dissimilar media. \cite{jie2021guided}. The same group later expanded their approach to simulate the dispersion curves through a 19-layer model of a LIB including graphite, copper, \ch{LiCoO2}, aluminum, separator, electrolyte, and outer cell casing \cite{jie2023ultrasonic}. Using SOC-dependent material properties, those authors were able to extract the group velocity from the simulated dispersion curves as a function of SOC and found that the group velocity increased with SOC, in agreement with the experimental results of Refs.~\cite{ladpli2018estimating, popp2019state, zhao2021state}. Jie et al.~additionally created a 2D finite-element simulation of this 19 layer model in COMSOL, solving for the dispersion curve using Biot's theory and found good agreement between this simulation and their global matrix model \cite{jie2023ultrasonic}. The same research group also showed a preliminary theoretical model of longitudinal guided wave propagation in the axial direction for cylindrical cells showing some calculations for the influence of the SOC and circumferential order on the dispersion curves \cite{Gao2022Cylindrical}. A more recent study by the same group reported an analytical acoustic model based on state-vector formalism and Legendre polynomial method that also considered the coupling of the electrochemical behavior of the LIBs to the mechanics (see scheme on Fig.~\ref{fig:GuidedWaveModeling}.A) through the mechanical stress that will be generated on the graphite electrode during charge/discharge, which they named the diffusion-induced stress (DIS) \cite{Gao2024TheoreticalGuidedWaves}. That research used COMSOL to simulate the electrodes changes in the mechanical properties during charge/discharge and the distribution of the lithium concentration for different SOC to calculate the DIS and calculate the dispersion curves for different SOC, as shown in Fig.~\ref{fig:GuidedWaveModeling}.B. This theoretical framework can be further used to analyze the propagation of guided waves in batteries with different SOH by including the simulation of battery degradation.

Koller et al.~made use of The Dispersion Calculator, version 1.11.2 by Armin Huber to calculate the dispersion curves up to \SI{15}{\kilo\hertz} for a 200 layer model of the cell \cite{koller2022determination}. Their model predicted a group velocity of the A0 mode of \SI{76}{\metre\per\second} at the frequency of interest of \SI{8}{\kilo\hertz} \cite{koller2022determination}, which lies between the experimental A0 mode group velocity of \SIrange{60}{170}{\metre\per\second}. Siegel et al.~later expanded on this work by creating a 2-D finite element model of a 264-layer cell in COMSOL. In order to match experimentally measured group velocity of \SI{69}{\metre\per\second} they had to reduce the model inputs of Young's modulus for both the anode and cathode material by \SI{30}{\percent}, which resulted in a simulated group velocity of \SI{71}{\meter\per\second}\cite{siegl2023sensor}. This research group extended their work to simulate the dispersion curves of a cell with SOC-dependent Young's modulus of the anode active material, finding that the group velocity should increase for the A0 Lamb mode with increasing SOC \cite{koller2023ultrasonic}. 

Most of the work on modeling ultrasonic wave propagation in LIBs has focused on elastic phenomena, leaving out thermoelastic and viscoelastic considerations that can also affect the ultrasonic signals, as has been shown in some of the reported research of this review. One of the only groups to consider thermoelastic effects was Lin et al.~\cite{Lin2024ThermoelasticGW,Lin2025Thermoelastic} who proposed an approach to analyzing the properties of thermoelastic wave propagation in a multi-layered porous electrode. Further, most of the modeling approaches reported in the literature can be described as root-finding methods using a characteristic equation of frequency and wavenumber to find dispersion curves associated with unforced wave propagation. In this regard, approaches like the work of Zhang et al.~\cite{Zhang2024PropModelUGWSpectral}, which employed spectral methods instead of root-finding techniques, might be interesting to extend to the LIBs case. Spectral methods use spectral differentiation matrices and orthogonal polynomials to solve differential equations by numerical interpolation and have emerged as an alternative for root-finding techniques in recent years. Zhang et al.~\cite{Zhang2024PropModelUGWSpectral} used the spectral approach for the Lamb wave equation for multilayered porous media and demonstrated its application to well-known porous materials. Although they did not apply the method for solving guided wave dispersion curves in LIBs, the spectral methods have the potential for faster computation speed and a more straightforward encoding process compared to root-finding methods. However, their applicability and usability in modeling US waves remain to be proven.

\textbf{Data-driven models:} The number of data-driven approaches to model ultrasonic guided waves is fewer than for bulk waves. However, they are based on similar principles but usually employ different features from the ultrasonic signals for battery state estimation. Liu et al.~\cite{Liu2022EvalSOCGuidedNN} used an artificial neural network that relies on simple parameters like SA and TOF, demonstrating that these characteristics are related to SOC, yet are vulnerable to environmental noise and the complexity of propagating modes.

In contrast, Li et al.~built an ``adaptive fully connected neural network(FNN)-XGBoost model for battery state estimation'' using seven signal features from a guided wave experiment and found a root mean square error of \SI{2.24}{\percent} SOC \cite{li2022state}. They found that the TOFS increased with increasing SOC, which is in contrast to the experimental results of Refs.~\cite{ladpli2018estimating, popp2019state, zhao2021state}. Similar to this approach, Ji et al.\cite{Ji2025AIModelGuidedWaves} reported a multi-feature fusion technique that integrates both time and frequency domain features ranked using the XGBoost-SHAP (XGBoost - Shapley additive explanations) model. Their Particle Swarm Optimization Gaussian Process Regression (PSO-GPR) model achieved an accuracy of RMSE \SI{3.65}{\percent} and demonstrated the robustness of feature selection while mitigating the limitations of single-feature approaches. While the Adversarial Neural Network (ANN) model is simpler and more computationally efficient, it lacks the adaptability and precision of the PSO-GPR approach, which, despite its complexity, provides a more reliable and scalable solution for battery management systems (see scheme in Fig.~\ref{fig:GuidedWaveModeling}.C).
 
Another study by Tian et al.~ \cite{Tian2024StateOfCharge} employed a deep learning model named Long Short Term Memory (LSTM) to map ultrasonic features (signal local amplitude and TOF) to SOC with high precision, with TOF decreasing as SOC increased. The study compared various frequencies, finding that higher frequency signals, such as \SI{140}{\kilo\hertz}, resulted in significantly lower accuracy (RMSE reaching \SI{10.68}{\percent}) and that \SI{60}{\kilo\hertz} provided the highest accuracy (lowest RMSE was \SI{3.07}{\percent}), making it the optimal detection frequency for the conditions that they explored. This is likely due to the fact that higher frequency waves have shorter wavelengths and thus energy may be distributed in multiple guided wave modes, whose group velocity and amplitude decay may differ from single-mode propagation.

\subsection{Ultrasonic Techniques for Damage Detection}\label{sec:damageDetection}
While most experimental publications employing UT on LIBs have focused on its viability for SOC and SOH estimation, a benefit of ultrasonic inspection is its potential to additionally monitor damage evolution within the cells. Ultrasonic inspection of LIBs has been used to detect gas formation \cite{pham2020correlative,Xu2023UltraPhasedArray}, lithium plating \cite{bommier2020operando, wasylowski2023situ, Wasylowski2024Operando, Jie2025ANovel}, overcharge \cite{kirchev2023part3, wu2019ultrasonic, appleberry2022avoiding, li2022state, shen2023situ}, over-discharge \cite{chang2019real, davies2017state, li2022state}, thermal abuse \cite{mcgee2023ultrasonic, mcgee2024ultrasonic, chang2020understanding, pham2020correlative, zappen2020operando, kirchev2022part1, owen2022operando, kirchev2023part2, zhang2023state,Owen2024OperandoTemperature}, and delamination \cite{sood2013health}. Since thermal runaway is the most catastrophic and costly failure state of LIBs, damage detection centered on early detection of thermal runaway is the highest priority for the BMS \cite{Williams2024AReviewTR}. Since the primary cause of thermal runaway is high-temperature thermal abuse, ultrasonic detection of this damage condition will be discussed first. Overcharge and over-discharge are also common damage conditions which may lead to thermal runaway, so ultrasonic detection of these damage conditions will be discussed next. We conclude this section with a discussion of UT for general defect detection in cells including gas detection.

\subsubsection{Thermal Abuse and Temperature Effects}
Due to the fact that LIB performance degrades at low and elevated temperatures, researchers have conducted many studies on the ability of UT to detect changes in the temperature of cells. Most of the work to date has focused on moderate temperature changes, i.e.~temperatures ranging from \SI{-10}{\celsius} to \SI{65}{\celsius}, though some examples of UT at temperatures leading to thermal runaway have also been reported. Research studies on moderate temperature effects have consistently shown that TOF increases with temperature, while SA behavior is less repeatable across studies. Temperature primarily affects the elastic modulus of cell components, particularly that one of the viscoelastic separator as mentioned in Table \ref{tab:summaryUSDegradation}), and the SA variation can be affected by competing changes in acoustic impedance contrast between layers (which may either increase or decrease SA) and changes in viscous and viscoelastic losses (which tends to decrease SA with increasing temperature). Table \ref{tab:moderateTemp} reports a summary of key findings of works that studied this moderate temperature range.

\begin{table*}

\caption{Examples of research performed to study ultrasonic signals changes with moderate temperature changes (\SI{-10}{\celsius} to \SI{65}{\celsius}) }
\label{tab:moderateTemp}
\begin{adjustbox}{width = \textwidth}
\begin{tabular}
{p{3cm}>{\raggedright\arraybackslash}p{3cm}>{\raggedright\arraybackslash}p{7cm}>{\raggedright\arraybackslash}p{7cm}>{\centering\arraybackslash}p{3cm}}
\hline\hline
Wave type & Temp. range & Key metric change for $\uparrow$ T & Notable detail & Relevant refs.\\
\hline\hline
Bulk & Heated cells to \SI{39}{\celsius}& TOF increases & Narrow temperature range; SA not diagnostically useful (``small and irregular'') & Ke et al. (2022) \cite{ke2022potential}\\
\hline
Bulk&  \SI{-10}{\celsius} $\rightarrow$ \SI{60}{\celsius} (\SI{10}{\celsius\per\hour}) & TOF increases and SA decreases & Broadest moderate temperature range in the literature & Owen et al.~(2024) \cite{Owen2024OperandoTemperature}\\
\hline
Guided ($A_{0}$) & \SI{5}{\celsius} $\rightarrow$ \SI{45}{\celsius} & Positive cubic relationship of TOF with temp. & Nonlinear TOF-temperature relationship; rate controlled cycling & Popp et al.~(2019) \cite{popp2019state}\\
\hline
Bulk & \SI{20}{\celsius} $\rightarrow$ \SI{50}{\celsius} & TOF decreases (due to analyzing early arrival)& Attributed to density decrease of DMC electrolyte solvent; note: they attributed this to a decrease in density of the electrolyte with temperature, although density decrease at constant mass requires volume increase, i.e., cell expansion & Zhang et al.~(2023) \cite{zhang2023state}\\
\hline
Air-coupled \newline (thickness \newline resonances) & \SI{21}{\celsius} $\rightarrow$ \SI{55}{\celsius} & Resonance frequency shifts to lower values &  Non-contact method; acoustic fingerprint approach using resonance frequency spectrum &Fariñas et al.~(2024) \cite{Farinas2024Contactless} \\
\hline\hline
\end{tabular}
\end{adjustbox}
\end{table*}

The studies that were more focused on safety-critical temperature detection are those that covered the thermal-abuse regime ($>$ \SI{60}{\celsius}). For example, Chang et al.~studied the effect of temperature transitions from \SI{0}{\celsius} to \SI{60}{\celsius} by cycling LCO cells in temperature-controlled incubators \cite{chang2020understanding}. Their objective was not only to measure ultrasonic TOFS during temperature cycling, but also to investigate how various charge-discharge profiles at different temperatures would change battery performance. Some of the temperature-cycling profiles resulted in complete failure and extensive gas generation, while others did not. They showed that elevated temperatures (greater than \SI{20}{\celsius}) following lithium plating at low temperature (below \SI{10}{\celsius}) can induce significant gassing and failure. Acoustic signal attenuation effectively detects gassing, with the rate of drop in SA following an Arrhenius relationship to temperature shifts. Furthermore, the SA decrease was related to the magnitude of the temperature shift, and they attribute the reduction in SA to substantial gassing. During cycling at \SI{<60}{\celsius}, the TOFS showed a generally decreasing trend with cycle number, while the SA trend varied from cell to cell. For all temperature profiles they reported, the US signal eventually dropped to the noise floor. The larger the temperature changes, the more rapid the reduction of the signal amplitude to the noise floor \cite{chang2020understanding}. They also noted that the TOFS tended to increase following the temperature shifts until the signal dropped to the noise floor. The authors hypothesized that the observed changes may result from a combination of lithium plating on the anode surface during the previous low-temperature cycling and enhanced electrolyte decomposition due to the higher temperatures of the following cycling.

\begin{figure}[ht]
    \centering
    \includegraphics[width = \linewidth]{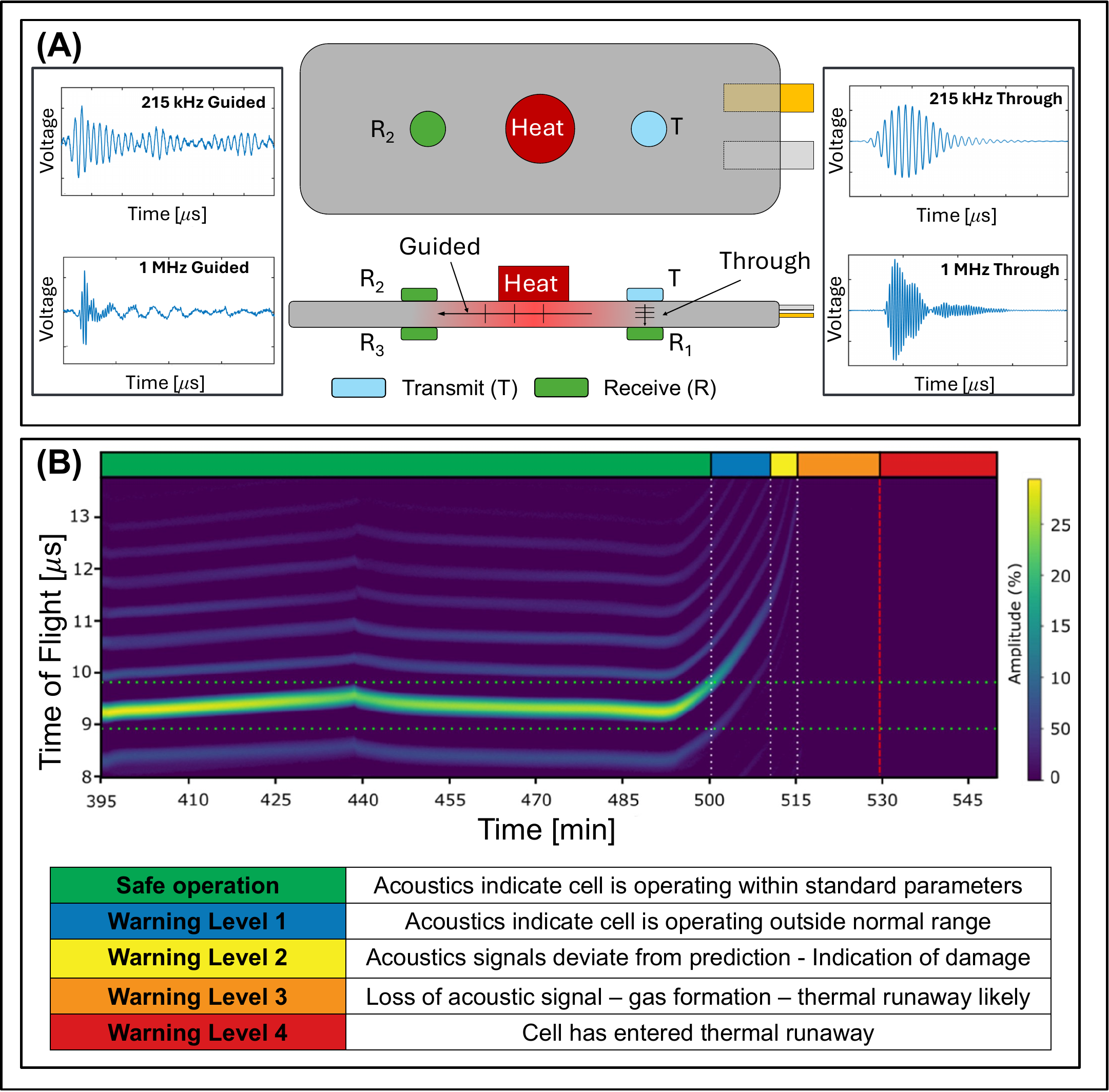}
    \caption{Examples of UT to detect thermal abuse in LIBs: \textbf{(A)} schematic illustration of UT performed on a cell exposed to localized thermal abuse. Adapted from Ref.~\cite{mcgee2023ultrasonic}, Copyright (2023), with permission from Elsevier. \textbf{(B)} Proposed framework to evaluate the risk of thermal runaway events depending on deviations of the TOF measurements from the expected values. Reproduced from Ref.~\cite{Owen2024OperandoTemperature}; licensed under a Creative Commons Attribution 4.0 International (CC BY 4.0) license (\href{https://creativecommons.org/licenses/by/4.0/}{https://creativecommons.org/licenses/by/4.0/}).}
    \label{fig:ThermalAbuse}
\end{figure}

Research has been conducted to investigate the effects of high-temperature loading and rapid temperature shifts on damage associated with thermal abuse, which can sometimes lead to failure. Zappen et al.~made use of a guided wave transducer configuration by using three transducers: two on the same cell face and one opposing the receiving transducer on the opposite cell face. Although they label the two paths as ``In-Plane" and ``Through-Plane", both receiving transducers should detect guided waves. They did not specify the exact frequency used for the guided wave, which could influence the energy content of the received signals \cite{zappen2020operando}. Their work considered the heating of an NMC pouch cell at \SI{100}{\percent} SOC from room temperature to \SI{110}{\celsius} while continuously acquiring guided wave UT measurements \cite{zappen2020operando}. They found that the signal intensity (akin to SA) decreased for both the ``In-Plane'' and the ``Through-Plane'' paths with a sudden decrease at \SI{90}{\celsius}, while what they refer to as the ``center of gravity'' of the signal in the time domain (akin to TOFS) increased slowly with temperature followed by a rapid increase at \SI{90}{\celsius}. It is not surprising that they found similar behavior for both observed paths since the transducer placement during these measurements meant that both measured the same guided wave, just at different locations. In a similar study, Pham et al.~heated LCO pouch cells charged to \SI{100}{\percent} SOC to TR while taking pulse-echo measurements using \SI{2.25}{\mega\hertz} center frequency pulses. They noted that the TOF increased with applied heat and that the ultrasonic signal was lost due to significant gas generation, though they did not report TOF values or temperature explicitly \cite{pham2020correlative}. Furthermore, the authors had mis-attributed each oscillation observed in the received waveform to distinct layers within the cell structure. This is unlikely for wavelengths of ultrasonic signals at \SI{2.25}{\mega\hertz} ($\approx$ \SI{670}{\micro\meter}) which are much larger than typical thicknesses of layers within each cell ($\approx$ \SI{10}-\SI{100}{\micro\meter}). 

Similar work by Kirchev et al.~monitored ultrasonic signals transmitted with a frequency of \SI{150}{\kilo\hertz} through model 18650 NMC811 cylindrical cells in an adiabatic Accelerating Rate Calorimeter (ARC) to either \SI{160}{\celsius} or TR depending on the cell SOC, and found that the ultrasonic signal strength decreased for all tested SOC with increasing temperature \cite{kirchev2023part2}. In a related study using ARC techniques but for a pouch cell, Feinauer et al.~\cite{Feinauer2023ARC} integrated data from \SI{150}{\kilo\hertz} ultrasonic transducers, a strain gauge, a thermocouple, and voltage and resistance measurements to monitor the thermal runaway process with cells heated up to \SI{350}{\celsius}. This work found that the transmitted ultrasonic signals can be used as an early detection metric of gas evolution, which leads to cell venting. They also showed that stable exothermic behavior (a key thermal event associated with self-heating during the thermal runaway process) can only be detected by the temperature and the overall disappearance of the ultrasonic signal. Fig.~\ref{fig:ThermalAbuse}.A shows a schematic of the experimental setup of McGee et al.~who performed two studies using UT of mechanically-confined NMC532 pouch cells exposed to localized high-temperature thermal abuse using contact ultrasonic transducers that generate \SI{250}{\kilo\hertz} guided waves and \SI{1}{\mega\hertz} through-thickness longitudinal waves\cite{mcgee2023ultrasonic, mcgee2024ultrasonic}. When testing for \textit{in situ} detection of thermal abuse, they found that the SA and TOFS of longitudinal bulk waves decreased and increased, respectively. In contrast, the SA increased for guided waves as temperature increased, and both of these indicators could provide an early indicator of failure \cite{mcgee2023ultrasonic}. The same authors then looked at the ability of UT measurements to detect localized thermal abuse in charge-discharge cycles that follow abuse and found that the SA and TOFS of the longitudinal bulk waves both increased globally in the three charge-discharge cycles following abuse at levels of \SI{100}{\celsius} or higher \cite{mcgee2024ultrasonic}.

Additional recent work by Owen et al.~showed the ability of UT to track averaged internal temperature fluctuations \cite{Owen2024OperandoTemperature} that can be used to design multi-stage warning systems based on the expected behavior of the TOF of ultrasonic signals during the cell cycling as shown in Fig.~\ref{fig:ThermalAbuse}.B. That work showed how the internal temperature can be estimated knowing the TOFS associated with SOC and temperature changes from previous tests, which can be a better metric compared to temperature measurements made on the surface of the cells, since LIBs are excellent thermal insulators.

\subsubsection{Overcharge or Over-Discharge}
While many researchers have broadly demonstrated the ability of UT to detect when a cell is overcharged, it is important to first explain that the term ``overcharge'' is very general and includes different types of overcharging scenarios. Overcharging can be carried out either through constant-current charging or constant-voltage charging. Constant-current charging delivers a fixed current to the terminals of the cell until it reaches a certain cut-off voltage or until the cell exceeds a specified temperature threshold, which, if inadequately defined, may lead to thermal runaway. In contrast, constant-voltage charging maintains the cell voltage at a constant level while applying a gradually decreasing current. For instance, as shown in Fig.~\ref{fig:ElectrodesVolumeChange}.A, volume changes in cathode materials can exhibit nonlinear behavior at high SOC approaching  overcharge conditions, behavior that may be detected with ultrasonic measurements and  motivates further investigation. Additionally, several researchers utilize slight levels of overcharging during testing, both with and without raised temperatures, to hasten aging mechanisms within the cells. Research that employs combined thermo-electrical loading complicates the separation of individual effects on the received ultrasonic signals.

Li et al.~conducted an experiment on an LCO pouch cell involving a slight overcharge test to \SI{107}{\percent} SOC and observed that the received ultrasonic signal using guided waves dropped to the noise floor \cite{li2022state}. In a similar work using also LCO pouch cells, Wu et al.~repetitively cycled a cell to an overcharge voltage of \SI{4.5}{\volt} in a laboratory oven held at a constant \SI{45}{\celsius} to accelerate the effects of aging \cite{wu2019ultrasonic}. They simultaneously acquired ultrasonic data at \SI{0}{\percent} SOC for each cycle and found that the TOFS increased with aging from \SI{7.6}{\micro\second} to \SI{8.4}{\micro\second}, while the SA decreased with increasing cycle number. Their sample size consisted of two cells, both of which exhibited a decrease in SA as they were subjected to aging caused by overcharging, though the extent of the SA decrease differed between cells \cite{wu2019ultrasonic}. They also performed a single overcharge to \SI{5}{\volt} on a cell that had previously been cycled 210 times. In this test, they found that the TOFS increased by \SI{0.2}{\micro\second} when the cell voltage reached \SI{4.5}{\volt}, when compared to the TOFS observed at the standard cutoff of \SI{4.2}{\volt}. They attributed this increase in TOFS, which is contrary to the normal reduction in TOFS during voltage increases in normal operating ranges, to tiny amounts of gas generation. The TOFS then increased rapidly following cell swelling, at which point the experiment was stopped. Wu et al.~noted a steadily decreasing SA as the cell was cycled to normal limits. However, the SA increased dramatically at a voltage of \SI{4.5}{\volt}, then decreased, followed by a leveling off before spiking a second time following cell swelling, though the authors did not specify the physical mechanism leading to the increase in SA \cite{wu2019ultrasonic}. 

Appleberry et al. performed experiments using bulk longitudinal waves across four different combinations of temperature and overcharging. Those authors conducted constant-current (CC) overcharge and constant-voltage (CV) overcharge at \SI{23}{\celsius}, and both CC and CV overcharge at \SI{65}{\celsius} \cite{appleberry2022avoiding}. Separately cells were cycled three times at \SI{18}{\celsius}, \SI{20}{\celsius}, \SI{22}{\celsius}, \SI{34}{\celsius}, \SI{40}{\celsius}, \SI{45}{\celsius}, \SI{50}{\celsius}, and \SI{65}{\celsius} to understand the temperature effect on US transmission. These authors utilized the root mean square of the time signal (similar to SA), the time of the peak in a Hilbert transform of the time signal (akin to TOF), and the maximum amplitude derived from a Fast Fourier Transform (FFT) of the time signal, finding that for CC overcharge, the two metrics analogous to SA declined in value as cell voltage increased, while the metric related to TOF remained relatively stable \cite{appleberry2022avoiding}.  

Shen et al.~investigated UT to detect overcharge \cite{shen2023situ}. Those authors overcharged pouch cells to \SI{4.8}{\volt} and \SI{5.0}{\volt}, finding massive reduction in signal energy when overcharged to \SI{5.0}{\volt}. These researchers verified through X-ray computed tomography (CT) that at this cell voltage, substantial electrolyte decomposition occurred, which caused the layers to separate \cite{shen2023situ}.

Davies et al.~performed a ``low impedance closed circuit over-discharge event'' on a LCO pouch cell that had previously been cycled 106 times to a SOH of \SI{>98}{\percent} \cite{davies2017state}. The SA of a bulk wave generated with a \SI{2.25}{\mega\hertz} transducer diminishes quite rapidly as the cell approaches a nearly depleted state, which the authors attribute to the breakdown of the electrolyte and the generation of gas within the cell. In the cycles following this over-discharge event the SA appears to return to normal while the TOFS appears to grow with each cycle \cite{davies2017state}. It is also notable to recall that the previously discussed work of McGee et al.~\cite{mcgee2024ultrasonic} utilized the rate of change of SA and TOFS in the subsequent cycles following a thermal abuse event as a sign of prior damage in the context of over-charge or over-discharge. Namely, the metric devised in that work might also potentially indicate past over-discharge events.

Kirchev et al.~performed longitudinal bulk wave UT on 18650 cylindrical cells as they were overcharged with different constant currents to a voltage exceeding \SI{4.8}{\volt} for cells rated to \SI{4.2}{\volt} \cite{kirchev2023part3}. The cells were fitted with charge interrupting devices (CID) that activated at differing SOC depending on the applied C-rate. They showed that the SA ``fluctuates significantly'' throughout the overcharging process, complicating the interpretation of results. Furthermore, they carried out constant-voltage overcharging of cells at voltages of \SIlist{4.6;4.7;4.8;4.9}{\volt} and observed variable SA trends for each cell tested, even within normal SOC \cite{kirchev2023part3}. To analyze their findings, they employed clustering techniques and mapped the SLDVity of the received waveforms, enabling them to use ultrasound as an early warning signal for overcharging.

\subsubsection{General Defect Detection Including Gas Generation}\label{sec:DefectDetection}
Defects that are detectable using UT can include manufacturing defects (e.g.~coating irregularities, contamination, or improper electrode assembly, as discussed in Sec.~\ref{sec:UTformanufacturing}) and in-use defects that arise due to cycling or abuse conditions (e.g.~gas formation, lithium plating, electrolyte depletion, or electrode ruffling). While much work for detecting damage within LIBs has focused on detecting thermal abuse or electrical abuse via UT measurements, there is also great need for tools that can identify mechanical defects within LIB cells. Additionally, defects of various sizes can form during normal charge-discharge cycling, or under non-dangerous but still sub-optimal operating conditions \cite{Qian2021RoleDefects}. The most commonly investigated damage condition for UT is gas formation \cite{li2019numerical, bauermann2020scanning, deng2020ultrasonic, wasylowski2022spatially}, but other types of ultrasonically detectable damage include metallic particle contaminants within an electrode layer \cite{Yi2021MetalDefect}, lithium plating on the anode surface \cite{wang2021theoretical, chang2021operando, wasylowski2022spatially,Wasylowski2024Operando,Jie2025ANovel}, electrolyte wetting or dry-out \cite{deng2020ultrasonic,Eldesoky2022LongTerm,Louli2020ElectrolyteDepletion}, or electrode ruffling where conductive layers may lose contact \cite{sood2013health}. Since these types of defects may occur randomly across the planar area of the cells, some groups have designed experiments where they can scan the cell by taking UT measurements across the cell surface \cite{chang2021operando, robinson2019spatially, deng2020ultrasonic, wasylowski2022spatially, li2019numerical, shen2023situ}. 

Wang et al.~incorporated a layer of lithium metal into their 13-layer COMSOL model to simulate lithium plating, and discovered that for a given frequency, the transmission coefficient through a cell decreases with the presence of lithium plating  \cite{wang2021theoretical}. Li and Zhou used air-coupled UT to identify a defect they called ``stomata,'' which is effectively a gas pocket within a particular layer \cite{li2019numerical}. In order to simulate a near-surface and near-bottom ``stomata'' they pasted copper ``flakes'' to the surfaces of the cell. Given the significant impedance mismatch between copper and air, it is unclear if this method of simulating damage is appropriate. Regardless, they found that the amplitude of the transmitted wave was reduced in the presence of a ``stomata \cite{li2019numerical}.'' In addition, they performed a scan of the entire cell and provided a heat map representation of the transmitted amplitude through the cell.

Bommier et al.~used longitudinal bulk UT on LiCoO$_2$/Graphite pouch cells to detect lithium plating on the graphite anode \cite{bommier2020operando}. They intentionally induced lithium plating by charge cycling cells at low temperatures and elevated C-rates (up to 4C). To determine a baseline TOFS at the end of charge, they used a fixed-capacity charging protocol at a C-rate of C/15, which did not induce lithium plating. They then compared this TOFS endpoint to the TOFS observed at various temperature and C-rate combinations and found that UT indicated lithium plating when the TOFS end point was more positive than baseline at low temperatures and high C-rates \cite{bommier2020operando}.

\begin{figure*}[ht]
    \centering
    \includegraphics[width = \linewidth]{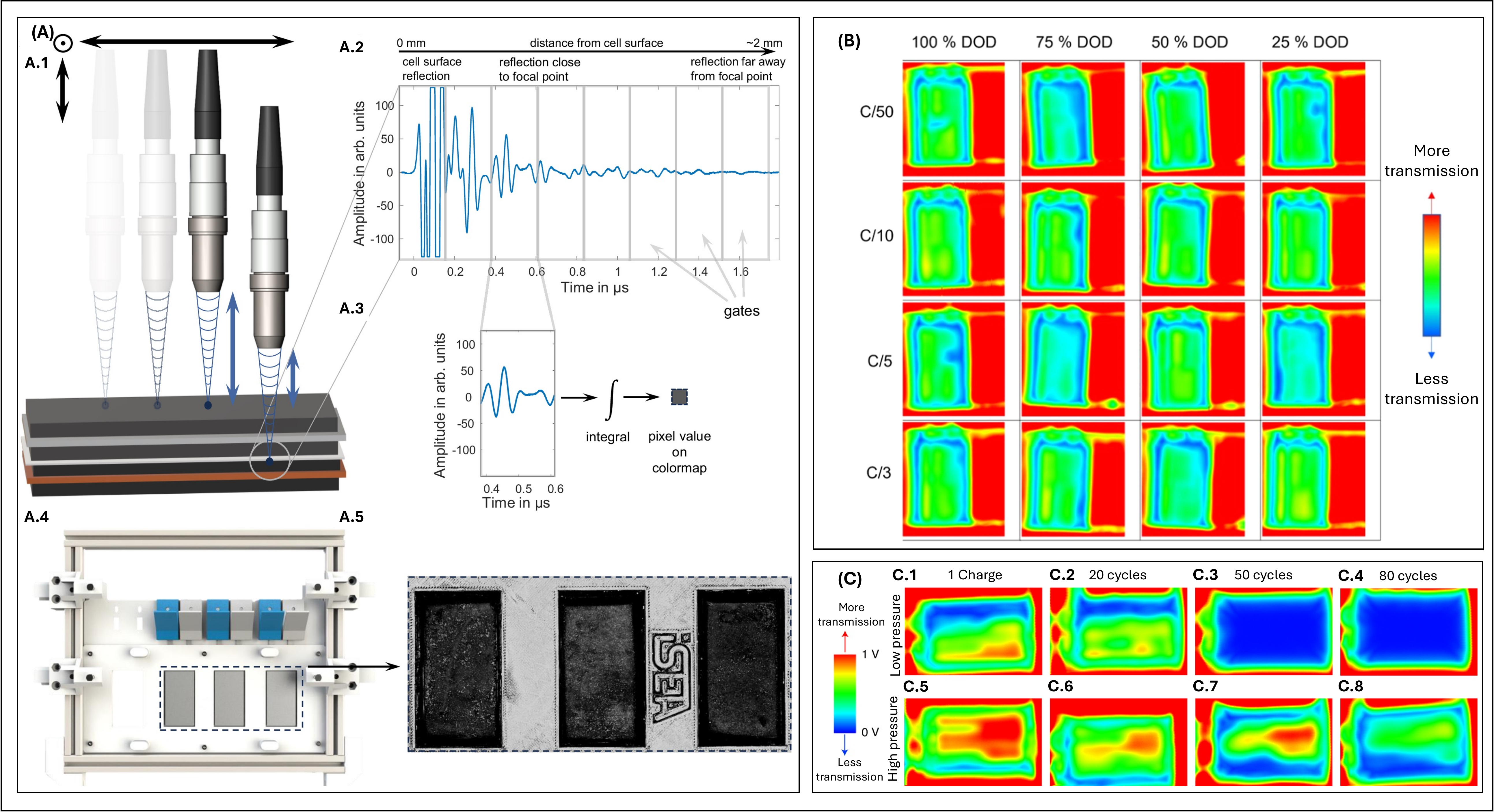}
    \caption{Ultrasonic imaging as a tool for LIBs inspection. \textbf{(A)} Example of a workflow and signal processing scheme to obtain images from mechanical scans of a focused ultrasonic transducer. Reproduced from Ref.~\cite{Wasylowski2024Operando}; licensed under a Creative Commons Attribution 4.0 International (CC BY 4.0) license (\href{https://creativecommons.org/licenses/by/4.0/}{https://creativecommons.org/licenses/by/4.0/}). Ultrasonic mapping in transmission mode for: \textbf{(B)} NMC811 pouch cells at different C-rates and discharge depth conditions (reproduced from Ref.~\cite{Eldesoky2022LongTerm}; licensed under a Creative Commons Attribution 4.0 International (CC BY 4.0) license (\href{https://creativecommons.org/licenses/by/4.0/}{https://creativecommons.org/licenses/by/4.0/})), and \textbf{(C)} NMC532 pouch cells at different pressures and cycles (reproduced from Ref.~\cite{Louli2020ElectrolyteDepletion}, Springer Nature, 2020, with permission from SNCSC.)}
    \label{fig:DefectDetection}
\end{figure*}

Bauermann et al.~conducted scanning acoustic microscopy (SAM) on coin and pouch cell batteries using immersion transducers with the cell in a tank of distilled water \cite{bauermann2020scanning}. The device they employed was a 500 HD$^2$ microscope (PVA TePla Analytical Systems GmbH) and the researchers used two different transducers with resonance frequencies of \SI{15}{\mega\hertz} and \SI{100}{\mega\hertz} in a pulse-echo setup. In order to test the ability of their SAM system to detect defects, they built coin cells with specific defects such as a cut anode or a cracked and poorly processed anode layer. The defects are clearly detectable via SAM compared to a scan of a defect-free coin cell \cite{bauermann2020scanning}. They identified areas of gas inclusions in their scans of a pouch cell, which they verified with a cell tear-down \cite{bauermann2020scanning}. Scanning ultrasonic imaging methods have also been employed in different works by Wasylowski et al.\cite{wasylowski2022spatially, wasylowski2023situ, Wasylowski2024Operando}. Wasylowski et al.'s first work \cite{wasylowski2022spatially} used scanning UT (using a \SI{1}{\mega\hertz}) on cells with different pressure loads to test the effect of pressure on capacity retention. They imaged their cells with UT and also SEM to confirm lithium plating on the anode as well as gas formation \cite{wasylowski2022spatially}. Subsequent research \cite{wasylowski2023situ} built on these results by introducing \textit{in-situ} tomography using a higher frequency transducer (\SI{25}{\mega\hertz}) and the reflected signal to obtain ultrasound images to achieve depth resolved electrode layer separation, allowing improved analysis of internal structural changes in aging cells. Finally, this research group applied SAM for \textit{in operando} visualization of lithium layer dynamics during fast charging, achieving \SI{75}{\micro\meter} spatial resolution and demonstrating real-time tracking of formation and stripping of lithium plating during cycling \cite{Wasylowski2024Operando} using a focused ultrasonic transducer of \SI{25}{\mega\hertz} at a specific internal layer as shown in the scheme in Fig.~\ref{fig:DefectDetection}.A. Key challenges in these studies include accurate deconvolution of overlapping wave reflections, signal attenuation in deeper layers, and optimization of transducer design for improved penetration and resolution. The use of ultrasonic tomography represents a promising technology to gain depth resolution in property and damage maps of LIBs, which potentially will give more insights in terms of defect location and shape \cite{Yi2021MetalDefect}. For example, quality assurance in manufacturing could benefit from further research using this technique. Although it is more accessible and cost-effective than its other techniques like X-ray CT, ultrasound tomography still must overcome resolution challenges due to LIBs' thin-layered structure and their ultrasonic attenuation.

Similar work by Deng et al.~employed ultrasonic scanning on NMC532/graphite and NMC622/graphite pouch cells using immersion focused transducers (\SI{2}{\mega\hertz} frequency, \SI{30}{\milli\meter} focal length) and a cell situated in a silicone oil bath \cite{deng2020ultrasonic}. They generated scans based on the peak-to-peak voltage of the received signals. The researchers created custom pouch cells without any electrolyte, discovering that a significantly lower amplitude signal is detected in these dry cells. They successfully tracked the electrolyte wetting process over time by analyzing the amplitude of a transmitted acoustic wave. Utilizing the same method, they monitored gas evolution within the cells by identifying regions with low transmission. In a subsequent experiment, cells were cycled 2,500 times at \SI{55}{\celsius} to either \SIlist{4.1;4.2}{\volt} or \SIlist{4.3}{\volt}. They observed a phenomenon they termed ``un-wetting,'' where the electrolyte no longer fully wetted the electrodes. To ensure that this was not misidentified as gas formation, they de-gassed the cells for verification. The same research group later applied ultrasonic mapping, using the same ultrasonic settings, to investigate two scenarios: (1) to track electrolyte distribution in \SI{4.20}{\volt} NMC811 cells cycled at \SI{40}{\celsius} for one year \cite{Eldesoky2022LongTerm} and (2) how ultrasonic transmission mapping provides insights into electrolyte depletion caused by increased lithium porosity over cycling  \cite{Louli2020ElectrolyteDepletion}. The objective of the first study was to determine whether electrolyte uptake and jelly roll ``unwetting'' had occurred. They found that electrolyte depletion was not a significant issue under these aging conditions for that specific cell, as shown in Fig.~\ref{fig:DefectDetection}.B. The second study investigated anode-free NMC532 cells, which clearly showed electrolyte depletion via reduced ultrasonic transmission, as can be seen in Fig.~\ref{fig:DefectDetection}.C.

Robinson et al.~performed ultrasonic scanning on both custom-built NMC622 pouch cells and purchased LCO pouch cells. For the custom-built cells, they used a \SI{10}{\mega\hertz}, 1D linear phased array probe in reflection mode to identify added defects such as the removal of one half of a complete electrode layer. They confirmed the presence of these defects through the variation of some peaks in the reflected signal and with X-ray CT \cite{robinson2020identifying}. They also performed measurements with a Panametrics pulser-receiver on the commercial cells both pulse-echo (\SI{1}{\mega\hertz}) and through-transmission (\SI{1}{\mega\hertz} transmit, \SI{5}{\mega\hertz} transducer as the receiver) configurations. The authors claim to have identified a defect in a commercial cell using both UT and X-ray CT that reduced the transmitted ultrasound SA to near zero. However, the defect they claimed to have observed is on the order of microns and, based on diffraction-based resolution limits, it is unlikely that such a small defect would completely block the transmission of an unfocused ultrasonic wave through a transducer with a diameter of \SI{12.7}{\milli\meter}. Although, the defect was small, they suggested that associated changes, such as small gas generation on the surroundings of the defect may explain the significant drop in the SA of the transmitted signal \cite{robinson2020identifying}. Other researchers have also explored the use of phased array probes to image in cells \cite{Xu2023UltraPhasedArray,Zhou2024Fast}. For example, Xu et al. \cite{Xu2023UltraPhasedArray} used ultrasonic phased array imaging to non-invasively track the appearance and evolution of gases in lithium-ion batteries using total focusing methods as the beamforming process to obtain ultrasonic images.

\subsection{UT inspection during the LIB life cycle} \label{sec:otherPartsLifecycle}

UT techniques are gaining increasing attention for use throughout the LIB life cycle, beyond their more explored use in condition and damage detection. This includes its use for electrode fabrication and interface engineering, battery performance enhancement, second-life battery screening, and charging protocol optimization. Overall,  ultrasound-based methods offer the ability to monitor and the potential to actively influence battery performance and degradation mechanisms. This section summarizes the various applications of UT to battery manufacturing/conditioning, second-life batteries, recycling, and other applications in the battery life cycle.

\subsubsection{UT for battery manufacturing and conditioning}\label{sec:UTformanufacturing}

Ultrasonic techniques offer valuable capabilities to monitor different processes within the battery manufacturing stage. Various studies have explored the integration of ultrasound in processes such as electrode manufacturing \cite{Zhang2021InSitu,Zhang2021Effective,Guk2024Calendering,Guk2025Cathode} and electrolyte-filling monitoring \cite{deng2020ultrasonic,Eldesoky2022LongTerm}, which is a critical step to achieve high performance in LIBs \cite{Qian2024InSitu,Li2025Sponge}. The same techniques discussed in Section \ref{sec:DefectDetection} for defect detection can be adapted to study electrolyte wetting during the filling process. McGovern et al.~highlight many such needs during the manufacturing of LIBs where UT could provide insight into the adhesion, porosity, and thickness of electrode coatings on the current collector films \cite{mcgovern2023review}. For instance, an \textit{in situ} ultrasound acoustic technique for measuring lithium-ion battery electrode drying dynamics was developed by Zhang et al. \cite{Zhang2021InSitu}. They continuously monitored the drying process, and revealed correlations between acoustic signal attenuation and the three-stage drying mechanism, linking binder distribution and solvent removal to electrochemical performance (as illustrated in Fig.~\ref{fig:OtherUsesUS}.C). The study demonstrated the technique's sensitivity to drying temperature, showing that lower temperatures promote more homogeneous binder distribution and improved electrochemical properties. As a continuation of research, they showed in another study \cite{Zhang2021Effective} measurements of electrodes with varying coating thicknesses (\SI{200}, \SI{250}, \SI{300}, and \SI{350}{\micro\metre}). Analysis of TOF signal shifts revealed three distinct drying stages: an initial stage with significant TOF decrease due to solvent evaporation, a second stage with milder changes, and a final stable stage. This TOF evolution, validated by concurrent gravimetric analysis, demonstrated the technique's potential for real-time process optimization by dynamically adjusting drying rates based on acoustic feedback.

Building on these capabilities, Guk et al. \cite{Guk2024Calendering,Guk2025Cathode} investigated the role of ultrasonic testing in battery manufacturing. They were focused on the impact of calendaring parameters on the microstructure of graphite anodes and NMC cathodes, demonstrating that ultrasonic testing could provide real-time insights of the electrode uniformity. The study proposed the potential integration of ultrasound as an in-line quality control method for battery production. For instance, Liminal\cite{liminalinsightsLiminalHome} and Titan AES \cite{titanaesHighResolution} are some examples of industrial efforts to develop commercial ultrasonic inspection facilities for battery manufacturing. 

As the final process within the manufacturing stage, LIBs will go through a very critical step of formation cycles, which occur at very low C-rates, where ultrasonic measurements can monitor the gas generated during interface formation of the initial SEI layer of the cells \cite{Chang2024Relating,Ponnekanti2025Formation}. The ability to monitor the formation process is critical, because it can take as long as three weeks and is estimated to account for \SI{48}{\percent} of the battery's total manufacturing cost \cite{GervilliMouravieff2024}.

In summary, although UT shows high potential for inline inspection during cell manufacturing, its current adoption within manufacturing process lags behind research efforts focused on the ``usage" stage of the life cycle shown in Fig.~\ref{fig:LIBUSlifecycle}. One of the main reasons for this lag is the coupled manufacturing requirements for high-speed inspection rates and high spatial resolution. These two requirements are challenging for UT methods individually and are more significant when paired. Specifically, UT has fundamental limits on penetration depth due to the high acoustic contrast for air-coupled UT which could otherwise enable high-throughput monitoring and high attenuation within the cell due to the multi-layered battery structure and attenuative properties of the electrode layers at ultrasonic frequencies \cite{mcgovern2023review, Zhang2021InSitu, deng2020ultrasonic}. Moreover, the wide variety of cell form factors and geometries makes it difficult to standardize the UT process. As a result, UT technology is presently limited to an auxiliary technique to traditional inspection approaches in manufacturing. However, UT provides unique information on subsurface structures such as coatings, porosity, and electrolyte distribution and is therefore highly valuable for battery inspection during manufacturing \cite{mcgovern2023review}. However, it is likely that transducer design improvements and custom data acquisition and data processing strategies will enable the integration of UT in the cell manufacturing to take advantage of the unique measurement capabilities afforded by ultrasonic waves.

\subsubsection{UT for second-life batteries}\label{sec:secondlife}

As described in Secs.~\ref{sec:stateDetection} and \ref{sec:damageDetection}, UT can be applied to monitor the change of mechanical properties of materials, the growth of gas pockets, and other forms of mechanical degradation like delamination of electrolyte layers. The increasing development of the EV market is generating a significant amount of retired LIBs. While these batteries may no longer meet the power demand of EVs, they normally have enough remaining capacity and functionality to be used in second-life applications that have less demanding usage requirements \cite{Hu2022AReviewSLB}. However, in order to evaluate cells for reuse in second-life applications, LIBs retired from high-power applications will need to be graded in terms of electrochemical performance and safety. This evaluation requires efficient and accurate methods for assessing their remaining state of health \cite{S2024SOHSLBs}. Ultrasonic inspection techniques are emerging as a promising non-destructive approach for this purpose, offering speed and cost-effectiveness when compared to traditional methods \cite{Zhu2021EndOfLife}. In addition, features such as gas pockets and delamination inside cells, which are not easily detected via electrochemical evaluation, may be identified using ultrasonic techniques. These features are key indicators of the potential hazards associated with localized high current densities that could cause local heating and risk of thermal runaway. Since these high-risk features may more easily be captured via UT, the use of ultrasound holds the potential to be cost-effectively scaled to meet the anticipated high throughput of end-of-life assessment of cells in the future.

A few studies have explored the correlation between UT and specific degradation mechanisms in LIBs. This includes the work of Williams et al.~\cite{Williams2025BatteryAge} who found a strong correlation between SOH reduction, which is attributed to loss of lithium inventory (LLI), and TOFS in ultrasonic signal responses. They observed consistent shifts in TOF, although the degradation rate was not perfectly linear, with fluctuations across charge cycles. Sun et al.\cite{sun2023ultrasonic} developed an acoustic framework to diagnose nonlinear aging characteristics under high-rate discharge conditions, proposing an $S$-value metric (defined in terms of the rise time and maximum amplitude of the envelope for a certain wave packet over the frequency of acoustic activity) that represents energy dissipation as the acoustic wave propagates. They argued that this could be a more robust metric compared to conventional TOF and SA metrics and identified cathode material collapse as a dominant factor for capacity fade under these high C-rate discharging conditions. Xie et al.~\cite{Xie2022Inhomogeneous} used ultrasonic diagnostics along with spatially-resolved deformation detection, to investigate the effects of lithium plating on inhomogeneous degradation in large-format batteries. They identified lithium plating as a primary mechanism for accelerated degradation, leading to inhomogeneous reactions and reduced performance.

Montoya-Bedoya et al.~\cite{MontoyaBedoya2021NonInvasive} used an ultrasound array to assess the influence of SOC and SOH on TOF and speed of sound in retired 18650 NMC cylindrical batteries. While clear correlations were not established in their preliminary study, they observed trends between ultrasound wave properties and the battery state. This work emphasizes the need for further research to improve measurement protocols and refine the understanding of the relationship between US signals and battery health in cylindrical cells. The study also notes the inherent complexity of ultrasound responses in second-life batteries due to the co-existence of multiple degradation mechanisms. Fordham et al.~\cite{Fordham2023Correlative} employed a combination of NDE techniques –- infrared thermography, ultrasonic imaging, X-ray CT, and synchrotron X-ray diffraction –- to analyze the aging of large-format pouch cells (with an LMO/NCA cathode blend and graphite anode) retired from an automotive application. They performed experiments on batteries vertically (``rotated'') and horizontally (``flat'') oriented within the EV's battery pack. Using a TOF window between \SI{26}{\micro\second} and \SI{38}{\micro\second}, they extracted the maximum SA and derived a percentage metric from the cell area exposed to high attenuation. They showed that ``rotated'' cells exhibited a higher area of high acoustic attenuation compared to ``flat'' cells, associated with higher degradation, highlighting the importance of cell orientation and location within a battery pack. This multi-modal approach is likely to yield improved understanding in future work and shows complementary insights into cell degradation, demonstrating the power of integrating different NDE methods.

Another recent study employed a 64-element ultrasonic array to obtain spectral-based quantitative parameters, including mid-band fit (MBf), spectral slope, and intercept, from circumferential waves propagating around cylindrical batteries \cite{MontoyaBedoya2024Quant}. That work demonstrated that the MBf parameter is highly sensitive to both SOC and SOH, and that it could be used to effectively classifying second-life batteries using this frequency-based ultrasonic parameter, even without knowledge of the prior usage history, thereby working to address one of the main challenges exposed by Zhu et al.~\cite{Zhu2021EndOfLife} for the use of acoustic monitoring to assess the SOH of used batteries. Notably, their method relied on measurements on pristine batteries of the same chemistry and manufacturer, which are not always available, to compare with the second-life batteries. However, it showcases a possible path to address this limitation.

One critical aspect of UT as currently employed is that they have been shown to be effective in a relative sense, by comparing observables from the same cell over time, rather than in an absolute sense, where one could infer the state or damage condition of a LIB based on measurements in arbitrary conditions and past loading history. However, in the case of second-life applications and recycling, batteries must be evaluated as a ``black-box'' without monitoring history for comparison. It is therefore of high importance to determine reliable ultrasonic signal signatures that are indicators of LIB state and damage type in the absence of previous history for UT to be a valuable tool for recycling applications and evaluation for second-life use.

The ability to integrate ultrasonic inspection into a comprehensive assessment procedure further enhances its value in optimizing the end-of-life decision-making process for retired cells from EV application \cite{Zhu2021EndOfLife}. Further research is needed to address existing limitations and fully extract the potential of ultrasonic inspection in promoting a sustainable circular economy for battery technology. The integration of advanced data analysis techniques and multi-modal NDE approaches will be crucial for achieving more accurate and reliable SOH estimations, facilitating the wider adoption of second-life battery applications \cite{Attia2025Challenges}.

\subsubsection{UT for recycling}\label{sec:recycling}

Presently, \SI{80}{\percent} of retired Li-ion batteries have been sent to landfills due to a lack of effective ways to reuse or recycle them \cite{Rasheed2024}. The increasing volume of toxic heavy metals (e.g., cobalt, nickel, manganese, among others) and flammable electrolytes coming from spent LIBs poses a significant environmental issue \cite{Zheng2018AMiniReview, Wang2022Prospects}. Inappropriate final disposal by land filling or incineration results in soil and water contamination, air pollution, and greenhouse gas emissions \cite{Mrozik2021Environmental}. In addition, the scarcity of critical materials such as cobalt and lithium \cite{Bauer2010USDOE} highlights the economic need of efficient recycling to ensure resource independence and minimize dependence on potentially unstable geopolitical sources \cite{Mossali2020LIBCircular}. Current battery recycling technologies, while evolving, often involve energy-intensive processes and may not achieve complete material recovery \cite{Mossali2020LIBCircular, Larouche2020Progress}, prompting the search for innovative and sustainable solutions. Here, a distinction must be made between ultrasonication, which has been used for recycling processes for a significant amount of time as a high-intensity source of energy (c.f.~Refs.~\cite{Lei2021Ultrasonication,Tong2023ReviewUltrasoudRecy,Song2024Exploring} ), and ultrasonic inspection, which is low power and does not damage cells and therefore offers a promising way to improve the inspection efficiency and safety associated with battery recycling processes. In the context of battery recycling, these capabilities translate into the potential to assess battery SOH, detect internal damage, and monitor the efficacy of recycling processes. 

For instance, in the recycling sector, UT methods can play a crucial role in the assessment of the state of spent LIBs before processing, helping to prevent thermal runaway incidents. The work of Wu et al.~\cite{Wu2022Avoiding} demonstrates the importance of discharging spent LIBs to a safe voltage ($<\SI{1.5}{\volt}$) before disassembly to mitigate thermal runaway risks. This pre-processing step is critical, as incompletely discharged cells retain significant energy that can lead to explosions or fires during handling and processing. UT can rapidly assess the SOC and voltage of individual cells, enabling efficient sorting and prioritizing for discharge. Furthermore, UT methods can detect internal defects such as cracks, delamination, or swelling, providing valuable information for optimizing the recycling process and ensuring safe handling. In addition, though outside the scope of this review and not covered here, some studies have explored the use of ultrasonic-assisted methods to enhance recycling processes. The reader can find more information in these articles \cite{Nshizirungu2024USAssisted,Kong2023AnImproved,Chen2024Chemical,Yan2022Improved,Zhao2020Ultrasonic,Huang2023Degradation}.

While the explicit application of ultrasonic inspection in monitoring battery recycling is not widely documented, the underlying principles and capabilities of ultrasonic techniques strongly suggest their potential utility. The existing literature reveals a clear trend towards the development of advanced NDE methods for LIB characterization, and ultrasonic inspection offers several advantages that could be leveraged in this context. For instance, the work by Bauermann et al.~demonstrates the successful application of SAM to visualize internal defects in coin and pouch-type battery cells. SAM, a high-resolution ultrasonic imaging technique, revealed defects such as electrolyte leakage, faulty electrodes, and gas accumulation with high accuracy \cite{bauermann2020scanning}. This capability is directly relevant to battery recycling, as it could enable the rapid and non-destructive sorting of spent batteries based on their condition. Batteries with minor defects might be candidates for second-life applications \cite{Iqbal2023ASurvey}, while those with significant damage would require different processing strategies \cite{Mossali2020LIBCircular}. Moreover, the ability of ultrasonic techniques to detect internal damage is crucial for enhancing the safety of battery dismantling and pre-treatment processes \cite{bauermann2020scanning}, minimizing the risk of thermal runaway or other hazards \cite{BravoDiaz2020Review}. 

The potential for integrating ultrasonic inspection into various stages of the battery recycling process warrants further investigation. For example, ultrasonic sensors could be incorporated into automated disassembly lines to guide the separation of various battery components \cite{Lander2023Breaking}. Real-time monitoring of the recycling process using ultrasonic techniques could allow process optimization and increase material recovery rates \cite{Mossali2020LIBCircular,Larouche2020Progress}. The ability to monitor material degradation during pyro-metallurgical or hydro-metallurgical treatments \cite{Larouche2020Progress} could provide valuable feedback for process control and efficiency improvements. In addition, ultrasonic inspection could play a role in quality control of recovered materials to ensure that recycled components meet the required specifications for reuse or re-manufacturing \cite{Mossali2020LIBCircular}. Future research should focus on optimizing process parameters, such as ultrasonic frequency and treatment duration, to maximize efficiency and further explore its integration with emerging green recycling technologies.

\subsubsection{Other uses of UT in the life cycle of LIBs}

The use of ultrasonic inspection techniques has expanded beyond conventional battery characterization and defect detection, finding innovative applications throughout the lithium-ion battery life cycle  and performance enhancement \cite{Huang2020Enabling,Huang2022Overcoming,Li2024Optimal,Im2023UltraInduced,Im2024UltrasoundEnabled}, demonstrating the versatility of this non-destructive technology.

\begin{figure*}[ht]
    \centering
    \includegraphics[width = \linewidth]{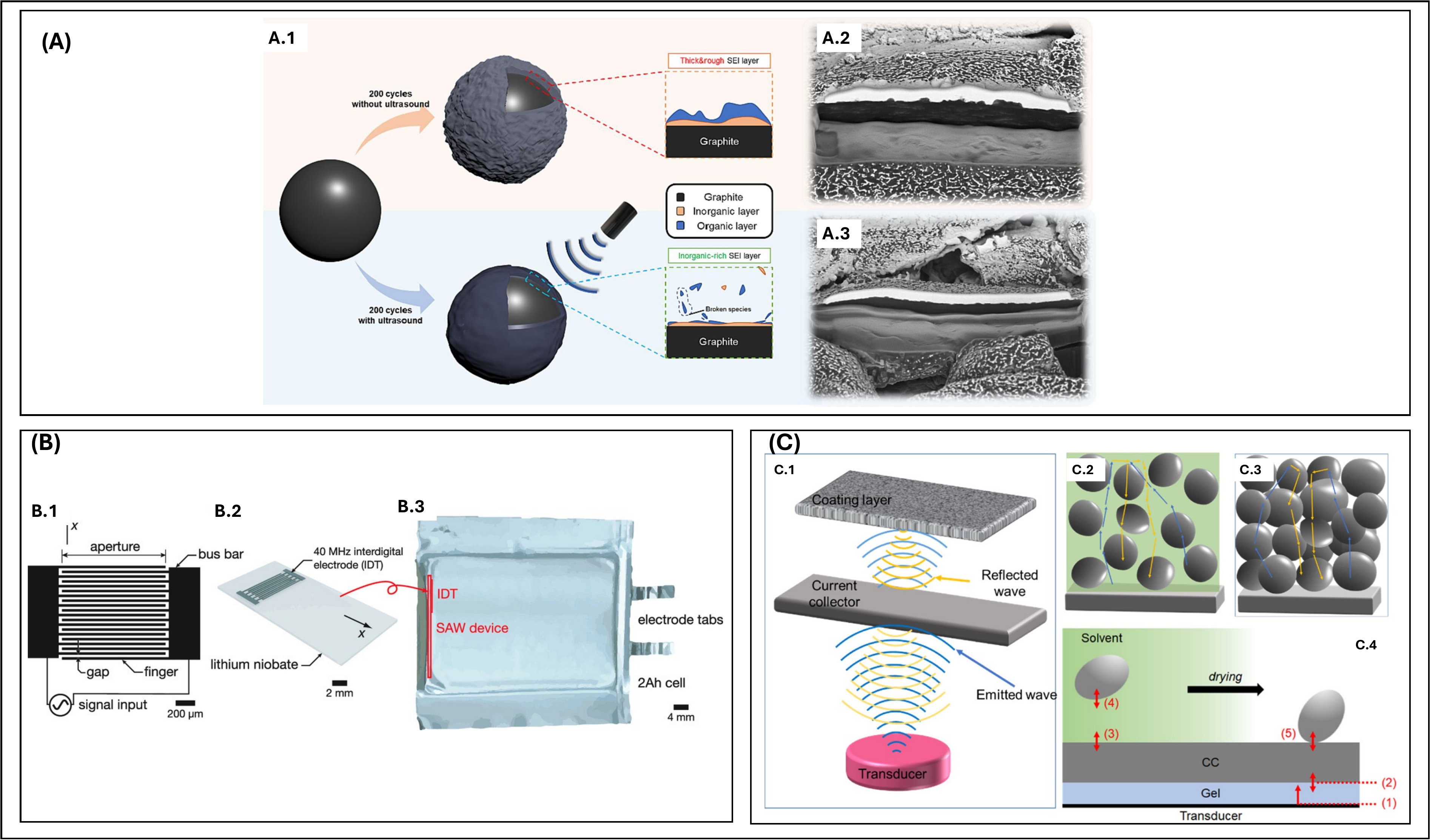}
    \caption{Non-conventional uses of ultrasonic testing for LIBs. \textbf{(A)} Influence of high intensity ultrasound exposure during cycling: (A.1) Illustrations of the graphite surface with (top) and without  UT (bottom) exposure after 200 cycles, and Focused Ion Beam Scanning Electron Microscopy (FIB-SEM) images of the anode with (A.2) and without (A.3) UT exposure. Reproduced from Ref.~\cite{Jie2023Enhanced}; licensed under a Creative Commons Attribution-NonCommercial-NoDerivatives 4.0 International (CC BY-NC-ND 4.0) license (\href{https://creativecommons.org/licenses/by-nc-nd/4.0/}{https://creativecommons.org/licenses/by-nc-nd/4.0/}). \textbf{(B)} Integration of a surface acoustic wave device into a pouch cell to enable fast charging through ion transport. Reproduced from Ref.~\cite{Huang2022Overcoming}; licensed under a Creative Commons Attribution 4.0 International (CC BY 4.0) license (\href{https://creativecommons.org/licenses/by/4.0/}{https://creativecommons.org/licenses/by/4.0/}). \textbf{(C)} Illustration of UT wave reflection in electrode coatings, exemplifying wave paths through wet (solvent-filled) and dried (solid) electrode. Adapted with permission from Ref.~\cite{Zhang2021InSitu} Copyright 2021 American Chemical Society.}
    \label{fig:OtherUsesUS}
\end{figure*}

Some authors have explored the use of ultrasound for battery interface engineering. Jie et al. \cite{Jie2023Enhanced} explored the role of ultrasound in modifying the SEI layer of lithium-ion batteries. They exposed coin cells to an ultrasound frequency of \SI{40}{\kilo\hertz} through a bath sonicator over the different electrochemical characterizations. They discovered that ultrasound-assisted treatment led to the formation of an inorganic-rich, thinner SEI layer, which enhanced charge transfer kinetics and improved cycling stability (as seen in Fig.~\ref{fig:OtherUsesUS}.A). This study highlights ultrasound’s potential for fine-tuning battery interfaces to optimize electrochemical performance. Another work in this area comes from Shpigel et al. \cite{Shpigel2018InSitu}, who developed an \textit{in situ} acoustic diagnostic technique to monitor particle-binder interactions in battery electrodes. In their study, an electrochemical quartz crystal microbalance with dissipation monitoring (crystal oscillating in the MHz-range frequency) was introduced to assess binder degradation and electrode stability. This method provides valuable insights into electrode integrity, aiding in the development of more durable battery materials.

On the side of battery operation, Li et al. \cite{Li2024Optimal} integrated ultrasound with a model predictive control (MPC) approach, and they were able to optimize battery charging protocols, preventing lithium plating and enhancing battery safety. Their research showed that ultrasound can help identify early-stage lithium deposition, a major cause of capacity loss and battery failure. An electrochemistry physics-based model was used to exploit the relationship between having a negative electrode potential of less than \SI{0}{\volt} and the generation of lithium plating. Thus, maintaining the negative electrode potential above \SI{0}{\volt} was the target for designing a model predictive control strategy for optimal charging. They used an ultrasonic array of \SI{64} elements with a center frequency of \SI{1}{\mega\hertz} to detect gas generated inside the battery, which the authors assumed to be associated with the lithium plating critical point. The voltage at which this critical point occurs for different temperatures was used for the model calibration. Another approach to improving the performance of rapid battery charging was reported by Im et al.~\cite{Im2024UltrasoundEnabled}, where they transmitted ultrasonic waves while charging the cells using a custom-designed multi-stage constant current protocol. This work built on their previous work \cite{Im2023UltraInduced} where they demonstrated that ultrasound waves of \SI{35}{\kilo\hertz} can reduce internal impedance by up to \SI{16.9}{\percent} at room temperature and increase the usable capacity by \SI{53.4}{\percent} at low temperature. Verified by complementary techniques, they showed an increased porosity in the SEI layer due to cavitation-induced disruptions, which facilitates the lithium-ion transport. These works demonstrate that ultrasound is not limited to passive monitoring; it can actively influence battery operation. In this sense, Huang et al. \cite{Huang2020Enabling} explored using surface acoustic waves (SAW) devices to enable rapid charging in lithium metal batteries. Their study demonstrated that a SAW device, operating at \SI{100}{\mega\hertz}, generates intense localized accelerations, inducing acoustic streaming—a form of electrolyte flow—within the battery's interelectrode gap. This electrolyte flow suppressed dendrite formation and facilitated uniform lithium deposition, allowing for significantly higher charging rates. As a continuation of this work, a more recent study \cite{Huang2022Overcoming} explored the use of a \SI{40}{\mega\hertz} SAW device as a tool to address intrinsic limitations in fast-charging LIBs with NMC532 chemistry, implemented as depicted in Fig.~\ref{fig:OtherUsesUS}.B. This enhanced electrolyte flow significantly improves lithium-ion transport, mitigating the formation of concentration gradients that are detrimental during fast charging. For instance, compared to batteries without SAW integration, the energy density doubled at a 6C charge rate and the batteries maintained at least \SI{72}{\percent} of their initial capacity after \SI{2000} cycles. However, further research is needed to fully elucidate the underlying mechanisms and optimize the SAW device design and integration for various battery configurations. Other more recent study \cite{Zhang2025Enhancing} has also explored the use of non-ultrasonic acoustic frequencies (between \SI{6}-\SI{11}{\kilo\hertz}) to enhance the performance of lithium metal batteries through stabilizing the SEI layer.

From material synthesis to performance optimization and manufacturing quality control, ultrasound promises advantages as a versatile and powerful tool in lithium-ion battery technology. The studies discussed here illustrate how ultrasound can be used both passively in diagnostics and monitoring, and actively for material processing and electrochemical performance enhancement. 

\section{Capability gap analysis}\label{sec:ChallengesOpportunities}

\subsection {Classification of cell-dependent ultrasonic metrics}\label{sec:cellDependence}

Ultrasonic wave propagation can differ for different LIB form factors due to evolving mechanical properties as discussed in Sec.~\ref{sec:fundamentals}. The presence of soft casings (i.e. pouch cells) versus rigid casings (i.e., prismatic or cylindrical cells) also affects UT and should therefore be considered. In addition, different cell geometries could support distinct propagating modes that can be used to detect changes within a cell. In this regard, while much of the research has focused on pouch cells, expanding studies to prismatic\cite{Wang2024ApplicationML,zhang2024exploring,Zhu2025Wavelet} and cylindrical \cite{hsieh2015electrochemical,MontoyaBedoya2024Quant,rohrbach2021nondestructive,Chen2024SOCCylindrical,Nguyen2024Ultrasonic} cells is crucial for a comprehensive understanding of how ultrasonic wave propagation varies with different cell geometries and internal structures and how this information can be exploited to enable ultrasonic testing.

Another factor that must be considered is cell-to-cell variability even within the same manufacturing batch. Variations on electrolyte filling, homogeneity of electrode coating, formation cycling, and other factors will introduce changes in the ultrasonic responses of nominally identical cells. While these subtle changes and the associated differences in ultrasonic signal metrics can be used to assess variations in standardized processes, they can also complicate the interpretation of the results when not all properties are presently not well-understood or documented.

Realizing the full potential of UT of LIBs requires the categorization of relationships between the internal state of cells and the ultrasonic metrics. This review and others provide a concise categorization based on existing knowledge, but much remains to be done, particularly as UT grows and new battery chemistries emerge. An ideal categorization would consist of UT under specific conditions and well-defined internal states using well-understood UT methods. In practice, this is much more difficult. For example, Chang et al.~demonstrated that fast changes in cell temperature can induce gas formation \cite{chang2020understanding}, which means researchers attempting to isolate thermal effects need to be careful in their experimental design. While isolating specific internal states may be difficult if not impossible, the impact of UT experiments can be amplified if authors report as much information as possible on experimental details such as the cell condition including ambient temperature, extent of calendar and cycle aging, exact specifications and compositions of cell components, and explicit statements regarding the type of wave propagation used. 

Given these challenges, a comprehensive approach that includes refining experimental design, standardizing reporting, and applying advanced analysis is crucial for wider use of UT of LIBs. Machine learning techniques trained on well-characterized ultrasonic data could provide insights by identifying patterns and correlations between specific degradation mechanisms and ultrasonic signals \cite{Liu2025ExplainableAI,Yang2025Review}. Additionally, researchers should validate their hypotheses of internal cell states with advanced imaging techniques such as X-Ray Diffraction (XRD) or SEM, as was shown in Refs.~\cite{wasylowski2022spatially,Wasylowski2024Operando}. With better reporting on cell condition, cell composition, and wave propagation researchers will better understand the lack of complete agreement between the observed ultrasonic signal characteristics and various cell states and devise methods to resolve ambiguity.

Addressing knowledge gaps in the relationship between constituent properties and cell state is urgent for the UT of LIB since the lack of generalizable acoustic properties for various cells is currently a significant bottleneck in the industrialization process. The key challenges include the lack of standardized test procedures, differences in reporting methods used by various researchers, and sensitivity in regards to cell composition parameters. This gap can be partly addressed through the success of material property parameterization efforts described in Sec.~\ref{sec:parameterizationGap}.

\subsection {Model improvement via material property parameterization}\label{sec:parameterizationGap}
Predictive models (i.e.~forward models) of wave propagation in complex media like LIB are central to accurate material characterization, damage detection, health monitoring, and other inverse problem applications.\cite{willcox2021imperative} Somewhat paradoxically, the development of accurate models requires detailed knowledge of LIB structure and constituent properties. This is necessary because physical models must accurately consider the dynamics for all appropriate length and time scales in order to correctly capture experimental observations and determine what changes as a function of loading history. Unfortunately, obtaining accurate measurements of the mechanical properties cell components is challenging for a number of reasons. The material properties of cell components are dependent on temperature \cite{xu2016coupled, love2011thermomechanical, love2016perspective}, pressure \cite{Li2022EffectExternalPressure}, presence and type of electrolyte environment \cite{chen2018evolution, sheidaei2011mechanical, xu2016coupled}, degree of lithiation \cite{koerver2018chemo, qi2010threefold, qi2014, xu2017mechanical}, electrodes' chemistry \cite{Stallard2022MechanicalCathode}, and orientation relative to manufacturing direction \cite{wang2020porosity, sheidaei2011mechanical, xu2016coupled}. These effects are further complicated by spatial variability in the plane of each component, meaning that individual measurements may not provide a complete picture of the material properties of each constituent. Because material properties of components are affected by the presence of the electrolyte environment\cite{de2019situ} and many cell components react in an air environment\cite{pan2020effect}, measurements need to either be taken \textit{in situ} or in an inert environment using specialized equipment, such as glove boxes. Future modeling efforts will therefore benefit from multi-scale material characterization approaches, where both macro and microscale effects are mechanically considered, documented, and integrated into multiphysics models. These efforts will lead to more reliable models and therefore improved inverse methods to monitor and characterize LIB. 

In addition, improved knowledge of microstructural variability in constituent material properties and LIB structure aid in understanding the expected range of ultrasonic signal metrics for any given battery state. Current knowledge of many material properties is very limited and therefore these parameters are often approximated as ``free variables'' in order to represent ultrasonic wave motion in LIB. This limited information often leads to unacceptably large uncertainty and a lack of uniqueness when using experimental data to infer changes in properties and structure. Improved knowledge of these parameters will reduce the number of unknown parameter values used in forward models and therefore improve the quality and reliability of wave physics models and inversion techniques they are built around. To improve the reliability of ultrasonic wave propagation models, open-access databases containing experimentally-validated material property datasets for different operating conditions should be created and maintained. Some initiatives on this approach of open-access databases for UT in LIBs are provided in Refs.~\cite{Galiounas2025Generalisation,Guk2026OpenData,Guk2026PulseEchoData}. These open-access resources would allow direct comparisons across different research works, thereby advancing the research of the entire community. 

Another key aspect of LIB modeling is the dynamic nature of most properties and structure, which evolve as a function of aging, mechanical stress, and electrochemical cycling. Thus, future studies should aim to integrate these changes into modeling frameworks to improve their effectiveness in evaluating battery performance and degradation in complex loading scenarios.

The current modeling capability gap reflects a fundamental dependence between the lack of accurate material properties for operational conditions and uncertainty introduced in wave propagation models, which ultimately limits the validity of UT-based state assessment. As mentioned above, the primary challenges associated with measuring material properties under different conditions, such as different SOC, are experimental meaning that gaps in material property knowledge and battery structure should be a key motivating factor for future experimental work. Given the centrality of modeling in inverse problems used to characterize constituent materials, map microstructure, identify damage, and monitor battery health, the urgency of addressing this gap is high relative to the other gaps if researchers want rapidly advance the field.

\subsection{Engineering challenges for \textit{in situ} applications}\label{sec:engineeeringChallenges}

Several engineering and technical challenges must be addressed for UT of LIB to pass the threshold from a fundamental and applied research topic to real-world applications. Most existing research involving UT of LIBs have relied on measurements of single cells. For this NDE methodology to be practical, it must be demonstrated in systems containing multiple cells. Experiments should be devised to test the practicality of UT of multi-cell modules or even full packs. As the community evaluates the potential challenges, there will be two main scenarios for the usage of UT: (\textit{i}) cases where batteries can be extracted from the application to do an ultrasonic measurement, and (\textit{ii}) integrated continuous monitoring of battery performance.  An example of the first scenario is rapid UT screening for second-life applications, by developing rapid UT screening that enables assessment of large quantities of cells in a short period of time. This would likely require air-coupled or immersion methods where coupling is repeatable, though novel contact methods may also be devised.  Depending on the scale of the application and with the current state of the contact UT, ultrasonic sensors will need to be located at each individual cell. Placing the transducer at each cell raises questions about economic feasibility and technical considerations, such as the size and specific location of the transducers. For small-scale form factors, such as cylindrical cells or small pouch cells, challenges might arise regarding miniaturization of commonly used lab scale (characteristic size of \SI{10}-\SI{25}{\milli\meter}) transducers. On the other hand, large pouch or prismatic format factors might benefit from ultrasonic guided wave approaches, since these waves can interrogate a larger cell volume. In addition, integrating ultrasonic monitoring into commercial BMS requires robust, cost-effective transducer designs and real-time ultrasonic data processing capabilities.

Further, UT should be adapted and validated to specific application. For instance, validating UT for EVs requires testing under real driving conditions that emulate load profiles\cite{popp2019state,owen2022operando,MontoyaBedoya2024Quant}. By testing under these conditions, researchers will gain important insights into the behavior of ultrasonic signals under real driving conditions. Practical implementation will also require the development of robust coupling methods for long-term operation since the most obvious approach, direct-contact or embedded piezoelectric transducers, require consistent adhesion over their lifetime.

Another engineering decision consideration is whether to employ single-element transducers or transducer arrays for a given application. Single-element transducers remain simple and affordable, yet arrays can deliver beam steering capabilities along with dynamic focusing for greatly enhanced spatial accuracy. Assessing these approaches, first through fundamental research studies and then through trade studies, will be necessary to optimize their use in large-scale LIB systems.

Another critical step in the integration of UT into the LIB life cycle the miniaturization of ultrasonic inspection systems to circuit board or chip-level as explored by Koller et al.~\cite{koller2023ultrasonic} and Popp et al.\cite{popp2019state}. Such technological advances will reduce cost and power requirements of UT systems and therefore represent critical steps towards the commercialization of ultrasound-based solutions by facilitating the integration into the LIBs value chain.

The capability of UT-based techniques to transition from academic research into industrial applications relies heavily on the adaptability to all cell types. As noted by Galiounas et al.~\cite{Galiounas2025Generalisation}, ultrasonic inspection has been shown to be a tool for identifying battery states within known cells, but it has not been demonstrated to generalize across a wider cell population. Further development is therefore required to identify features of ultrasonic signals that are common to all cell form factors thereby ensure accurate correlation of ultrasonic signal features with battery state indicators (SOC/SOH) for different cells. 

Addressing the engineering challenges described above is urgent for the adoption of UT-based technologies into the LIB industry. The primary obstacles are economic (cost per cell), regulatory (lack of standards for UT for LIBs), and technical (coupling robustness, transducers size, and signal processing time), all of which are significant. Some success with this problem depends on resolving the generalization challenge discussed in Sec.~\ref{sec:cellDependence}.

\subsection{Next-generation cells and electrochemical systems}

Ultrasonic inspection provides insights into electrode morphology, interfacial stability, and tracking battery state for operation in different conditions. One direction for future work on UT of electrochemical systems is the expansion of the work to the so-called next-generation cells, including fully solid-state cells \cite{schmidt2016situ, musiak2021real, huo2022evaluating, Sun2025ChemoMechanics}, cells incorporating lithium metal anodes \cite{musiak2021real, chang2021operando, huo2022evaluating, Chang2024Relating}, silicon anodes \cite{bommier2020SiGr, Sun2025ChemoMechanics}, other novel nano-engineered electrodes \cite{An2024Construction}, redox flow batteries \cite{chou2016novel, yan2023online}, sulfur electrodes testing for lithium-sulfur technology \cite{Sun2024SulfurCathodes}, sodium-ion batteries \cite{Laufen2024SIBUltrasound}, and lithium-ion capacitors \cite{oca2019lithium}. UT is a promising method for state detection for these next-generation cells for reasons, as one may anticipate from the utility demonstrated by UT for conventional LIB state detection. For example, silicon anodes experience a volume increase of \SI{300}{\percent} when fully lithiated, which should result in a measureable change in the received ultrasonic signals. For instance, Sun et al.~used ultrasound to investigate the chemo-mechanical dynamics of silicon anodes in solid-state batteries, showing that metastable $\mathrm{Li}_x\mathrm{Si}$ phases lead to significant and irreversible porosity changes during cycling \cite{Sun2025ChemoMechanics}.

In addition, the recent work by Chang et al.~\cite{Chang2024Relating} explored the use of acoustic transmission measurements to track the effective stiffness of anode-free cells, which are cells that are produced without an anode, but a lithium metal anode is formed after the first time it is charged \cite{Huang2022AnodeFreeSSB}. The study found that a faster formation rate (C/3) yielded comparable cycling performance and cell stiffness changes to a slower rate (C/10), suggesting that the traditionally-used slow formation processes may be unnecessary for anode-free cells. This contrasts with previous studies using \textit{in situ} volume and pressure measurements that found only linear relationships with SOC \cite{Louli2017Pressure,Louli2019Pressure}, highlighting the superior sensitivity of ultrasonic inspection to detect subtle, nonlinear changes associated with complex internal processes. Related to this work, Webster et al.\cite{Webster2024USLithiumMetalNasa} explored local resonant ultrasonic spectroscopy on lithium metal cells. They tracked the performance with this technique during aging for cells cycled at 0.1C, 0.2C, and 0.5C and observed shifts to lower resonance frequencies as the battery ages with an inhomogeneous shift for different regions of the battery. These resonances are detected by tracking a minimum in the reflected spectrum associated with thickness-mode resonances of the cells. They quantitatively tracked shifts in these resonances, assessed changes in the local bulk properties with aging, and created resonance maps obtained by scanning the samples. Future research should focus on the complex task of mapping the shift in resonance location to quantitative values of effective elastic moduli, which is a significant outstanding challenge as discussed in Sec.~\ref{sec:parameterizationGap}. While the work of Webster et al.~exploited thickness-mode resonances, resonant ultrasonic spectroscopy is a well-known technique that has been used to compute all components of the elastic modulus tensor in other applications \cite{Balakirev2019RUS} and therefore have strong potential to provide additional information on changes in LIBs.

Beyond LIBs, Sun et al.~applied \textit{operando} acoustic transmission to sulfur electrodes in lithium-sulfur batteries and found that wave damping correlates with sulfur phase transitions, enabling real-time detection of structural evolution and demonstrated how ultrasonic wave damping is highly sensitive to solid-liquid-solid transformations in sulfur cathodes \cite{Sun2024SulfurCathodes}. Another work by Laufen et al.~\cite{Laufen2024SIBUltrasound}, has recently implemented ultrasonic diagnostics for sodium-ion batteries, which may suffer more particle cracking and volumetric changes than lithium-ion technology due to the larger size of sodium ions compared to lithium-ions. However, that work only showed the ability of the technique to detect gas generation after aging in this new battery technology.

Despite its advantages, ultrasonic inspection faces technical challenges on next-gen batteries. Signal interpretation remains complex, particularly in differentiating lithium plating from gas evolution, as both induce similar acoustic signatures \cite{chang2021operando, huo2022evaluating}. Spatial and temporal resolution limitations prevent nanometer-scale imaging of solid-electrolyte interphases, making it difficult to precisely characterize early-stage degradation in next-generation chemistries \cite{Sun2025ChemoMechanics}. 

Addressing this this capability gap is not as urgent as the previous cases discussed here, its importance is growing due to rapid commercialization efforts of solid-state and silicon-anode technologies\cite{yang2025addressing}. This gap is likely to be addressed in part by UT research on conventional LIB. Namely, progress made in addressing the engineering challenges identified in Sec.~\ref{sec:engineeeringChallenges} and concerted efforts to collect material property data as discussed in Sec.~\ref{sec:parameterizationGap} will inform implementation of UT for next gen cells.

\section{Summary and outlook}\label{sec:conclu} 

Ultrasonic testing of LIBs is a developing field at the intersection of materials science, electrochemistry, and acoustics. Despite notable advancements showcasing the potential of UT techniques for battery assessment, the field is still in its infancy. The physical sources of observed acoustic signatures, the generalizability of results across cell chemistries and form factors, and the understanding of wave propagation in operational cells are just a few of the many fundamental questions remain open. This review presented a comprehensive summary of the UT research for different portions of the LIB life cycle. The review provided a broad overview of the fundamentals of ultrasonic wave propagation in LIBs as multilayer structures and presented the challenges and opportunities for this technique as a non-destructive tool that provides information on LIB behavior that is unique from traditional measurements to assess LIB health and safety.

As emphasized throughout this work, UT methods have emerged as powerful diagnostic tools for energy storage devices, which can provide rapid and cost-effective insight into material changes within LIB cells. UT has been shown to accurately determine the SOC and SOH of cells and to identify various unfavorable phenomena within cells that may pose safety risks or cause accelerated capacity degradation. 

However, much work must be done to quantitatively link observed ultrasonic signals to changes in material properties and structure, which is central to understanding LIB health. Improvements in our ability to reliably map property and structure changes to observables remains a fundamental challenge that requires more extensive material characterization and that is likely to be improved using multi-modal diagnostics such as concurrent UT and X-ray CT imaging. Improved UT methods will open further opportunities for rapid identification of important degradation mechanisms in cells, provide more accurate lifetime predictions, and function as a valuable tool for estimating the risk of cells undergoing hazardous failure. Modeling and inversion is a key enabler for achieving higher confidence in UT methods, since it provides interpretable and quantifiable correlations between known physical changes within LIB cells and the observed UT metrics. Modeling and inversion methods requires a significant effort to link existing electrochemical and system models with elastic and poroelastic wave modeling, which is a topic that is still in its infancy in the field of UT for LIB.

In this sense, notable challenges and opportunities exist for the application of ultrasonic monitoring and characterization of LIBs and are summarized below.

\begin{enumerate}
\item The effect of cell form factor on ultrasonic wave propagation and testing methodology requires further investigation to tailor methodologies to form factor and application.
\item The development of reliable ultrasonic signal metrics that indicate LIB state and damage type in the absence of previous history in order to compare cells with different histories is a challenge that must be addressed in order for UT to be a valuable tool for recycling applications and evaluation for second-life use.
\item Temperature and C-rate variations during operation introduce changes in ultrasonic signal interpretation. Accounting for these changes requires robust models that consider electrochemical and thermal histories in order to use UT in monitoring systems that assure stable operating conditions and SOH estimates.
\item Further research is needed to fully document and parametrize the mechanical properties of LIB constituent materials. This research will greatly aid in the creation of elastodynamic and poroelastic models that relate ultrasonic signals to physical phenomena within the cells.
\item Machine learning approaches have shown promise in correlating ultrasonic signals with LIB state metrics. However, their generalization across cell chemistries and even across batches of the same cell remains an open question. Further, machine learning methods that employ interpretable, physics-based constraints should be prioritized.
\item Integrating UT with complementary NDT methods, such as electrochemical impedance spectroscopy, X-ray CT, infrared thermography, among others, is a promising path toward the development of more interpretable and robust diagnostic and prognostic methods.
\item Standardizing UT measurement protocols, reporting conventions and usage history, and benchmark datasets are crucial for comparing studies and gaining industry acceptance. 
\item Community-driven initiatives for open-access databases of validated UT measurements across cell chemistries, form factors, and operating conditions would significantly accelerate progress.

\end{enumerate}

\section*{Supplementary Material}\label{sec:SM}
\noindent See Supplementary Material for a mathematical description and additional context regarding ultrasonic wave propagation in LIBs.

\begin{acknowledgments}
\noindent The authors acknowledge support from grant AR0001723 under the ARPA-E Jumpstart Opportunities to Unleash Leadership in Energy Storage (JOULES) program, managed by Halle Cheeseman. J.~S.~Lee acknowledges support from Chungnam University for sabbatical research visit. S.~Montoya-Bedoya acknowledges support from the Foreign Fulbright Colombia Minciencias program.
\end{acknowledgments}

\section*{Author declarations}
\subsection*{Conflict of interest}
\noindent The authors have no conflicts to disclose.

\section*{Data availability}
\noindent The data that support the findings of this study are available within the review article and supplementary material.

\bibliographystyle{ieeetr}
\bibliography{References.bib}


\end{document}